\documentclass[11pt,a4paper]{article}

\pdfoutput=1
\usepackage{jheppub}
\usepackage{simplewick}
\usepackage{thumbpdf}
\usepackage{amsmath}
\usepackage{amssymb}
\usepackage{array}
\usepackage{cancel}
\usepackage[hang,small]{caption2}
\usepackage{subfigure}
\usepackage{verbatim}
\usepackage{slashed}
\usepackage{graphicx}
\usepackage{url}
\usepackage{amsfonts}
\usepackage{multirow}
\usepackage{rotating}
\usepackage{adjustbox}
\usepackage{verbatimbox}
\usepackage{wasysym}

\makeatletter
\def\maketag@@@#1{\hbox{\m@th\normalfont\normalsize#1}}
\makeatother

\newenvironment{rcases}
  {\left.\begin{aligned}}
  {\end{aligned}\right\rbrace}

\newcommand{\p}{\partial}
\newcommand{\Tr}{{\text{Tr}}}
\newcommand{\tr}{{\text{\,tr}}}
\newcommand{\sym}{${\cal N}=4$ SYM  }

\newlength{\blockwidth}
\AtBeginDocument{\settowidth{\blockwidth}{0000}}

\title{Exact OPE data for short operators  in  double-scaled \(\gamma\)-deformed \(\mathcal{N}=4\) SYM}

\author[a,b]{Vladimir Kazakov,}
\author[c]{Enrico Olivucci,}
\author[a,b]{Michelangelo Preti}
 
\affiliation[a]{Laboratoire de Physique Th\'{e}orique de l'\'{E}cole Normale Sup\'{e}rieure, 24 rue Lhomond,
F-75231 Paris Cedex 05, France}
\affiliation[b]{Universit\'{e} Paris-VI, PSL Research University, Sorbonne Universit\'{e}s, UPMC Univ. Paris 06, CNRS, 75005 Paris, France } 
\affiliation[c]{II. Institut f\"ur Theoretische Physik, Universit\"at Hamburg, Luruper Chaussee 149, 22761
Hamburg, Germany}

\emailAdd{kazakov@lpt.ens.fr}  
\emailAdd{enrico.olivucci@desy.de} 
\emailAdd{michelangelo.preti@lpt.ens.fr}

\abstract{We study the Feynman graph structure and compute certain exact four-point correlation functions   in chiral CFT\(_4\)  proposed by \"{O}.~G\"{u}rdo\u{g}an and one of the authors as a double scaling limit of \(\gamma\)-deformed  \(\mathcal{N}=4\) SYM theory. We give full description of  bulk behavior of large Feynman graphs: it shows a generalized ``dynamical fishnet" structure, with a dynamical exchange of bosonic and Yukawa couplings.   We  compute certain four-point correlators in the full chiral CFT\(_4\), generalizing recent results for a particular one-coupling version of this theory -- the bi-scalar "fishnet" CFT. We sum up exactly the corresponding Feynman diagrams, including both bosonic and fermionic loops, by  Bethe-Salpeter method. This provides explicit  OPE data for various twist-2 operators with spin, showing a rich analytic structure, both in  coordinate and coupling spaces.        } 

\title{Exact four-point correlators of \(\gamma\)-twisted \(\mathcal{N}=4\) SYM in the double-scaling limit}

\title{ Generalized Fishnets and Exact Four-Point Correlators in Chiral CFT\(_4\)}

\date{}
\begin{document}

\maketitle

\section{Introduction}
\label{intro}

Quantum conformal field theories in various space-time dimensions attracted recently a considerable attention, due to their phenomenological importance in physics,  for  subjects  ranging from  the description of critical phenomena to the fundamental interactions beyond the Standard Model, but also due to their beautiful mathematical structure allowing to get a deep insight into the basic features of   Quantum Field Theory and, via AdS/CFT duality, of Quantum  Gravity. In spite of the considerable simplifications in the properties of  CFTs w.r.t. the massive QFTs,  the non-perturbative structure of strongly interacting CFTs in \(d>2\) dimension is very complicated and in general not very well studied analytically. A considerable progress in this direction has been achieved due to the conformal bootstrap methods \cite{Rattazzi:2008pe,ElShowk:2012ht} based on the basic properties of CFTs following from the conformal symmetry, such as crossing symmetry in various channels for the four-point correlation functions. But this approach stays to a great extent ``experimental", based on heavy numerical computations  rather than on  explicit analytic formulation of the final results. A great progress in the  understanding of analytic structure of CFTs in \(d>2\) dimensions has been achieved for  various superconformal QFTs, often due to the AdS/CFT correspondence. In a special case  -- the \sym-- the analytic study of OPE data was greatly advanced  due to the planar  integrability of the theory~\cite{Beisert:2010jr}. In particular, the spectral problem -- exact, all-loop calculation of anomalous dimensions of local operators -- found its  ultimate formulation in terms of the Quantum Spectral Curve (QSC)~\cite{Gromov:2013pga,Gromov:2014caa} -- a system of algebraic relations on Baxter-type \(Q\)-functions, supplied by analyticity properties and Riemann-Hilbert monodromy conditions (see recent reviews~\cite{Gromov:2017blm,Kazakov:2018hrh}). 

The integrability appears to persist for a class of 3-parameter \(\gamma\)-deformations of the \(R\)-symmetry of \sym~\cite{Beisert:2005if,Frolov:2005dj,Kazakov:2015efa} if one tunes the so-called double-trace terms, generated by the RG of the model, to their critical, in general complex values~\cite{Sieg:2016vap,Grabner:2017pgm}. The  \(\gamma\)-deformed \sym appears to be a family of non-sypersymmetric and non-unitary four-dimensional CFTs labeled by 't~Hooft coupling \(g\) and three \(\gamma\)-deformation     angles \(\gamma_j,\,\,j=1,2,3\). The OPE data of \sym has been studied in numerous papers, using the  integrability properties, as well as AdS/CFT correspondence for the strong coupling regime \(g\to\infty\), or a direct Feynman graph calculus at weak coupling \(g\to 0\). Apart from the spectral problem,  an impressive progress has been achieved in a more difficult problem of computation of structure constants and correlation functions~\cite{Basso:2015zoa,Fleury:2016ykk,Eden:2016xvg,Coronado:2018cxj}, as well as of \(1/N_c^2\) corrections~\cite{Bargheer:2017nne}. However, the efficient all-loop solution of these problems is still hindered by outstanding technical complexity.  We also have to admit that integrability of \sym\ is still a somewhat mysterious phenomenon, not very well understood, especially on the CFT side of this AdS\(_5\)/ CFT\(_4\) duality. 

In 2015, one of the authors and \"O.~G\"urdogan proposed a family of non-unitary,  non-supersymmetric CFTs~\cite{Gurdogan:2015csr}, based on a special double scaling limit of \(\gamma\)-deformed \sym\ combining weak coupling limit of small 't~Hooft coupling, \(g\to 0\), and strong imaginary twist, \(\gamma_j\to i\infty \), with three finite effective couplings \(\xi_j=ge^{-i\gamma_j/2}\). The gauge fields and the gaugino decouple in this limit and one is left with three complex scalars and three complex fermions with certain chiral structure of interactions (see the Lagrangian of the theory~\eqref{chiFT4},\eqref{fullL}). These  CFTs, on the one hand, helps to   shed some light on the origins of integrability in \sym, and on the other hand, the double scaling limit significantly facilitates the computations of  interesting physical properties, such as OPE data and certain multi-loop Feynman graphs, 
revealing   rich and instructive dynamical properties of the theory. It was further studied  in~\cite{Caetano:2016ydc}, in particular, by the asymptotic Bethe ansatz methods. This full three-couplings double scaled version of~\sym was dubbed in~\cite{Caetano:2016ydc} the chiral CFT, or, shortly, \(\chi\)CFT.  We will employ this name in what follows.

In the single coupling reduction, \(\xi_1=\xi_2=0,\,\,\xi_3\ne 0\), the theory reduces to two interacting complex scalar matrix fields~(see eq.\eqref{bi-scalarL}). The planar Feynman graphs for typical  physical quantities in such a bi-scalar theory appear to have, at least in the balk, the fishnet structure where the massless scalar propagators form a regular quadratic lattice~\cite{Gurdogan:2015csr}. This theory will be called in what follows the bi-scalar, or fishnet CFT. The fishnet graphs of simple shape, such as a torus,  appear to represent an  integrable statistical mechanical system~\cite{Zamolodchikov:1980mb}.  Remarkably,  there exists also an  integrable generalization of the Fishnet CFT to any dimension \(d\)~\cite{Kazakov:2018qbr}.
          
 Many results recently obtained for the biscaler fishnet CFT, would be too difficult to achieve for the analogous quantities in the full \(\gamma\)-deformed \sym. Among the studied quantities are  anomalous dimensions of the operators \(\Tr[\phi_1^L]\)  dominated by   wheel-type fishnet graphs. They were computed explicitly, in terms of MZV values, at two wrappings (up to \(2L\) loops for any \(L\)~\cite{Ahn:2011xq,Gurdogan:2015csr}) and, iteratively, to\ any loop order for \(L=3\).   Another remarkable example of exact computations, unique in \(d>2 \) CFTs,  are the all-loop four-point correlation functions of the shortest protected operators
 \cite{Grabner:2017pgm,Gromov:2018hut,Kazakov:2018hrh}.  The biscalar fishnet CFT gives a unique opportunity of the study the single-trace multi-point correlators and of the related  exact planar scalar amplitudes, revealing their explicit and well-defined Yangian symmetry~\cite{Chicherin:2017cns,Chicherin:2017frs}. One is even able to compute exactly, using the above mentioned exact four-point correlators, the simplest non-planar (\(\sim1/N_c^2\)) scattering amplitude~\cite{Korchemsky:2018hnb}~(see also~\cite{Ben-Israel:2018ckc} for the perturbative study of this amplitude).   

All this shows that  this integrable  theory resulting from the double-scaling limit  of planar \(\gamma\)-deformed \sym allows a unique insight   into the non-perturbative structure of strongly interacting CFTs and a closer look at them could reveal many general properties of CFTs in \(d>2\) dimensions. It is also worth mentioning the existence of 3 dimensional analogue of these CFTs, obtained by a similar limit from the three-dimensional \(\gamma\)-deformed ABJM model~\cite{Caetano:2016ydc} dominated by fishnet graphs with regular triangular structure, as well as  the  \(6d\) version of fishnet CFT~\cite{Mamroud:2017uyz}, where the fishnet graphs have a regular hexagonal structure. The ``bulk'' integrability of all three cases of  regular fishnet planar graphs was predicted in~\cite{Zamolodchikov:1980mb}. 

Whether as a big progress already has been done in the  study of the  bi-scalar fishnet CFT has been done, little is known about the most general version of the   double-scaled  \(\gamma\)-deformed \sym\ mentioned above. Until very recently, apart from the original formulation~\cite{Gurdogan:2015csr} and the study, in ~\cite{Caetano:2016ydc}, of asymptotic Bethe ansatz equations for anomalous dimensions in certain sectors of this very interesting theory, as well as the computations of related unwrapped and single-wrapped Feynman graphs,  no serious attempts had been undertaken, at least until very recently, to understand deeper the physical properties and the Feynman graph structure of the full \(\chi\)CFT. It is worth noticing that, unlike the bi-scalar fishnet CFT, the reasons for the integrability of this model remain mysterious.

A few days before the completion of the current paper, a very interesting study of
the one loop perturbative properties of this \(\chi\)CFT was undertaken~\cite{Ipsen:2018fmu}, and especially of its two reductions: the bi-scalar fishnet CFT, as well as   \(\beta\)-deformed \({\cal\ N}=1\) supersymmetric case, when all three couplings are equal~\cite{Caetano:2016ydc}. The paper explores an interesting subject of study of non-unitary spin chains, having a rich and complicated structure of the spectrum, including the Jordan cells as specific multiplets of states. The Jordan multiplets leading to the logarithmic behavior in non-unitary CFT's~\cite{Gurarie:1993xq}, are noticed and studied perturbatively in the fishnet CFT~\cite{Caetano:TBP,Gromov:2017cja}. The notion of one-loop integrability in \(\chi\)CFT appears to be quite different from the one-loop integrability of its mother theory -- the \sym~\cite{Minahan:2002ve,Beisert:2003yb}. 

The non-unitarity of the studied  \(\chi\)CFT represents an obvious drawback from the point of view of the physical interpretations: the presence of complex  OPE data violating various basic quantum-mechanical axiomes and usual analyticity constraints.  On the other hand, the non-unitary theories are  curious objects  in themselves, having interesting OPE properties, such as a logarithmic behaviour of certain correlators \(\chi\)CFT is an example of  logarithmic CFTs). In addition, they share many basic common features with unitary CFTs and help to understand their general features.  

We attempt in this paper to answer some of the questions posed above about the \(\chi\)CFT. First of all, we will give the complete   description of the  bulk structure of Feynman graphs (far from their boundaries defined by the particular underlying physical quantities). It appears to be much richer than in the fishnet CFT, though much simpler than in the full  \sym conserving a certain lattice regularity. A pictorious way to describe  these graphs is to  introduce the regular triangular lattice and the to do all possible Baxter moves of all three types of lines, as shown on Fig.\ref{fig:patched_fishnet}. These lines should represent sequences  of bosonic and fermionic propagators and the mixed  intersections (where  both  bosonic and fermionic propagators meet) should be disentangled, in a unique way, into pairs of Yukawa vertices). These configurations should be summed up, so that the collection of such graphs could be called the ``dynamical fishnet". The integrability of these graphs, or the sum of them, remains to be proved, though we demonstrate it in this paper in a simpler case of the two-coupling reduction of \(\chi\)CFT~(see eq.\eqref{DSprime}), with two bosonic and one Yukawa coupling.  

  Then we will compute exactly the 4-point correlation functions of certain short, protected scalar operators, similar those obtained in fishnet CFT~\cite{Grabner:2017pgm,Gromov:2018hut,Kazakov:2018hrh}. For that we identify all the graphs contributing these quantities and sum them up using the Bethe-Salpeter approach helped by the conformal invariance. In comparison  to the fishnet CFT, the two-coupling dependence of these correlators in the full double-scaled CFT reveals a rich  phase structure in the coupling space.  We study in detail the related perturbative expansions of these correlators, as well as their strong coupling limits.

\section{Feynman graphs and correlators of  \(\chi\)CFT  -- the strongly \(\gamma\)-deformed \(\mathcal{N}\)=4 SYM theory }\label{sec:chiCFT}

In this section, we will study the generic structure of planar Feynman graphs and discuss their integrability properties, in the full three-coupling chiral CFT (\(\chi\)CFT)   proposed in \cite{Gurdogan:2015csr}~(see also ~\cite{Caetano:2016ydc} for more details).   

This CFT was obtained as a double scaling limit of \(\gamma\)-twisted \sym
 described above. It is defined by the Lagrangian for three complex scalars and three complex fermions transforming in the adjoint representation of \(SU(N_c)\):
\begin{equation}\label{chiFT4}
  {\cal L}_{\phi\psi}=N_c\Tr\left(-\frac{1}{2}\p^\mu\phi^\dagger_j\p_\mu\phi^j
  +i\bar\psi^{\dot\alpha}_{ j}
  (\tilde\sigma^{\mu})^\alpha_{\dot\alpha}\p_\mu \psi^j_{\alpha }\right)
+{\cal L}_{\rm int}\,,
\end{equation}
where
the sum  is taken with respect to all doubly repeated indices, including \(j=1,2,3\), and the interaction part is 
\begin{equation}
\begin{aligned}
     \mathcal{L}_{\rm int} ={}N_c
\,\Tr\Bigl[\xi_1^2\,\phi_2^\dagger \phi_3^\dagger &
\phi^2\phi^3\!+\!\xi_2^2\,\phi_3^\dagger \phi_1^\dagger 
\phi^3\phi^1\!+\!\xi_3^2\,\phi_1^\dagger \phi_2^\dagger \phi^1\phi^2\!+\!i\sqrt{\xi_2\xi_3}(\psi^3 \phi^1 \psi^{ 2}+ \bar\psi_{ 3} \phi^\dagger_1 \bar\psi_2 )\\
& +i\sqrt{\xi_1\xi_3}(\psi^1 \phi^2 \psi^{ 3}+ \bar\psi_{ 1} \phi^\dagger_2 \bar\psi_3 )
 +i\sqrt{\xi_1\xi_2}(\psi^2 \phi^3 \psi^{ 1}+ \bar\psi_{ 2} \phi^\dagger_3 \bar\psi_1 )\,\Bigr].  
\label{fullL}\end{aligned}
\end{equation}
We suppressed in the last equation the spinorial indices assuming the scalar product of both fermions in each term. We will refer to this theory as \(\chi\)CFT theory.

The double scaling procedure and the derivation of this action from \(\gamma\)-deformed \sym\  can be found in papers~\cite{Fokken:2013aea},\cite{Caetano:2016ydc}.
In the next sections, we will study the four-point functions  obtained by point splitting of fields in coinciding points, in the two-point correlation functions of local operators of three types: 
\begin{align} \label{operators}
\text{Tr}[\phi_j^2(x)]\quad (j=1,2,3),\qquad  \text{Tr}[\phi_j\phi_k(x)]\quad (i>j),\qquad \text{Tr}[\phi_j\phi_k^\dagger(x)]\quad (i\ne j).
\end{align}     

Since the Lagrangian \eqref{fullL} depends on three arbitrary couplings, For some particular values of these couplings,  interesting reductions of this \(\chi\)CFT emerge. For example, in the limit $\xi_1\rightarrow 0$, one fermion decouples and we obtain the following action~~\cite{Caetano:2016ydc}
  \begin{equation}
\label{DSprime}
     \mathcal{L}_{\rm int} ={}N_c
\,\Tr\Bigl(\xi_3^2\,\phi_1^\dagger \phi_2^\dagger 
\phi^1\phi^2+\xi_2^2\,\phi_3^\dagger \phi_1^\dagger 
\phi^3\phi^1
 +i\sqrt{\xi_2\xi_3}(\psi^2 \phi^1 \psi^{ 3}+ \bar\psi_{ 2} \phi^\dagger_1 \bar\psi_3 )\,\Bigr).  \\
\end{equation}
We will refer to this theory as \(\chi_0\)CFT theory.
Another interesting case of~\eqref{fullL} occurs when all three  couplings are equal $\xi_1=\xi_2=\xi_3=\xi$ and corresponds to the doubly-scaled $\beta$-deformed SYM \cite{,Leigh1995,Lunin:2005jy}. It has the following interaction Lagrangian~\cite{Caetano:2016ydc}
\begin{equation}
\begin{split}\label{betadef}
     \mathcal{L}_{\rm int} &={}\xi^2N_c
\,\Tr\Bigl(\phi_2^\dagger \phi_3^\dagger 
\phi^2\phi^3+\phi_3^\dagger \phi_1^\dagger 
\phi^3\phi^1+\phi_1^\dagger \phi_2^\dagger \phi^1\phi^2\Bigr)\\
 &+i\xi N_c \Tr\Bigl(\psi^3 \phi^1 \psi^{ 2}+ \bar\psi_{ 3} \phi^\dagger_1 \bar\psi_2 
 +\psi^1 \phi^2 \psi^{ 3}+ \bar\psi_{ 1} \phi^\dagger_2 \bar\psi_3 
 +\psi^2 \phi^3 \psi^{ 1}+ \bar\psi_{ 2} \phi^\dagger_3 \bar\psi_1 \,\Bigr). 
\end{split}
\end{equation}
In this case, one  supersymmetry is left unbroken, as in the original \(\beta\)-deformed \({\cal\ N}=1\) SYM.

Most of the papers on this relatively young subject were devoted to the abovementioned single coupling reduction  of this model:   \(\xi_1=\xi_2=0,\quad\ \xi_3\equiv\xi\ne 0\), i.e.  the \textit{bi-scalar}, fishnet CFT defined by the action     ~\cite{Gurdogan:2015csr}:      
\begin{equation}
    \label{bi-scalarL}
    {\cal L}_{\phi}= \frac{N_c}{2}\Tr
    \left(\p^\mu\phi^\dagger_1 \p_\mu\phi^1+\p^\mu\phi^\dagger_2 \p_\mu\phi^2+2\xi^2\,\phi_1^\dagger \phi_2^\dagger \phi^1\phi^2\right)\,.
  \end{equation}   
Our paper is devoted to generalization of some of these results of \cite{GrabnerGromovKazakovKorchemsky,Gromov:2018hut} to the full chiral model --  \(\chi\)CFT \eqref{fullL} and to its various limits presented above.    
This represents a step forward, w.r.t. the bi-scalar model \eqref{bi-scalarL}, in understanding the non-perturbative structure of physical quantities of the full \sym.

We will also describe the general bulk structure of the underlying planar graphs.
\begin{figure}[!t]
     \hspace{-0.8mm}  
     \subfigure
   {\includegraphics[width=3.0cm]{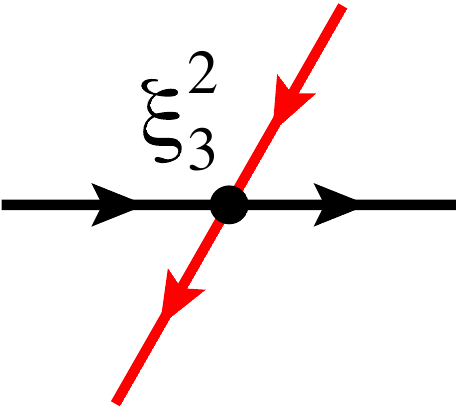}}
    \hspace{27.7mm} 
     \subfigure
   {\includegraphics[width=1.82cm]{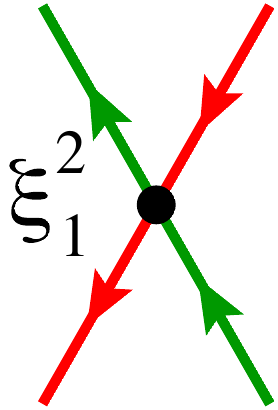}}
    \hspace{30.1mm} 
     \subfigure
   {\includegraphics[width=3.0cm]{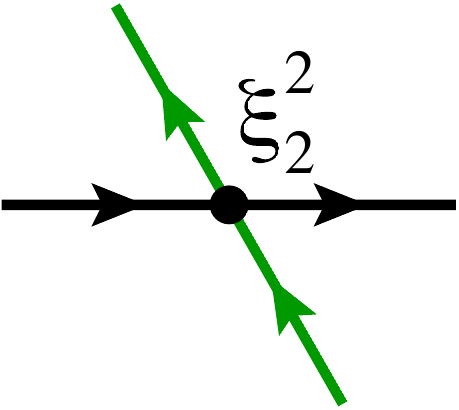}}\\
     \subfigure
   {\includegraphics[width=4.0cm]{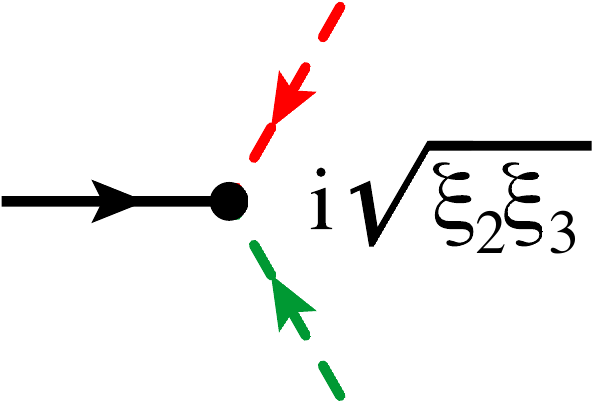}}
    \hspace{13.8mm} 
     \subfigure
   {\includegraphics[width=4.0cm]{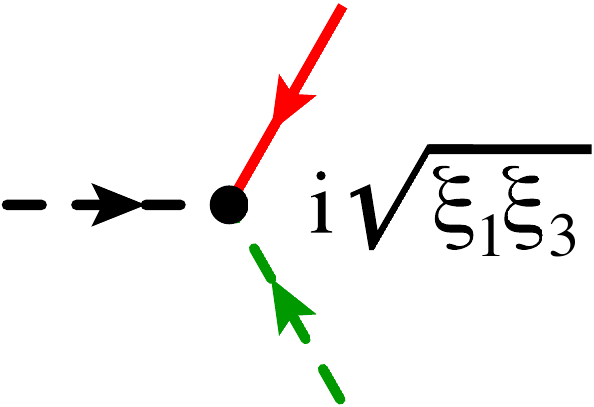}}
    \hspace{13.8mm} 
     \subfigure
   {\includegraphics[width=4.0cm]{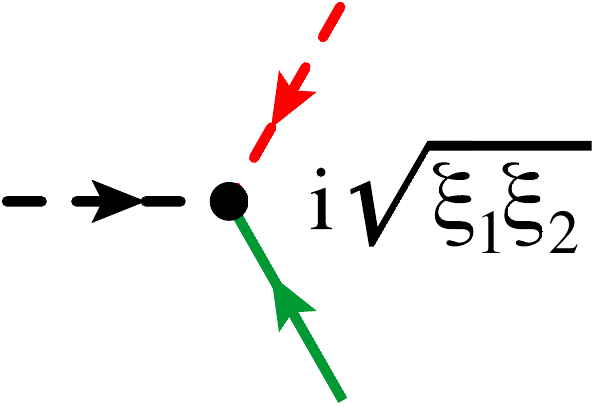}}
 \caption{The chiral vertices of the DS theory \eqref{fullL}. The graphs of the first line represent the quartic scalar interactions and the ones in the second line are the Yukawa interactions. Tick solid lines and dashed lines represent scalar and fermionic propagators respectively. Colors stand for the various "flavour" of the particles $\phi^i$ and $\psi^i$: black for $i=1$, red for $i=2$ and green for $i=3$. Arrows symbolize the fixed orientation (chirality) of the vertices and, according to our notation, it points always to the fields with bars or daggers. The second chirality of Yukawa interactions \textit{i.e.} the one with $\bar{\psi}_i\rightarrow\psi^i$ and $\phi^\dagger_i\rightarrow\phi^i$ with $i=1,2,3$, can be represented as the second line of vertices with flipped arrows.}
  \label{fig:vertices}
 \end{figure}
Indeed, one interesting feature of those models is the drastic simplification of their weak coupling expansions in terms of Feynman diagrams in the planar limit. 
In general, any diagram of \(\chi\)CFT  can be built as a collection of the vertices in Fig.\ref{fig:vertices}, connected by scalar and fermionic propagators\footnote{Apart from the double-trace vertices~\cite{Tseytlin:1999ii,Dymarsky:2005uh} whose role will be discussed below}. The arrows indicate the fixed orientation (chirality) of the interactions, i.e. in a  propagator it is directed from a field to its hermitian  conjugate. An essential feature of \eqref{fullL} is the absence of the hermitian conjugate of every interaction vertex, or of the vertices obeying the reality condition. The chirality of this theory makes it non-unitary and plays a crucial role for the underlying conformality and integrability in the 't Hooft limit. In fact, in the absence of the hermitian conjugate vertices, all the graphs which could renormalize the couplings and the mass are non-planar. As  a consequence,  we will see  in sec.\ref{sec:section2} that  the planar weak coupling expansion of physical quantities w.r.t. interactions \eqref{fullL} in \(\chi\)CFT   is dominated, at least in the bulk and for high enough perturbative order, by a specific class of planar diagrams having a kind of a lattice structure, much more rigid than the structure of graphs in the original \sym. This lattice structure is richer, and more ``dynamical" than in the bi-scalar theory where the unique regular square fishnet structure dominates at any order in perturbation theory. In the full  \(\chi\)CFT, due to the presence of Yukawa interactions and quartic scalar vertices, there are more planar graphs contributing at each perturbative order, but the chirality still dramatically reduces their number.
We can dubb the structure of full \(\chi\)CFT graphs as ``dynamical fishnet".
\subsection{Double-trace interactions and conformal symmetry}

The $\gamma$-deformed $\mathcal{N}=4$ SYM theory and its doubly-scaled version are not conformal in a strict sense, not even in the planar limit \cite{,Fokken:2013aea}.  Indeed, the renormalization group calculations show~\cite{Fokken:2014soa} that the new, scalar double-trace interactions are generated
\begin{equation}\label{doubletraces}
\mathcal{L}_{\text{dt}}\!=\!(4\pi)^2\!\sum_{j=1}^3\!\!\left[\alpha_{1,j}^2\text{Tr}[\phi_j\phi_j]\text{Tr}[\phi_j^\dagger\phi_j^\dagger]\!+\!
\alpha_{2,j}^2\text{Tr}[\phi_j\phi_{j_+}^\dagger]\text{Tr}[\phi_j^\dagger\phi_{j_+}]\!+\!
\alpha_{3,j}^2\text{Tr}[\phi_j\phi_{j_+}]\text{Tr}[\phi_j^\dagger\phi_{j_+}^\dagger]\right]\, ,
\end{equation}  
where in our notation $j_{+}=j+1$ with the constraint $3_{+}=1$. The double-trace couplings \(\alpha_{k,j}\) generically flow with the scale. They are needed to renormalise the 2-point correlators of the local operators \(\text{Tr}[\phi_j\phi_{j}],\,\text{Tr}[\phi_j\phi_{j_+}^{\dagger}]\) and \(\text{Tr}[\phi_j\phi_{j_+}]\) respectively. For any of these planar correlators only one double-trace term contributes, that is the \(\beta\)-function of each \(\alpha_{k,j}\) depends only on couplings \(\{\xi_1,\xi_2,\xi_3\}\) and  \(\alpha_{k,j}\) itself. Due to permutation symmetry of flavour indices \(j=1,2,3\) in the Lagrangian \eqref{chiFT4}, the functions \(\beta_{\alpha_{k,j}}\) show the same symmetry in the coupling dependance, namely
\begin{align}
\beta_{\alpha_{k,j}}(\alpha_{k,j},\xi_j,\xi_{j_+},\xi_{j_-})\,=\,\beta_{\alpha_{k,j'}}(\alpha_{k,j'},\xi_{j'},\xi_{j'_+},\xi_{j'_-})\, ,
\end{align}
thus will drop in what  follows the specification of subscript \(j\) in double-trace couplings. The double-trace terms \eqref{doubletraces} appear in the theory already at one-loop renormalization and the $\beta$-functions associated to the couplings $\alpha_k^2$ are not zero.
In $\gamma$-deformed \sym the one-loop $\beta$-function  associated to the double-trace interaction $\alpha_1^2\,\text{Tr}[\phi_j\phi_j]\text{Tr}[\phi_j^\dagger\phi_j^\dagger]$ of \eqref{doubletraces} is \cite{Fokken:2014soa}\begin{equation}\label{betaphi2}
\beta_{\alpha_k}= \frac{g^4}{\pi^2}\sin^2\gamma_k^+\sin^2\gamma_k^- +4^3\pi^2\alpha_k^4\,+{\cal O}(g^6,\alpha_k^6)\, ,
\end{equation}
where \(\gamma^{\pm}_k\) are linear  combinations of the deformation parameters \(\gamma_j\) of the theory defined in \eqref{gammapm}.
Let us turn to the theory  \eqref{fullL} with the double-trace terms \eqref{doubletraces}. In contrast to the bi-scalar theories, where the invariance under exchange
\begin{align}
\begin{cases}
\phi_j \longrightarrow \phi_{j_+}\\
\phi_{j_+} \longrightarrow \phi_{j}^{\dagger}
\end{cases}
\end{align}
allows to identify \(\alpha_2\) and \(\alpha_3\), the presence of Yukawa interactions in \(\chi\)CFT specifically breaks this symmetry, and operators \(\text{Tr}[\phi_j\phi_{j_+}^{\dagger}]\) and \(\text{Tr}[\phi_j\phi_{j_+}]\) show different behaviour.
 When only one $\alpha_k$ coupling is running, the corresponding $\beta$-function has the following form
\begin{equation}\label{betageneral}\beta_{\alpha_k}=a(\bold\xi)+b( {\xi})\,\alpha_k^2+c({\xi})\,\alpha_k^4\, ,
\end{equation} 
where $a,b,c$ are functions of the couplings ${\xi}=\{\xi_1,\xi_2,\xi_3\}$. This quadratic behavior of $\beta$ as a function of $\alpha_k^2$ was encounter for the first time in \cite{Dymarsky:2005uh} as an example of non-supersymmetric orbifold theories with double-trace interactions and established in \cite{,Pomoni:2008de} for a generic deformed theory in the 't Hooft limit. 
If the running coupling $\alpha_k$ is associated to the double trace interaction $\text{Tr}\mathcal{O}\text{Tr}\mathcal{O}^\dagger$ of length-two scalar operators $\mathcal{O}$, the functions $a$, $b$ and $c$ are related to the normalization coefficient of the two-point function of $\mathcal{O}$, the contribution of the single-traces to the anomalous dimension of $\mathcal{O}$ and the coefficient of the induced double-trace terms.

To make the theory conformal at the quantum level, one needs to tune the double-trace couplings to a fixed point. In the original \(\gamma\)-deformed \sym, the 't~Hooft coupling \(g^2\) \ is not running, so the critical (conformal) point for double-trace couplings  can be computed imposing the vanishing of their \(\beta\)-functions. In the case of  a single running coupling, \eqref{betaphi2}  has the following fixed points
\begin{equation}\label{fix1}
\alpha^2_{k\star}=\pm \frac{ig^2}{8\pi^2}\sin\gamma_k^+\sin\gamma_k^-+\mathcal{O}(g^4)\, .
\end{equation} 

Similarly,  the coupling constants $\xi_i$ of the theory \eqref{fullL} are not running in the 't Hooft limit and one can fine-tune the double-trace couplings $\alpha_i^2$ to critical values in terms of their $\xi_i$ dependence, imposing the vanishing of the underlying $\beta$-function \eqref{betageneral} as follows
\begin{equation}\label{fixedpoint}
\beta_{\alpha_k}\overset{!}{=}0\qquad\Rightarrow\qquad(4\pi)^2\alpha^2_{k\star}=-\frac{b\pm\sqrt{b^2-4ac}}{2c}\,.
\end{equation}
   At the two fixed points \eqref{fixedpoint}, it is possible to write the anomalous dimension  $\gamma_\star$  of the operator $\mathcal{O}$ in terms of the discriminant of $\beta_{\alpha_k}=0\)~\cite{Pomoni:2008de} 
\begin{equation}\label{gammafixedpoint}
4\gamma^2_{\mathcal{O}\star}=b^2-4a c\,.
\end{equation}
 At the fixed points for all double-trace couplings \eqref{doubletraces} of \(\gamma\)-deformed \sym, the theory becomes a genuine non-supersymmetric CFT. This conformal theory appears also to be integrable ~\cite{Kazakov:2015efa,Grabner:2017pgm,Frolov:2005dj} and its spectrum of anomalous dimensions can be treated by such a powerful tool as quantum spectral curve (QSC)~\cite{Gromov:2013pga,Gromov:2014caa,Kazakov:2015efa}. The same statements hold for the double-scaling limit of the \(4D\) \(\chi\)CFT theory \eqref{fullL}, to which we have to add the double-trace Lagrangian \eqref{doubletraces}.
  Integrability of the full \(\chi\)CFT is still a conjecture, as it is for the full \(\gamma\)-deformed \sym. It was demonstrated explicitly  only for the simplest reduction of  \(\chi\)CFT -- the bi-scalar  CFT~\eqref{bi-scalarL}, where the fishnet planar graphs have an iterative regular lattice structure~\cite{Gurdogan:2015csr}, shown to be integrable long ago by A.Zamolodchikov ~\cite{Zamolodchikov:1980mb}~(see also~\cite{Gromov:2017cja}). We extended the proof of integrability to a larger, two-coupling sector of \(\chi\)CFT in Sec.\ref{wheels}, by methods of conformal \(SU(2,2)\) quantum spin chain. In the case of \(\chi\)CFT we also have good chances to prove full integrability on the level of planar Feynman graphs since, as we show below, these graphs preserve a certain rigid lattice structure.  
   
The obvious physical defect of such CFTs is the loss of unitarity. Indeed, as it will be clear with the explicit example below, the discriminant of the equation $\beta_{\alpha_i}=0$ is negative, inducing complex values for the fixed points \eqref{fixedpoint} and anomalous dimension \eqref{gammafixedpoint}. Moreover in the AdS/CFT context, this fact can be interpreted as the presence of true tachyons in the bulk on the string theory side \cite{,Pomoni:2008de}.

The one-loop anomalous dimension of the length-two operator \(\text{Tr}[\phi_j\phi_j]\) in \(\gamma\)-deformed \sym at the fixed point is~\cite{Fokken:2014soa}
\begin{equation}\label{an1}
\gamma_{\phi_j\phi_j\star}=\mp \frac{ig^2}{2\pi^2}\sin\gamma^+_j\sin\gamma^-_j+\mathcal{O}(g^4).
\end{equation}
Notice that both the fixed points \eqref{fix1} and the anomalous dimensions \eqref{an1} are complex conjugate, as expected. Those relations are actually valid in the full $\gamma$-deformed \sym theory, but in the double-scaling limit under analysis it is simple to obtain some predictions for the one-loop $\beta$, the associated critical points and the anomalous dimensions. In particular we have
\begin{equation}\label{predictionspectrum}
\gamma_{\phi_j\phi_j\star}\overset{\text{DS limit}}{=}\mp2i(\xi_{j_+}^2\!\!-\xi_{j_-}^2)+\dots
\qquad\text{and}\qquad
\alpha^2_{1\star}\overset{\text{DS limit}}{=}\pm\,i\frac{\xi_{j_+}^2\!\!-\xi_{j_-}^2}{2}+\dots\,.
\end{equation}   
In Sec.\ref{sec:spectrumphi1phi1} and Sec.\ref{sec:feynman11} we will verify these results computing the exact spectrum of the operator  $\text{Tr}[\phi_j\phi_j]$ with the Bethe-Salpeter method, and the first order of the fixed point $\alpha^2_{1\star}$ using Feynman diagrams.

Non-unitary CFTs are usually logarithmic~\cite{Gurarie:1993xq}, i.e. with an interesting, logarithmic behavior of certain correlators. The \(\gamma\)-deformed \sym~ and its double-scaled version -- the \(\chi\)CFT  \eqref{fullL} (and its reductions mentioned above) are not exceptions: they show the same logaritmic properties due to the non-hermiticity of their dilatation operators~\cite{Caetano:TBP,Gromov:2017cja}.~  
  
\subsection{The bulk structure of large planar graphs}\label{sec:section2}

\begin{figure}[!t]
 \centering
   {\includegraphics[width=.6\textwidth]{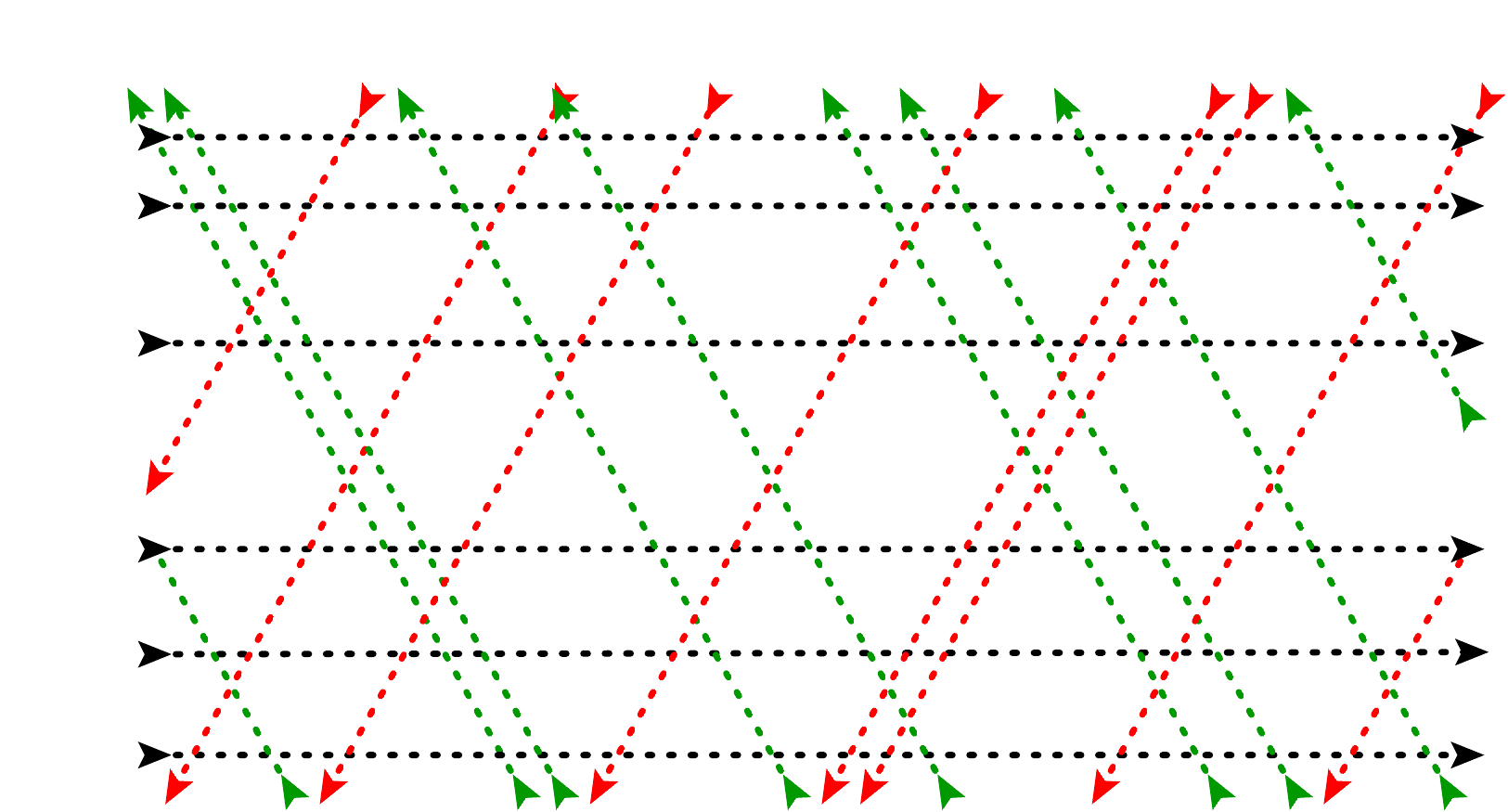}}
  \caption{
  General ``dynamical" fishnet bulk structure  of a planar graph  for 3-coupling chiral  CFT \eqref{fullL}.  Dotted lines represent scalar or fermionic propagators (the rules for the choice of propagators will be explained below and demonstrated in Fig.\ref{fig:bulk_DSlimit2}) The colors and directions of the lines stand for the three ``flavours" of the particles $i=1,2,3$ with the same notation as we used in Fig.\ref{fig:vertices}. The intersections correspond to six different effective vertices that  can be written in terms of the usual ones following the map given in Tab.\ref{tab:vert}.}
  \label{fig:patched_fishnet}
 \end{figure}
 
Let us try to describe the general structure of an arbitrarily big Feynman graph in the bulk, far from the boundaries. The generic picture is illustrated on Fig.\ref{fig:patched_fishnet}. The theory \eqref{fullL} contains 3 complex scalars $\phi^i$ and 3 complex fermions $\psi^i$ labelled by $i=1,2,3$. We chose to represent scalar propagators with thick solid lines and fermionic propagators with dashed lines (see Fig.\ref{fig:vertices}), while the label denoting their \(U(1)^{\otimes 3}\) flavour (see App.\eqref{charges_table}) is mapped into colours: \((1,2,3) \equiv (black, red, green)\). In Fig.\ref{fig:patched_fishnet}, coloured dotted lines in a particular direction represent a generic propagator, both scalar or fermionic. In this framework, a set of parallel lines represents any combination of fermionic and scalar propagators of a given flavour.
\begin{table}[!t]
\begin{center}
\begin{adjustbox}{width=14cm}
\begin{tabular}{c||c|c||c|c||c|c||}
\cline{2-7}
&  \multicolumn{2}{c||}{\includegraphics[scale=.4]{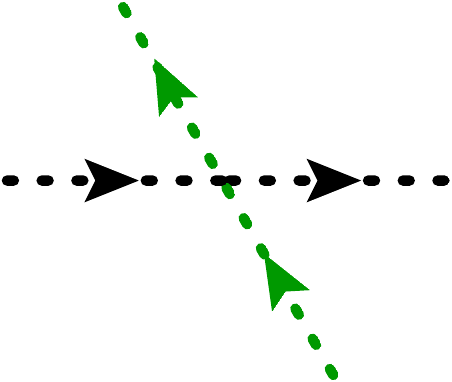}}&\multicolumn{2}{c||}{\includegraphics[scale=.4]{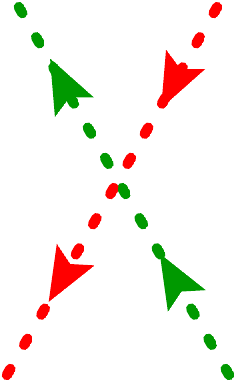}}&\multicolumn{2}{c||}{\includegraphics[scale=.4]{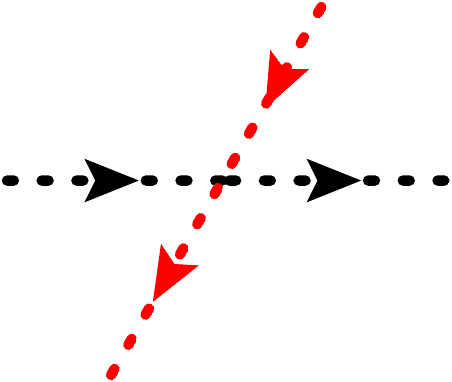}} \\
\cline{2-7}
 & Effective & Real &Effective & Real &Effective & Real \\ \cline{1-7}
\multirow{2}{*}{ \addvbuffer[2ex]{\begin{turn}{90}4-Scalar\end{turn}}}&
  \multicolumn{2}{c||}{$\xi_2^2$}&\multicolumn{2}{c||}{$\xi_1^2$}&\multicolumn{2}{c||}{$\xi_3^2$} \\
  &\includegraphics[scale=.3]{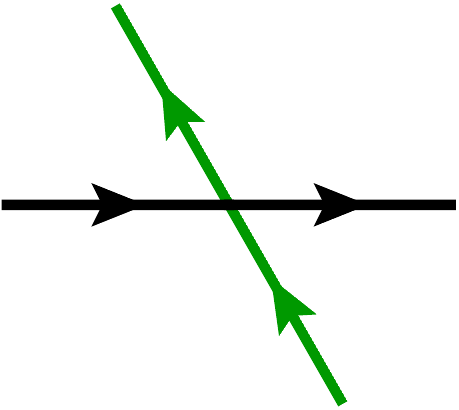} &\includegraphics[scale=.3]{V3Bos}&\includegraphics[scale=.3]{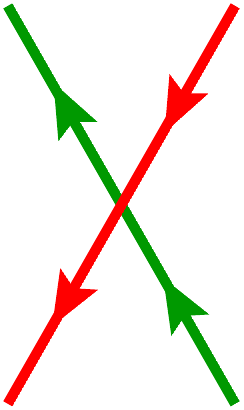} &  \includegraphics[scale=.3]{V1Bos}&\includegraphics[scale=.3]{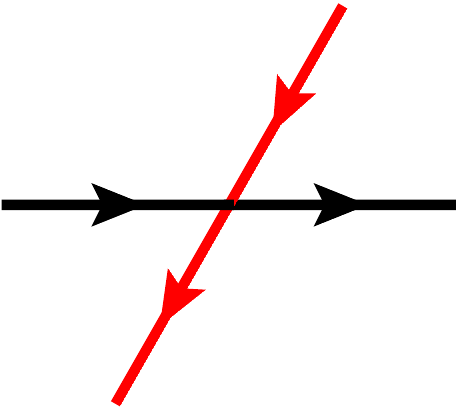} &  \includegraphics[scale=.3]{V2Bos} \\
\cline{1-7}
\multirow{2}{*}{ \addvbuffer[2.5ex]{\begin{turn}{90}4-Fermion\end{turn}}}&
  \multicolumn{2}{c||}{$\xi_1\xi_3$}&\multicolumn{2}{c||}{$\xi_2\xi_3$}&\multicolumn{2}{c||}{$\xi_1\xi_2$} \\
 & \includegraphics[scale=.3]{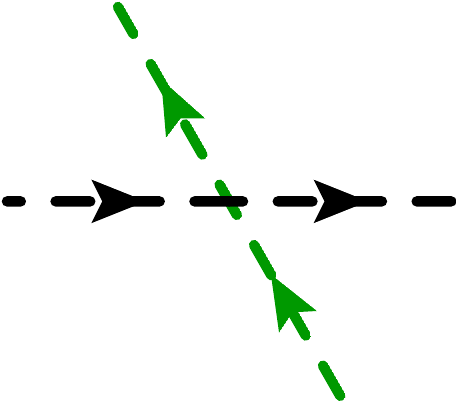} & \includegraphics[scale=.3]{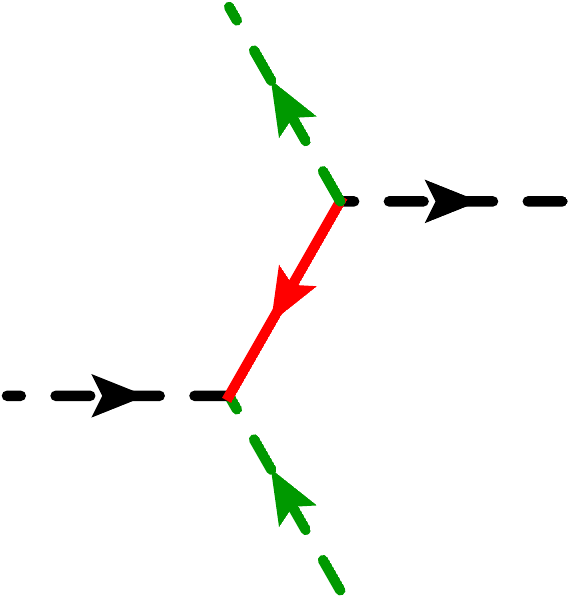}&\includegraphics[scale=.3]{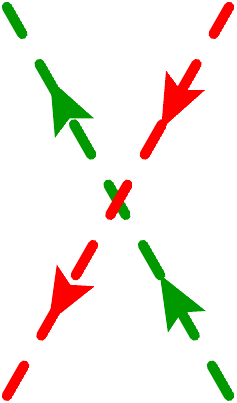} & \includegraphics[scale=.3]{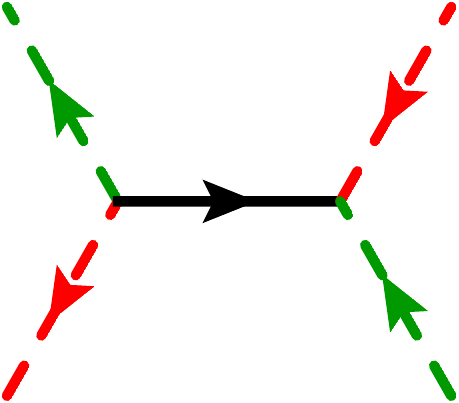}&\includegraphics[scale=.3]{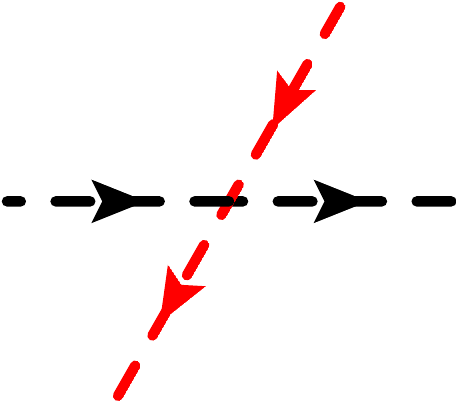} & \includegraphics[scale=.3]{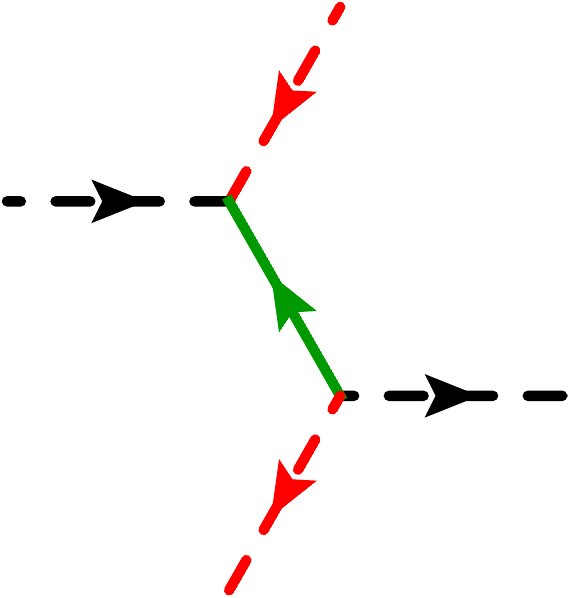}  \\
  \cline{1-7}
\multirow{4}{*}{ \addvbuffer[11ex]{\begin{turn}{90}Crossing\end{turn}}}&
  \multicolumn{2}{c||}{$\xi_2\xi_3$}&\multicolumn{2}{c||}{$\xi_1\xi_2$}&\multicolumn{2}{c||}{$\xi_1\xi_3$} \\
 & \includegraphics[scale=.3]{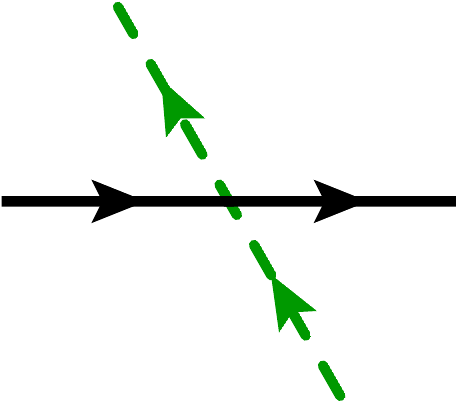} & \includegraphics[scale=.3]{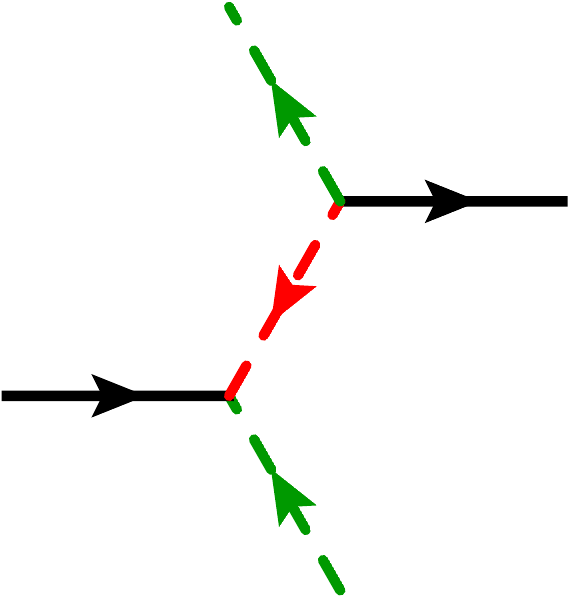} &\includegraphics[scale=.3]{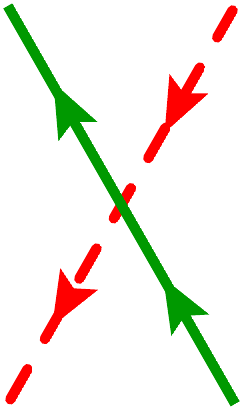} & \includegraphics[scale=.3]{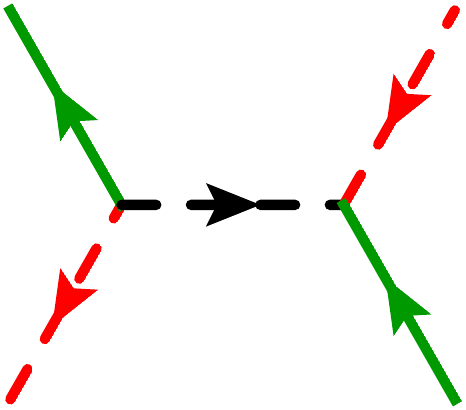} &\includegraphics[scale=.3]{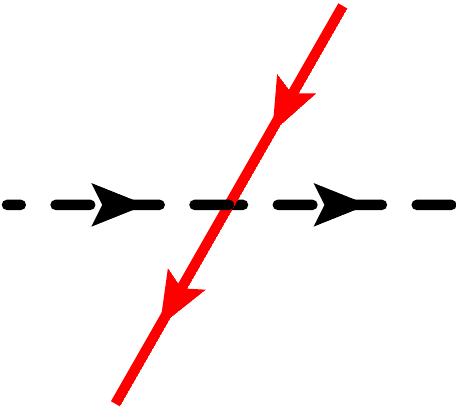} & \includegraphics[scale=.3]{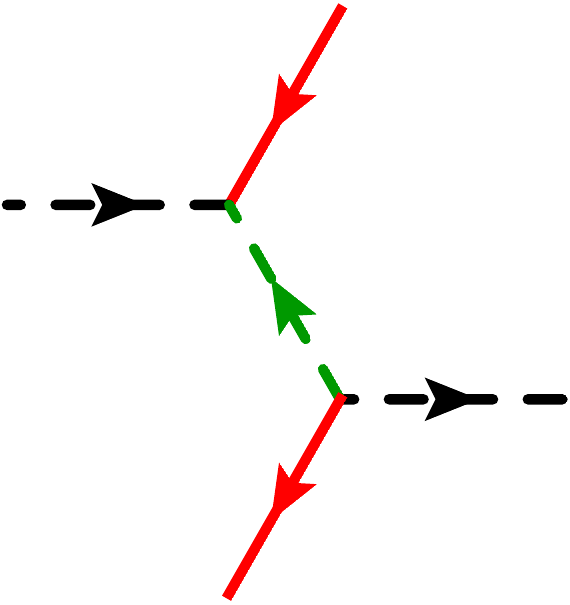} \\
  \cline{2-7}
  &  \multicolumn{2}{c||}{$\xi_1\xi_2$}&\multicolumn{2}{c||}{$\xi_1\xi_3$}&\multicolumn{2}{c||}{$\xi_2\xi_3$} \\
  &\includegraphics[scale=.3]{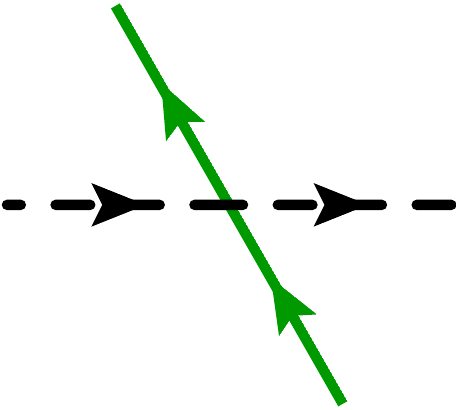} & \includegraphics[scale=.3]{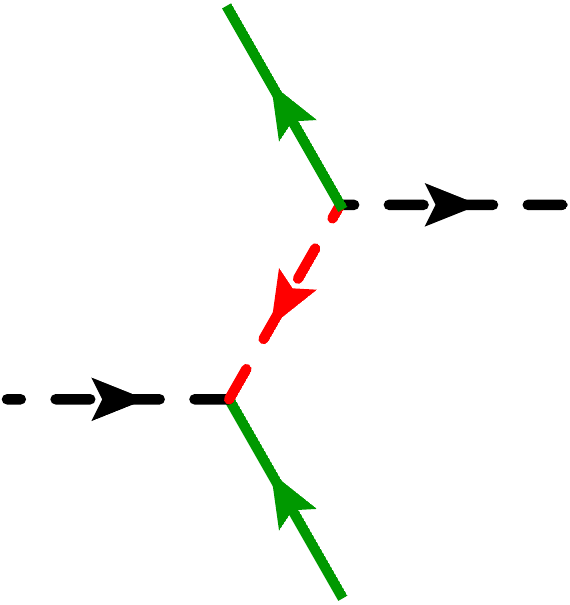}&\includegraphics[scale=.3]{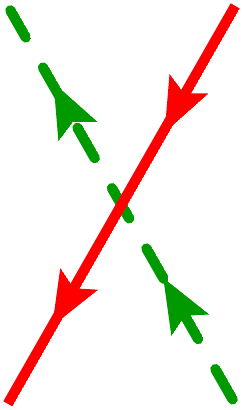} & \includegraphics[scale=.3]{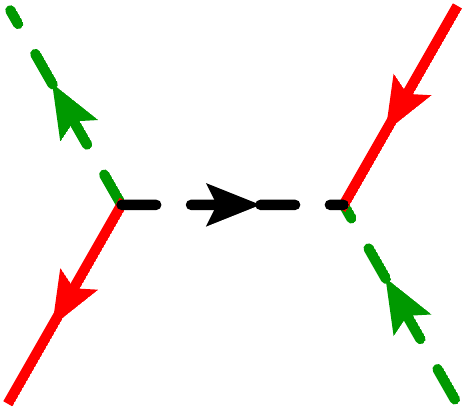}&\includegraphics[scale=.3]{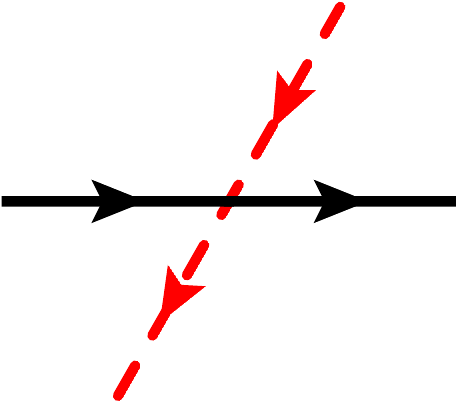} & \includegraphics[scale=.3]{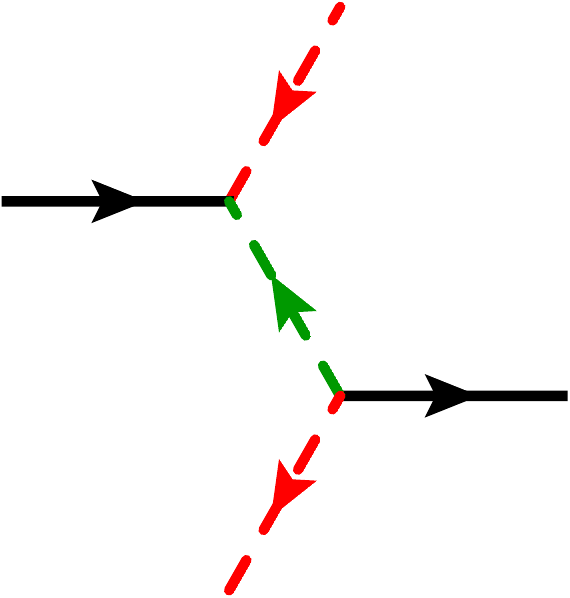} \\
\cline{1-7}
\multirow{4}{*}{ \addvbuffer[9ex]{\begin{turn}{90}Scattering\end{turn}}}&
  \multicolumn{2}{c||}{$\xi_2\sqrt{\xi_1\xi_3}$}&\multicolumn{2}{c||}{$\xi_1\sqrt{\xi_2\xi_3}$}&\multicolumn{2}{c||}{$\xi_3\sqrt{\xi_1\xi_2}$} \\
  &\includegraphics[scale=.3]{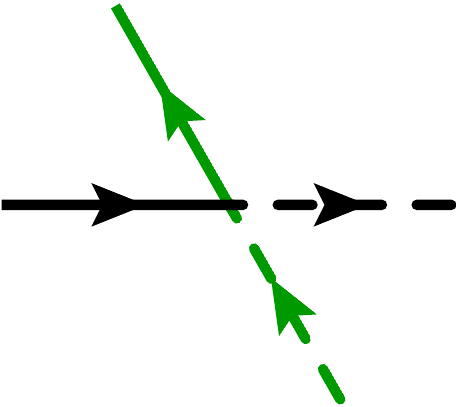} & \includegraphics[scale=.3]{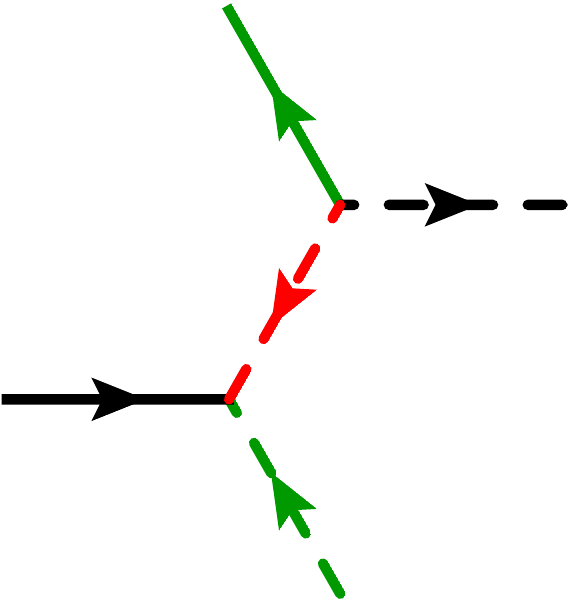}&\includegraphics[scale=.3]{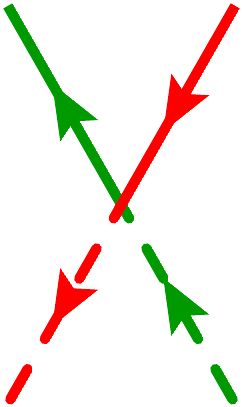} & \includegraphics[scale=.3]{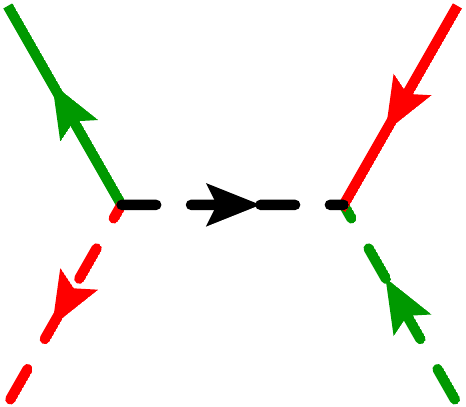}&\includegraphics[scale=.3]{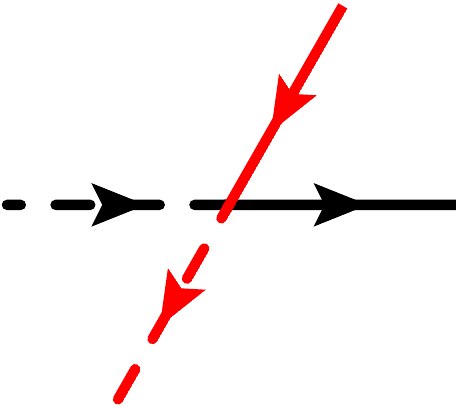} & \includegraphics[scale=.3]{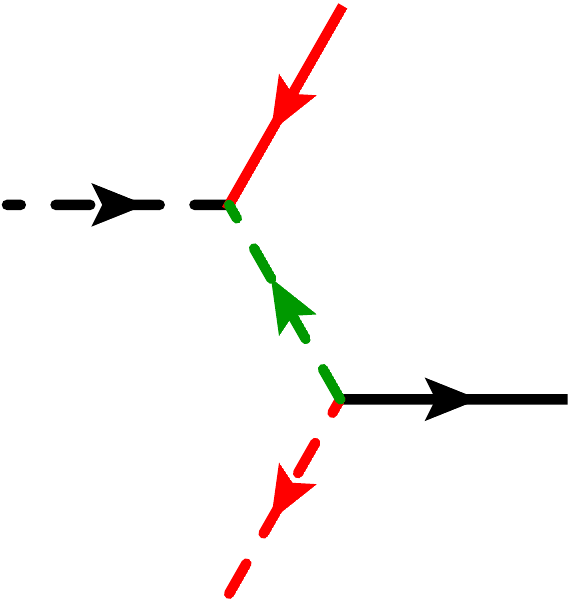} \\
  \cline{2-7}
    &  \multicolumn{2}{c||}{$\xi_2\sqrt{\xi_1\xi_3}$}&\multicolumn{2}{c||}{$\xi_1\sqrt{\xi_2\xi_3}$}&\multicolumn{2}{c||}{$\xi_3\sqrt{\xi_1\xi_2}$} \\
  &\includegraphics[scale=.3]{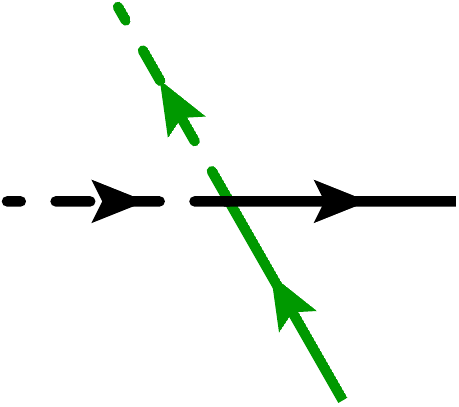} &\includegraphics[scale=.3]{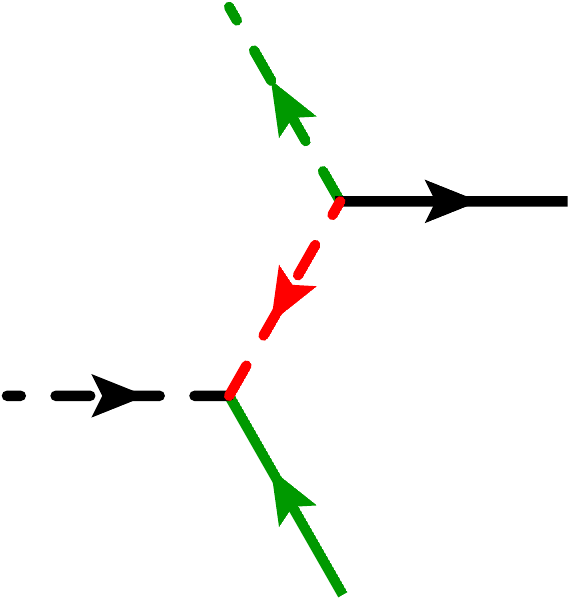}& \includegraphics[scale=.3]{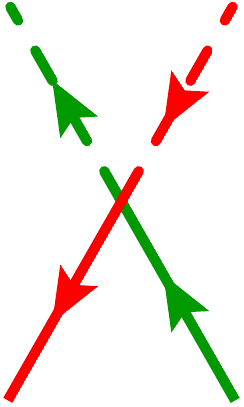} & \includegraphics[scale=.3]{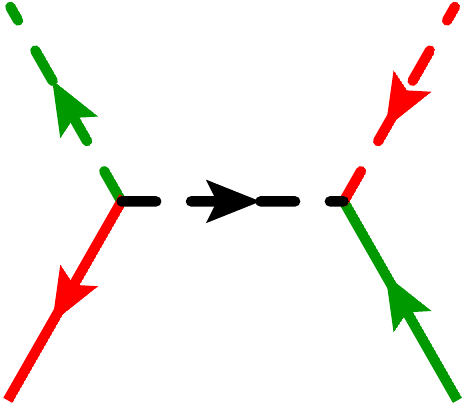}& \includegraphics[scale=.3]{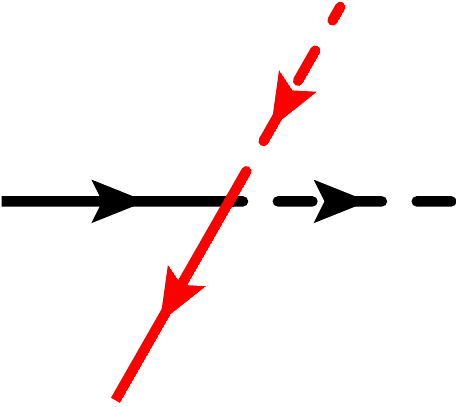} & \includegraphics[scale=.3]{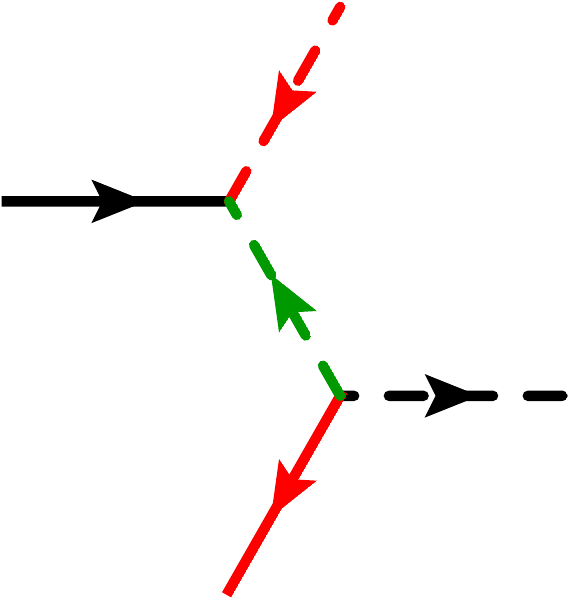} \\
\cline{1-7}
\end{tabular}
\end{adjustbox}
\end{center}
\caption{Substitution rules for the effective vertices appearing in the fishnet bulk structure of Fig.\ref{fig:patched_fishnet} in terms of the Feynman rules of Fig.\ref{fig:vertices}.
Any effective vertices is associated with a combination of the coupling constants $\xi_i$ with $i=1,2,3$ of order $\xi^2$.}
\label{tab:vert}
\end{table}

This system of three dotted lines forms a lattice which combines the features of both regularity and irregularity. Any such lattice can be obtained from the regular triangular lattice (or a more general Kagom\'{e} lattice) by arbitrary Baxter moves of all lines: displacements in the direction orthogonal to the line, i.e. conserving its direction. 

  The  links of the resulting lattice are propagators while nodes are quartic \textit{effective interactions}.  These interactions are of three kinds, depending on which lines are crossing  and which propagators enter the corresponding crossing (effective vertex).  They can represent a set of $\phi^4$ or various  Yukawa vertices, according to the rules listed in Tab.\ref{tab:vert}. 
Indeed in this framework, a quartic vertex involving fermions can be though of as a couple of Yukawa vertices, or similarly, as a split quartic vertex in which we have added a propagator in the remaining direction, according to the rules in Tab.\ref{tab:vert}. The quartic interaction can involve four scalars, four fermions or two of each. Moreover, we chose the directions of the arrows to be consistent with the Feynman rules in Fig.\ref{fig:vertices}. Depending on the orientation of the mixed interactions we will refer to them as \textit{crossing} or \textit{scattering} interactions as in Tab.\ref{tab:vert}.

Given three sets of parallel lines crossing each other with quartic interactions, the resulting irregular lattice is formed by a finite set of convex polygons. The smallest possible $n$-gon is a triangle and the largest one is a hexagon. Those convex polygons can be constructed locally by the abovementioned moves of lines in two or three different directions:
\begin{itemize}
\item \textit{2 directions (colors)}: We can discard the lines in one of the directions. The local interaction of lines with only two directions (colors) forms a square lattice as in \cite{,Zamolodchikov:1980mb,Gurdogan:2015csr}. Since we are considering three colors, we can have three different squares depending on their directions.
\item \textit{3 directions (colors)}: In this case there are more possibilities to build convex polygons. Indeed let's start with the crossing of three lines with three different directions. Locally, they form a triangle that can have two different orientations. Adding another line, parallel to one of the previous three, and cutting the triangle, we will end up with a square. Since we can add a line of any color and there are two possible triangle orientations, we can draw 6 different squares. Iterating this cutting procedure by adding one and two lines we obtain pentagons and the hexagon.
\end{itemize}    
In the following table we recap all the possible $n$-gons and their multiplicity, that is the number of different ways (i.e.: not superposable by simple translation and scaling) the same polygon can appear in the graph.
\begin{center}
\begin{tabular}{l*{4}{c}}
$n$-gon              & $\vartriangle$ & $\Square$ & $\pentagon$ & $\hexagon$  \\
\hline
Multiplicity$\qquad$ & 2 & 9 & 6 & 1  
\end{tabular}
\end{center}
It follows that for a given set of lines, the resulting lattice can be seen as a tiling of the plane with 18 different tiles.

\begin{figure}[!t]
 \centering
   {\includegraphics[width=16cm]{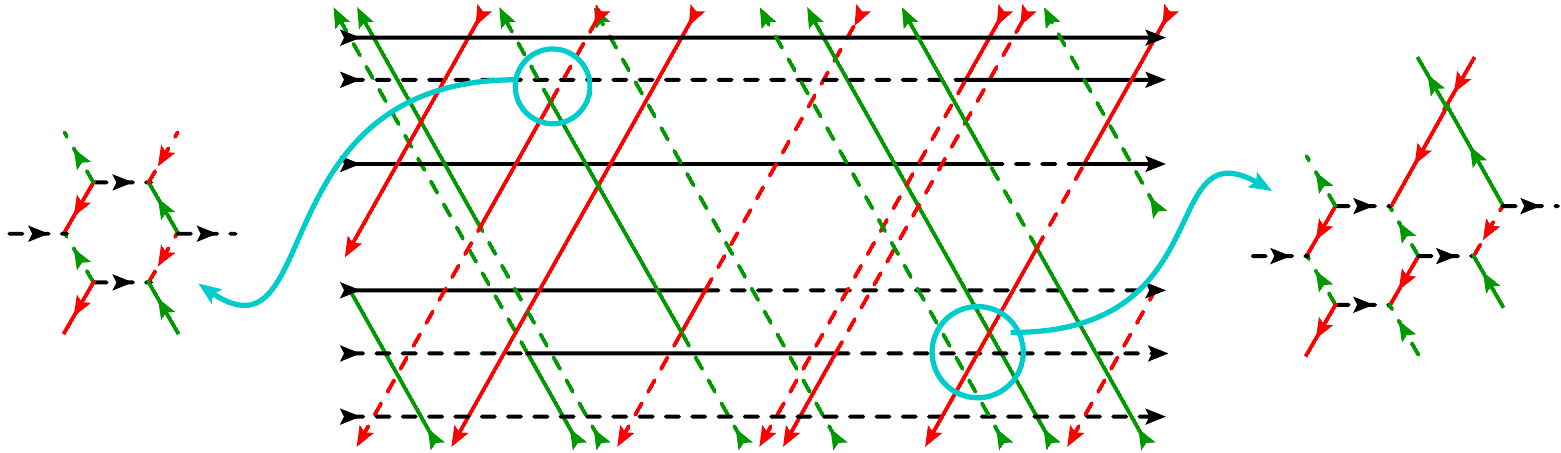}}
  \caption{
 One of the possible configurations in terms of effective vertices of Tab.\ref{tab:vert}  for the bulk topology represented in Fig.\ref{fig:patched_fishnet}. 
 The diagrams at the two sides of the figure represents the parts of the graph in the light-blue circles in terms of real vertices of Fig.\ref{fig:vertices} according with the rules given in Tab.\ref{tab:vert}. We stress that given a set of effective vertices, the translation in real vertices is unique.        }
  \label{fig:bulk_DSlimit2}
 \end{figure}

The structure of the fishnet bulk is very rich, indeed once the topology of the lattice is defined as in Fig.\ref{fig:patched_fishnet}, some information is lost, as any quartic dotted-vertex can be associated to six different physical vertices, as listed in Tab.\ref{tab:vert}. 
The number of possible Feynman diagrams $N_d$ which can be associated to a given close $n$-gon, defined by $n$ quartic dotted-vertices, can be computed  considering first all possible combinations of fermionic and scalar propagators for the edges of the polygon and then cancel out those vertices which does not fit in any configuration. After this tedious
combinatorics we obtain the following table
\begin{center}
\begin{tabular}{l*{4}{c}}
$n$-gon $\quad$             & $\vartriangle$ & $\Square$ & $\pentagon$ & $\hexagon$  \\
\hline
$N_d$ & 28 & 82 & 244 & 730  
\end{tabular}
\end{center}
This result can be written in the following compact formula
\begin{equation}
N_d(n)=1+3^n\,.
\end{equation}
Now we can estimate the number of Feynman diagrams for a given topology of the dotted-fishnet bulk. This number has the sum of all the $N_d$'s for all the polygons as an upper bound and we can estimate its order of magnitude.  Then the number of possible Feynman diagrams for the topology of the fishnet bulk given in Fig.\ref{fig:patched_fishnet} is around $1.5\times 10^4$. Moreover, since any vertex is associated with a combination of the couplings $\xi_i$ with $i=1,2,3$ of order 2, we know that the diagram in Fig.\ref{fig:patched_fishnet} is of order $\xi^{234}$.  One of those configurations is represented in Fig.\ref{fig:bulk_DSlimit2}.  
\label{sub:bulk}
\subsection{Single-trace correlation functions}

We can realize the above mentioned bulk graphs (with fixed coordinates of external legs) as a single-trace operator of the form: \begin{align}\label{Ktr}
K(x_1,x_2,\dots,x_M)=\Tr\left[\chi(x_1)\chi(x_2)\dots\chi(x_M)\right],\\ \chi\in\{\phi_j,\phi_j^\dagger,\psi_j^\alpha,\bar\psi_j^{\dot\alpha}\},\qquad\,(j=1,2,3;\,\,\alpha,\dot\alpha=1,2\,),
\end{align}  i.e. each \(\chi(x)\) under the trace is one of  18 fields of the \(\chi\)CFT model \eqref{chiFT4}-\eqref{fullL}.  Of course \eqref{Ktr} must have zero overall \(R\)-charge, to have a non-zero answer. This implies a condition on the elementary fields under trace, namely if we define \(n_j\) and \(m_j\) as the differences between the number of \(\phi_j\), respectively \(\psi_j\) and the conjugated fields, the mentioned condition reads
\begin{align}
n_j + 2 m_j - \sum_{k\neq j} n_k =0\,,\qquad j=1,2,3\,. 
\end{align}
\begin{figure}[!t]
 \centering
   {\includegraphics[width=16cm]{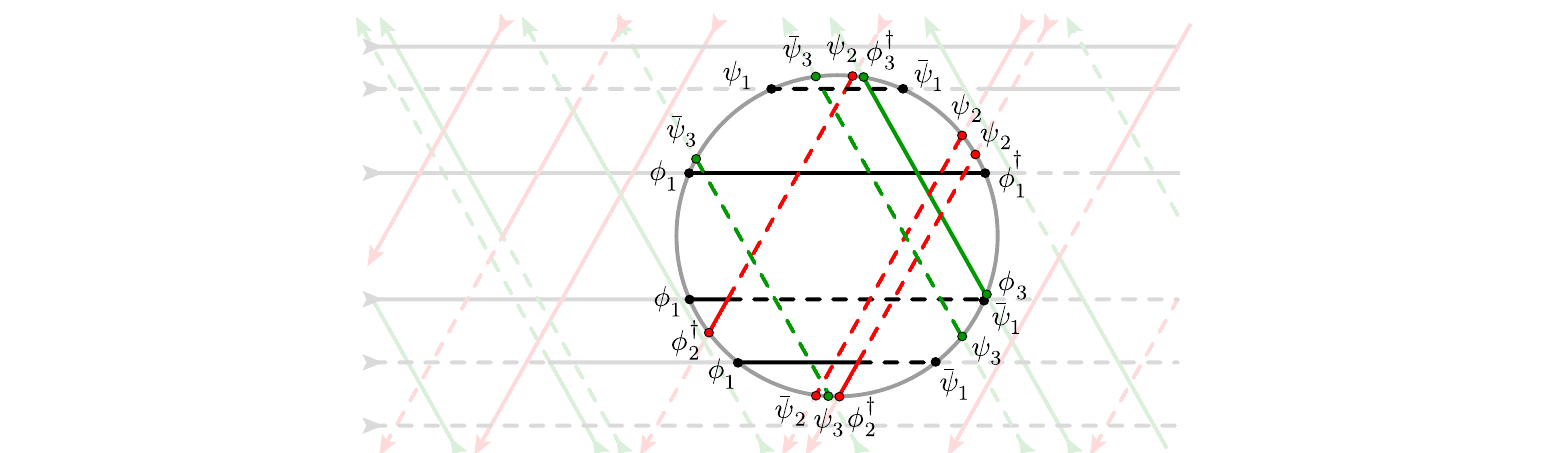}}
  \caption{
 The result of drawing a disc on the lattice of Fig.\ref{fig:bulk_DSlimit2} can be interpreted as one planar graph contributing to an \(n\)-point functions of the kind \eqref{Ktr}, drawn in terms of \emph{effective} vertices. In this example we present \(\Tr[\phi_1  \bar \psi_3 \psi_1 \bar \psi_3 \psi_2 \phi_3^{\dagger} \bar \psi_1 \psi_2 \psi_2 \phi_1^{\dagger} \phi_3 \bar \psi_1 \psi_3 \bar \psi_1 \phi_2^{\dagger} \psi_3 \bar \psi_2 \phi_1 \phi_2^{\dagger}\phi_1](x_1\dots x_{20})\),  and the graph is of order \(\xi^{42}\).
As it results from Tab.\ref{tab:vert}, each effective vertex can be replaced in a univocal way in terms of structure made of real vertices. }
  \label{fig:bulk_disc}
 \end{figure}
To describe the Feynman graph content of this quantity, let us remind that a similar single-trace correlator in bi-scalar fishnet CFT~\cite{Chicherin:2017cns,Chicherin:2017frs},  consisting only of scalar fields, was given by a single fishnet graph of the disc topology where the disc was cut out across the edges of a regular square lattice. The ends of the cut edges represented external fixed coordinates and the integrals were taken over all vertices inside the disc.  Similarly, for each of the quantities \eqref{Ktr} there exist a collection of graphs of the disc shape  cut out of the lattice of the type drawn on Fig.~\ref{fig:bulk_DSlimit2}. The types of external legs  -- the cut edges along the boundary --  define the species of fields from the set \(\chi\) following in the same order under the trace in \eqref{Ktr}.  We present an example in Fig.\ref{fig:bulk_disc}, where the disc is drawn on the concrete realization of the lattice as given in Fig.\ref{fig:bulk_DSlimit2}.
A big difference w.r.t. the bi-scalar single-trace correlators is that in the full \(\chi\)CFT such a quantity is defined by the sum of all graphs with the same order of fields on the boundary (same sequence of external legs) which are related to each other by the orthogonal moves of three types of parallel lines described in the previous subsection (as example, see Fig.\ref{fig:multA} (right)). Furthermore, even at fixed topology, one can change the interaction vertices inside the graph, namely switching some dashed (fermionic) lines to solid (scalar) lines and vice-versa (Fig.\ref{fig:multA} (left)). This corresponds to different realizations of a disc segment of the dotted-lattice in Fig.\ref{fig:patched_fishnet} with boundary conditions fixed by the external legs. The number of possible graphs can be estimated by considerations of the previous subsection.
\begin{figure}[!t]
 \centering
   {\includegraphics[trim={0 .15cm 0 0},clip,width=7.85cm]{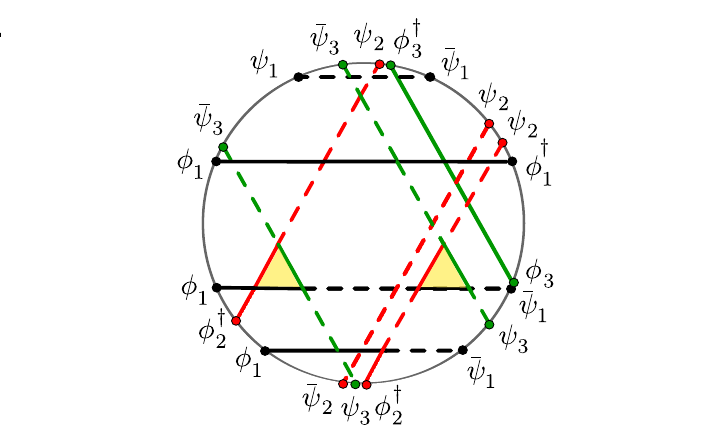}}
      {\includegraphics[trim={0 .05cm 0 0},clip,width=6.5cm]{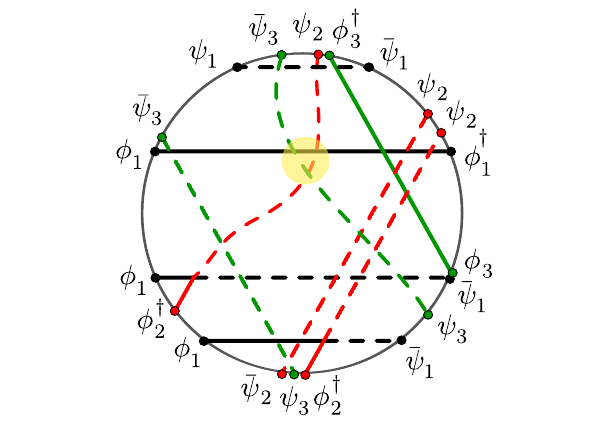}}
  \caption{
 Other possible planar graphs for the same \(20\)-point function of Fig.\ref{fig:bulk_disc} at order \(\xi^{48}\). On the left, the yellow triangles have different edges w.r.t. Fig.\ref{fig:bulk_disc}. On the right, one red-dashed line has been moved down-right, changing the topology w.r.t. Fig.\ref{fig:bulk_disc} in the highlighted region. 
 }
  \label{fig:multA}
 \end{figure}
This single-trace correlator can be used to define the scattering amplitudes via Lehmann-Symanzik-Zimmerman procedure, by going to the dual momentum space and taking on-shell external momenta, in the spirit of the papers
\cite{Chicherin:2017cns,Chicherin:2017frs}.
It is worth noticing that not all the planar single-trace correlators are obtained out of this procedure. Indeed certain external states can be cut out only drawing a circle on the actual Feynman graph (see Tab.\ref{tab:vert}) where all propagators are explicitly drawn. Moreover, for a given correlation function, there are lower order graphs in the coupling which cannot be cut out of the planar lattice, but from two or more sheets of such lattice as explained in \cite{Chicherin:2017cns}.\\
It would be interesting to show the Yangian invariance of these single-trace correlators, in the same spirit as it was done in ~\cite{Chicherin:2017cns,Chicherin:2017frs} for bi-scalar case. Namely, to define the monodromy (``lasso") around the boundary for which each of these graphs, or sum of all graphs, is an eigenfunction. This is one of the ways to show the integrability of the full \(\chi\)CFT. In this paper we will limit ourselves by the proof of integrability of the model \eqref{DSprime} which will be given in the next subsection.

\subsection{Integrability of Wheel graphs in \(\chi\)CFT}
\label{wheels}
\begin{figure}[t]
 \centering
   {\includegraphics[width=6.2cm]{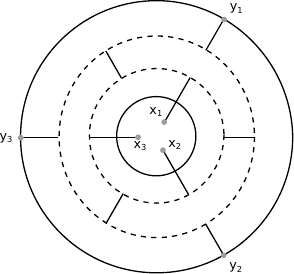}}
 \caption{An example of bulk of planar diagram appearing in the perturbative expansion of \(\langle \Tr[\phi_j^3](x)\Tr[\phi_j^3]^\dagger(y)\rangle\). It mixes together a square lattice structure of quartic scalar interactions and the ``brick-wall" domain made by Yukawa interactions. This case corresponds to the operatorial expression \(\hat{\mathcal{H}}_B^{(3)}(\hat{\mathcal{H}}_F^{(3)})^2\,{\mathcal{H}}_B^{(3)}(x_1,x_2,x_3|y_1,y_2,y_3)\).}
  \label{fig:wheel_brick}
 \end{figure}

\begin{figure}[t]
 \centering
   {\includegraphics[scale=1.05]{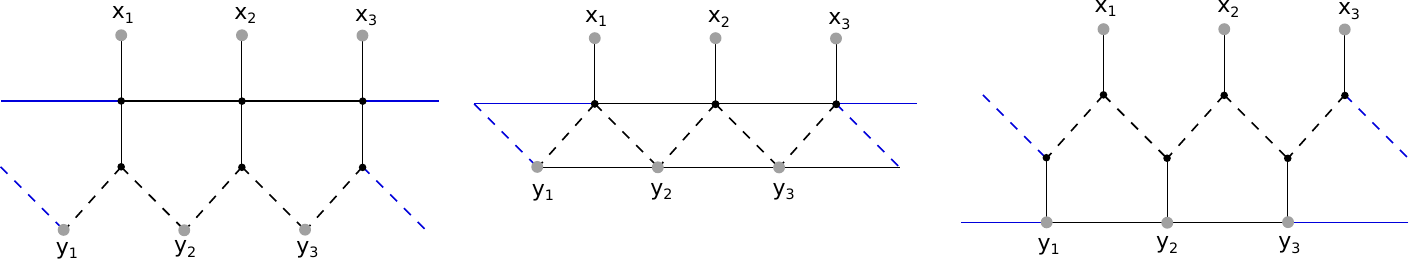}}
 \caption{Proof of the commutation relation \([\hat{\mathcal{H}}_B^{(L)},\hat{\mathcal{H}}_F^{(L)}]=0\) at \(L=3\). Gray blobs are external coordinates, black dots are integration points and we denoted lines which coincide due to periodic b.c. with blue. Left: \(\hat{\mathcal{H}}_B^{(L)}\mathcal{H}_F^{(L)} (x_1,x_2,x_3|y_1,y_2,y_3)\). In the middle: the result of integration over Yukawa vertices. Right: \(\hat{\mathcal{H}}_F^{(L)}\mathcal{H}_B^{(L)} (x_1,x_2,x_3|y_1,y_2,y_3)\) as result of opening triangles with single \(y_j\) vertex in the middle figure.}
  \label{fig:wheel_comm}
 \end{figure}

A statement of integrability, milder than the lattice integrability of the bulk of large planar graphs, can be made for the scaling dimension of \(\Tr[\phi_j^L]\) operators at any \(L\). These operators, protected in the original \(\mathcal{N}=4\; \text{SYM}\) due to supersymmetry, are described in the planar limit of bi-scalar theory by a perturbative expansion in globe-like fishnet graphs \cite{Gurdogan:2015csr} with an integrable square-lattice bulk \cite{Zamolodchikov:1980mb}. These graphs can be built up by the action of an integral ``graph-building" kernel \(\hat{ \mathcal{H}}_B^{(L)}\)
\begin{align}
\label{HBanyL}
[\hat{\mathcal{H}}_B^{(L)} \Phi](x_1,\cdots,x_L)  =\,\frac{1}{\pi^{2L}}\int \,\prod_{k=1}^L \frac{d^4 y_k}{(x_k-y_k)^2(y_k-y_{k+1})^2}  \Phi(y_1,\cdots,y_L),\qquad y_{L+1}\equiv y_1\, .
\end{align}
It represents one of the conserved charges generated by the transfer matrix of the integrable quantum \(SU(2,2)\) spin chain of \(L\) sites in the scalar \((\Delta,J_1,J_2)=(1,0,0)\) representation~\cite{Gromov:2017cja}. 
Similarly, in the two-coupling version \eqref{DSprime} of \(\chi\)CFT the perturbative expansion can be described by graphs which, in spite of more complicated structure (see Fig.\ref{fig:wheel_brick}), can be still built by integrals of motion of the conformal spin chain. Namely, every planar graph in the \(\xi_j\) expansion is a certain permutation of multiple action of operators \(\hat{\mathcal{H}}_B^{(L)}\) and \(\hat{\mathcal{H}}_F^{(L)}\), where the latter operator is responsible for fermionic loops contribution. As we will see, the order in the permutation doesn't matter, since any fermionic loop can be moved through scalar wrappings, due to their commutativity, and this fact lays at the basis of   integrability of these graphs.
The action of \(\hat{\mathcal{H}}_F^{(L)}\) reads
\begin{eqnarray}
&&[\hat{\mathcal{H}}_F^{(L)} \Phi](x_1,\cdots,x_L)  =\,\int \,\prod_{k=1}^L {d^4 y_k\,d^4 z_k}\; \mathcal{H}_F^{(L)}(x_1 \cdots x_L|y_1 \cdots y_L) \Phi(y_1,\cdots,y_L)\\
&&\mathcal{H}_F^{(L)}(x_1 \cdots x_L|y_1 \cdots y_L) =\frac{\tr[\sigma_{\mu_1}\bar{\sigma}_{\nu_1} \cdots \sigma_{\mu_L}\bar{\sigma}_{\nu_L}]}{(4\pi^3)^{2L}}\!\!\int \!\!\prod_{k=1}^L \frac{d^4 z_k }{(x_k-z_k)^2}\frac{(z_k-y_{k})^{\mu_{k}}(y_{k}-z_{k+1})^{\nu_{k}}}{ |z_k-y_{k}|^4 |z_{k+1}-y_{k}|^4},\notag
\end{eqnarray}
and it builds up an integrable ``brick-wall'' domain \cite{Chicherin:2017frs}. 
Its commutation with \(\hat{\mathcal{H}}_B^{(L)}\) can be proven directly by star-triangle relation \eqref{uniqferm}, as shown in Fig.\ref{fig:wheel_comm}.

In order to show that \(\hat{\mathcal{H}}_F^{(L)}\) is a conserved charge of the conformal scalar spin chain, we should prove its commutation with the transfer matrix at any value of the spectral parameter \(u\),
\begin{align}
\label{charge}
[\hat{\mathcal{H}}_F^{(L)},\mathbb{T}^{(L)}(u)]=0\,.
\end{align}
For this purpose, we rewrite the kernel integrating out \(z_k\) variables
\begin{align*}
\mathcal{H}_F^{(L)}(x_1 \cdots x_L|y_1 \cdots y_L) =\frac{\tr[\sigma_{\mu_1}\bar{\sigma}_{\nu_1} \cdots \sigma_{\mu_L}\bar{\sigma}_{\nu_L}]}{(2\pi)^{4L}}\prod_{k=1}^L\, \frac{(y_k-x_{k})^{\mu_{k}}(x_{k}-y_{k+1})^{\nu_{k}}}{(x_k-y_{k})^2 (x_{k}-y_{k+1})^2 (y_{k}-y_{k+1})^2}\, ,
\end{align*}
and we recall the definition of \(\mathbb{T}^{(L)}(u)\)
\begin{eqnarray}\label{trans}
\mathbb{T}^{(L)}(u) &&= \Tr_0[{R}_{10}(u)\,{R}_{20}(u)\cdots {R}_{L0}(u)],\qquad {R}_{j0}(u)\,\in \, \text{End}(L^2(x_j)\otimes L^2(x_0))\qquad \notag\\
\label{R}
[{R}_{ij}(u) \Phi](x_i,x_j) &&=\frac{4^{2u}}{\pi^4}\frac{\Gamma\left(u+2\right)^2}{\Gamma\left(-u-1\right)\Gamma\left(-u+1\right)}\int {d^4 x _{i'}d^4 x_{j'}\,\Phi(x_{i'},x_{j'})\over (x_{ij}^2)^{-u-1} (x_{ji'}^2)^{1+u}(x_{ij'}^2)^{3+u}(x_{i'j'}^2)^{-u+1}},
\notag\\\end{eqnarray}
where \({R}_{ij}(u)\) is the \(R\)-operator of the scalar conformal chain. It satisfies the Yang-Baxter equation \cite{Derkachov:2010zz} 
\begin{align}
\label{YBE}
{R}_{ij}(u){R}_{ik}(v){R}_{jk}(v-u)={R}_{jk}(v-u){R}_{ik}(v){R}_{ij}(u)\,.
\end{align}
Then operator \eqref{HBanyL} coincides with \(4^{-2L}\mathbb{T}^{(L)}\) in the limit \(u\to -1\), as pointed out in~\cite{Gromov:2017cja}, since the first propagator under the integral in~\eqref{trans} disappears and the last one effectively becomes a \(\delta\)-function. 

\begin{figure}[t]
 \centering
   {\includegraphics[scale=1.7]{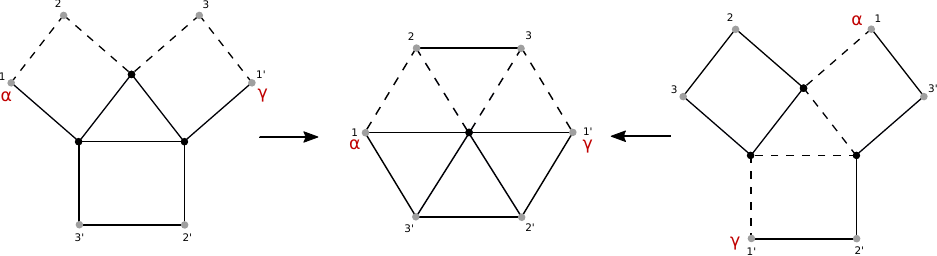}}
 \caption{Graphical representation of relation \eqref{tildeYBE} of Yang-Baxter type. The squares represent the kernels of \(R(v-u)_{23}\) (solid lines) and \((\tilde{R}_{12})^{\alpha}_{{\beta}}(u)\), \((\tilde{R}_{13})^{\beta}_{{\gamma}}(v)\) (solid and dashed lines). Black dots are integration points, while gray blobs are external coordinates. Figures on the left and on the right are respectively the L.H.S. and R.H.S. of \eqref{tildeYBE}. Both sides can be transformed in the hexagonal object in the middle. First the triangle is opened into a star integral using \eqref{unique} (left side) or \eqref{uniqferm} (right side). Doing so, each of the three black dots will become the end of only three lines. Then integration can be performed by \eqref{unique},\eqref{uniqferm} and leads to the hexagon.}
  \label{fig:wheel_YBE}
 \end{figure}

Now we introduce the transfer matrix for the brick-wall domain\footnote{Here we  implicitly mean the trace over spinorial indices of the fermionic loop.}
\begin{equation}
\label{transF}
\mathbb{T}^{(L)}_F(u) = \Tr_0[\tilde{R}_{10}(u)\tilde{R}_{20}(u)\cdots \tilde{R}_{L0}(u)],\quad \tilde{R}_{j0}(u)\in \text{End}(L^2(x_j)\otimes L^2(x_0)\otimes\mathbb{C}^2)
\end{equation}
\begin{equation}
\label{tildeR}
[(\tilde{R}_{ij})^{\alpha}_{\beta}(u)\Phi](x_i,x_j)\!= \frac{4^{2u}}{\pi^4}\frac{\Gamma\left(u+2\right)^2}{\Gamma\left(-u\right)\Gamma\left(-u+1\right)}\!\int\! d^4 x_{i'} d^4 x_{j'} \frac{(\sigma_\mu)^{\alpha \dot \alpha} (\bar \sigma_\nu)_{\dot \alpha \beta}\, \,x_{ij'}^{\mu}\,x_{i'j}^{\nu}\;\Phi(x_i',x_j')}{(x_{ij}^2)^{-u} (x_{ji'}^2)^{1+u}(x_{ij'}^2)^{3+u}(x_{i'j'}^2)^{-u+1}}, \notag
\end{equation}
and we check, similarly to the above scalar case, that \(\lim_{u \to -1}\mathbb{T}^{(L)}_F(u)\,=\,\hat{\mathcal{H}}_F^{(L)}\). The final step to prove \eqref{charge} is to show that
\begin{align}
\label{transFcomm}
[\mathbb{T}^{(L)}(u),\mathbb{T}_F^{(L)}(v)]=0 \qquad \forall u,v\,\, ,
\end{align}
which will be done by means of a Yang-Baxter type relation
\begin{align}
\label{tildeYBE}
\tilde{R}^{\alpha}_{ij~\beta}(u)\tilde{R}_{ik~\gamma}^{\beta}(v){R}_{jk}(v-u)={R}_{jk}(v-u)\tilde{R}^{\alpha}_{ik~ \beta}(v)\tilde{R}_{ij~\gamma}^{\beta}(u)\, .
\end{align}
graphically represented in Fig.\ref{fig:wheel_YBE}.
Indeed \eqref{transFcomm} follows immediately from \eqref{tildeYBE}.
 First of all we can introduce the monodromy operators \begin{align}
\tilde{\Omega}_{0~\beta}^{(L)~\alpha}(u)= \left[\tilde{R}_{01}(u)\cdots\tilde{R}_{0L}(u)\right]^{\alpha}_{~\beta}\qquad\text{and}\quad
{\Omega}_{0}^{(L)}(u)= {R}_{01}(u)\cdots {R}_{0L}(u)\, ,
\end{align}
then iterating \eqref{tildeYBE} we can write
\begin{align}
\left[\tilde{R}_{00'}(u)\tilde{\Omega}_0^{(L)}(v)\right]^{\alpha}_{~\beta}{\Omega}_{0'}^{(L)}(u-v)&={\Omega}^{(L)}_{0'}(u-v)\left[\tilde{\Omega}_0^{(L)}(v)\tilde{R}_{00'}(u)\right]^{\alpha}_{~\beta}\,,
\end{align}
and we finally trace over space \(L^2(x_0)\otimes L^2(x_{0'})\) and over spinorial indices getting
\begin{eqnarray}\notag
\Tr_{0,0'}\left(\tilde{\Omega}^{(L)}_{0}(v){\Omega}^{(L)}_{0'}(v-u) \right)&&=\Tr_{0,0'}\left(\tilde{R}_{00'}(u)^{-1}{\Omega}^{(L)}_{0'}(v-u)\tilde{\Omega}^{(L)}_0(v)\tilde{R}_{00'}(u)\right)\,\\\notag\\
\Tr_{0}\left(\tilde{\Omega}^{(L)}_0(v)\right)\,\Tr_{0'}\left({\Omega}^{(L)}_{0'}(v-u) \right)&&=\Tr_{0'}\left({\Omega}^{(L)}_{0'}(v-u) \right)\,\Tr_{0}\left(\tilde{\Omega}^{(L)}_0(v)\right)\, ,
\end{eqnarray}
which is equivalent to \eqref{transFcomm}. Our derivation straightforwardly shows that from the point of view of integrability the regular square lattice and the brick-wall lattice built by Yukawa vertices can be combined into the same integrable structure and form a mixed lattice. This concludes the demonstration of integrability of the two-coupling model  \eqref{DSprime}. The proof of integrability of the full \(\chi\)CFT~\eqref{fullL} is a more tricky exercise and we leave it for the future.

\section{Bethe-Salpeter equation for four-point correlators and conformal data}\label{sec:bethe}

In this and the next sections of this paper, we will exploit conformal symmetry and the Bethe-Salpeter method to 
obtain the exact 4-point correlations functions \begin{equation}\label{Ggeneral}
G_{\mathcal{O}_1\mathcal{O}_2}(x_1,x_2|x_3,x_4)=\langle\text{Tr}[\mathcal{O}_1(x_1)\mathcal{O}_2(x_2)]\text{Tr}[\mathcal{O}^\dagger_1(x_3)\mathcal{O}_2^\dagger(x_4)]\rangle\,,
\end{equation}
where the operators $\mathcal{O}_i$ are protected operators  in the double-scaled $\gamma$-deformed $\mathcal{N}=4$ SYM  theory, the \(\chi\)CFT.
 Then we will extract from it the OPE data, anomalous dimensions and structure constants,  for length-2 unprotected operators exchanged in the s-channel of \eqref{Ggeneral}.
In the current section, we present the generalities of conformal Bethe-Salpeter approach, generalizing the one applied in~\cite{Grabner:2017pgm,Kazakov:2018qbr,Gromov:2018hut} to the bi-scalar fishnet CFT, to sum up the  Feynman graphs for these quantities in \(\chi\)CFT. 
 
 At the fixed point \eqref{fixedpoint} and in the planar limit, the correlation function \eqref{Ggeneral} is a finite function of the couplings $\xi_{i}$ with $i=1,2,3$.
 The correlation functions can be remarkably written as a geometric sum of primitive divergencies in the perturbative expansion.
For this reason, we will study those diagrams with the help of the Bethe-Salpeter equation. In the following we will review the Bethe-Salpeter method pointing out how to extract the spectrum and the OPE data from the four-point functions \eqref{Ggeneral}.
In Sec.\ref{sec:section2}, we presented the \textit{bulk} fishnet structure of large planar diagrams in the general double-scaled $\gamma$ deformed $\mathcal{N}=4$ SYM theory. In this section we will focus on the correlation functions defined by \eqref{Ggeneral} for matrix (untraced) operators with bare dimensions $\Delta_{\mathcal{O}_1}$ and $\Delta_{\mathcal{O}_2}$. Since to preserve the renormalizability of the theory we have to supplement it with double-trace counter-terms \eqref{doubletraces}, diagrams in the perturbative expansion of \eqref{Ggeneral} will take the following \textit{chain} structure
\begin{equation*}
\vcenter{\hbox{\includegraphics[width=12 cm]{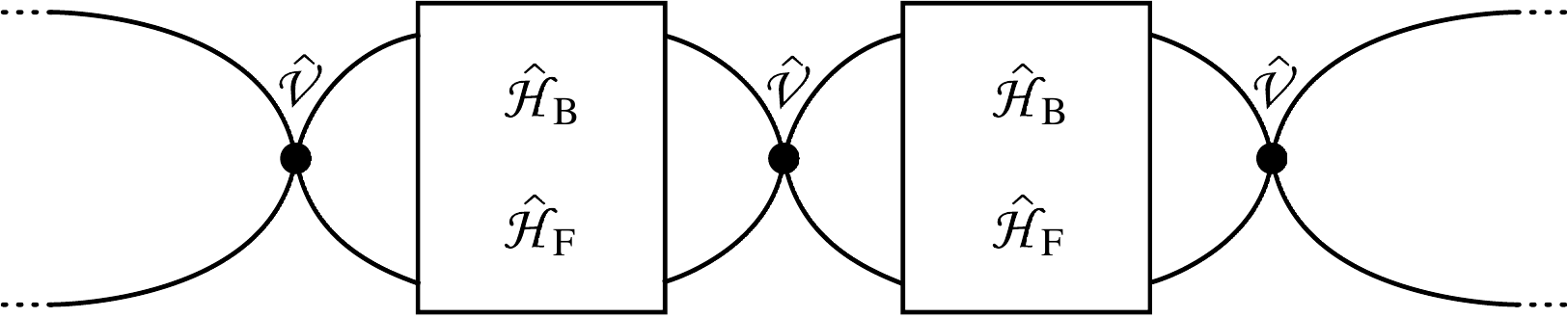}}}
\end{equation*}
where the black dots are insertions of the double-trace operator\footnote{Such insertions should always split a graphs, and its color structure, into two disconnected pieces.} and the links of the chain are periodically repeating configurations of propagators (a special case of the topologies presented in Sec.\ref{sec:section2}) generated by the kernel of integral operators. We will refer to this set of operators as \textit{Hamiltonian graph-building operators} $\hat{\mathcal{H}}_i$. 
In the family of theories we are considering, $\hat{\mathcal{H}}_i$ Hamiltonians can be of three different kind: the double trace operator $\hat{\mathcal{V}}$, the bosonic operator $\hat{\mathcal{H}}_B$ and the fermionic operator $\hat{\mathcal{H}}_F$.
The operators $\hat{\mathcal{V}}$, $\hat{\mathcal{H}}_B$ and $\hat{\mathcal{H}}_F$  separately produce divergent integrals. However, at the fixed point, their combination is finite due to conformal symmetry (see Sec.\ref{sec:feynman}).
These integral operators commute among themselves and they are diagonalized by the same basis of conformal triangles -- the 3-point correlators of the protected operators with  un protected operator with spin, described below. 

The correlation function \eqref{Ggeneral} can be written in general as a geometric series of a linear combination of the Hamiltonian graph-building operators as follows
\begin{equation}\begin{split}\label{Ggeom}
\hat{G}_{\mathcal{O}_1\mathcal{O}_2}&=\left(\frac{c_B}{x^2_{34}}\right)^{\Delta_{\mathcal{O}_1}+\Delta_{\mathcal{O}_2}-D}\sum_{\ell=0}^\infty
\hat{\mathcal{H}}_B(\chi_\mathcal{V}\hat{\mathcal{V}}+\chi_B\hat{\mathcal{H}}_B+\chi_F\hat{\mathcal{H}}_F)^\ell\\
&=\left(\frac{c_B}{x^2_{34}}\right)^{\Delta_{\mathcal{O}_1}+\Delta_{\mathcal{O}_2}-D}\frac{\hat{\mathcal{H}}_B}{1-\chi_\mathcal{V}\hat{\mathcal{V}}-\chi_B\hat{\mathcal{H}}_B-\chi_F\hat{\mathcal{H}}_F}\,,
\end{split}\end{equation} 
where $c_B=1/(4\pi^2)$ is the normalization factor of the free scalar propagator, $\chi_\mathcal{V}$, $\chi_B$ an $\chi_F$ are combinations of the couplings $\alpha_i$ and $\xi_i$ with $i=1,2,3$, which will be  introduced later (see Sec.\ref{sec:phi1phi1}).  The spacetime dimension $D$ in this paper is always taken to be $D=4$. \footnote{It is possible to generalize the bi-scalar fishnet theory to any integer dimension $D$, as in~\cite{Kazakov:2018qbr}, at the cost of losing locality. It is not evident that such a generalization is possible for the full \(\chi\)CFT.} 
The correlation function \eqref{Ggeneral} can be obtained from the operator $\hat{G}_{\mathcal{O}_1\mathcal{O}_2}$ as follows 
\begin{equation}
G_{\mathcal{O}_1\mathcal{O}_2}(x_1,x_2|x_3,x_4)=\langle x_1,x_2|\hat{G}_{\mathcal{O}_1\mathcal{O}_2}|x_3,x_4\rangle\,,
\end{equation}
where the Hamiltonian operators are represented by the corresponding integration kernels such that
\begin{equation}
\!\langle x_1,x_2|\hat{\mathcal{H}}_i^n|x_3,x_4\rangle\!=\!\!\int \!\prod_{k=1}^{2n}\!d^4y_k \mathcal{H}_i(x_1,x_2|y_1,y_2) \mathcal{H}_i(y_1,y_2|y_3,y_4)...\mathcal{H}_i(y_{2n-1},y_{2n}|x_3,x_4)\,.
\end{equation}
In order to compute the correlators $G_{\mathcal{O}_1\mathcal{O}_2}$, given the set of Hamiltonian graph-building operators $\hat{\mathcal{H}}_i$,
we need to compute their eigenvalues and decompose $\hat{G}_{\mathcal{O}_1\mathcal{O}_2}$ over a complete basis of their eigenfunctions.

To compute the eigenvalues of $\hat{\mathcal{H}}_i$, we can use the fact that these integral operators transform covariantly with respect to the \(\left(1,0,0\right)\otimes \left(1,0,0\right)\) conformal spin chain generators.\footnote{In particular, defining the inversion $I[x_i^\mu ]=x_i^\mu/x_i^2$, we have, for a conformal triangle \(\Phi_{x_0}(x_1,x_2)\)in the representation \((1,0,0)\otimes (1,0,0)\),
$I[\Phi_{x_0}(x_1,x_2)]=\Phi_{x_0}(x_1/x_1^2,x_2/x_2^2)= U \Phi_{x_0}(x_1/x_1^2,x_2/x_2^2)$, and \(U= x_1^2 x_2^2 x_0^{\Delta-S}\).~ We can check that for every integral operator: $I[\hat{\mathcal H}_i]=U \hat {\mathcal H}_i U^{-1}$.} 
This property completely fixes their eigenstates to be the \textit{conformal triangle} $\Phi_{\Delta,S,x_0}(x_{1},x_{2})$, the three-point function of two (scalar) operators in $x_1$ and $x_2$ with an operator $O_{\Delta,S}(x_0)$ with scaling dimension $\Delta$, spin $S$ at the position $x_0$
\begin{equation}\begin{split}\label{Phi-def}
\Phi_{\Delta,S,x_0}&(x_{1},x_{2}) = \langle{\Tr[\mathcal{O}_1(x_1) \mathcal{O}_2(x_2)] O_{\Delta,S}(x_0)}\rangle\\
=& (x_{12}^2)^{p-\tfrac{\Delta_{\mathcal{O}_1}+\Delta_{\mathcal{O}_2}}{2}}
(x_{10}^2)^{\tfrac{\Delta_{\mathcal{O}_2}-\Delta_{\mathcal{O}_1}}{2}-p}
(x_{20}^2)^{\tfrac{\Delta_{\mathcal{O}_1}-\Delta_{\mathcal{O}_2}}{2}-p}
\left(\frac{2(nx_{02})}{x_{02}^2}-\frac{2(nx_{01})}{x_{01}^2}\right)^S\,,
\end{split}\end{equation}
where $p=\frac{\Delta-S}{2}$ and $n^\mu$ an auxiliary light-cone vector. 
In the case $S=0$, the conformal triangle is composed by simply three scalar propagators that we can graphically represents as follows
\begin{equation}\label{eigenfunction}
\Phi_{\Delta,0,x_0}(x_1,x_2)\;\equiv\vcenter{\hbox{\includegraphics[width=2.7cm]{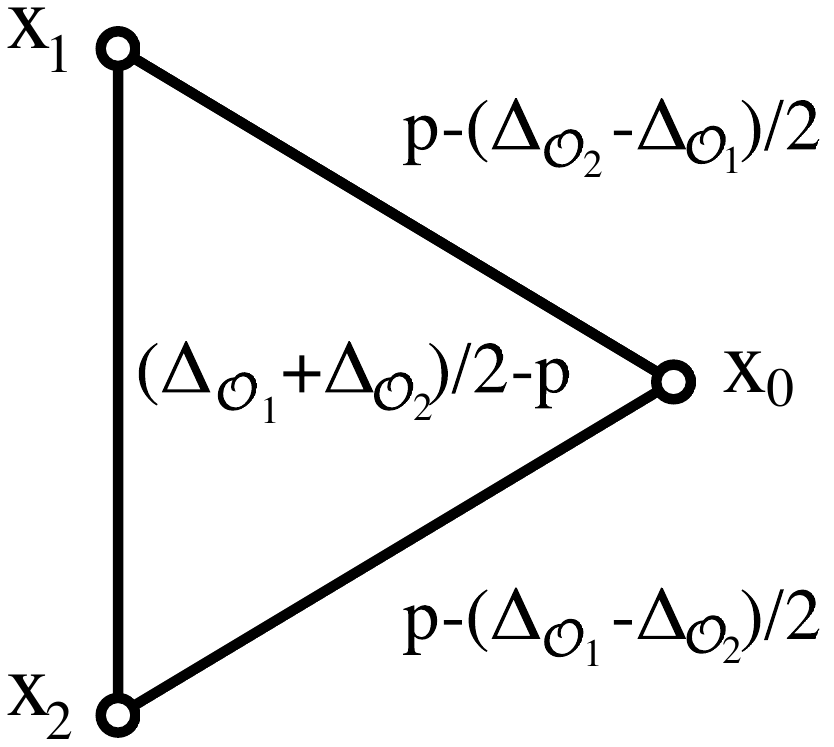}}}\,.
\end{equation}
Finally, given the eigenstate \eqref{Phi-def}, we can compute the spectrum of the Hamiltonian operators $\hat{\mathcal{H}}_i$ as follows
\begin{equation}\label{eigengeneral}
\left[\hat{\mathcal{H}}_i\,\Phi_{\Delta,S,x_0}\right](x_1,x_2)  \equiv\!\int d^4y_1d^4y_2 \,\mathcal{H}_i(x_1,x_2|y_1,y_2)\,\Phi_{\Delta,S,x_0}(y_1,y_2)={h_i}_{\Delta,S}\,\Phi_{\Delta,S,x_0}(x_1,x_2),
\end{equation}
where ${h_i}_{\Delta,S}$ is the eigenvalue. More specifically, given the Hamiltonians operators defined in \eqref{Ggeom}, we have
\begin{align}\label{eigen1}
{}& \left[\hat{\mathcal V}\; \Phi_{\Delta,S,x_0}\right](x_1,x_2)  = {h_\mathcal{V}}_{\Delta,S} \; \Phi_{\Delta,S,x_0}(x_1,x_2),
\\\label{eigen2}
{}& \left[\hat{\mathcal{H}}_B \;\Phi_{\Delta,S,x_0}\right](x_1,x_2)  =  {h_B}_{\Delta,S}\;\Phi_{\Delta,S,x_0}(x_1,x_2),
\\\label{eigen3}
{}& \left[\hat{\mathcal{H}}_F \;\Phi_{\Delta,S,x_0}\right](x_1,x_2)  =  {h_F}_{\Delta,S}\;\Phi_{\Delta,S,x_0}(x_1,x_2).
\end{align}
In Sec.\ref{sec:phi1phi1} and Sec.\ref{sector2}, we will verify that \eqref{Phi-def} diagonalizes these Hamiltonians and perform a direct computation of the eigenvalues. \\The scaling dimension appearing in \eqref{Phi-def} is defined as \cite{Dobrev:1977qv}
\begin{equation}
\Delta=2+2i\nu\,,
\end{equation}
with $\nu$ a non-negative real number. For such values of $\Delta$, the state $\Phi_{\Delta,S,x_0}$ belongs to the principal series of type-{I} irreducible  representations \((\Delta,S,0)\) of the conformal group labelled by $\nu$ and the discrete compact spin $S$ and satisfies the orthogonality condition \cite{Fradkin:1978pp,Dobrev:1977qv}
 \begin{align}\label{states}
& \int\!  \frac{d^4x_{1}d^4x_{2}\overline{ \Phi_{\Delta',S',x_{0'}}}(x_{1},x_{2})\Phi_{\Delta,S,x_0}(x_{1},x_{2}) }{(x_{12}^2)^{4-\Delta_{\mathcal{O}_1}-\Delta_{\mathcal{O}_2}}}= 
 (-1)^S c_1(\nu,S) \delta(\nu-\nu')\,\delta_{S,S'}\delta^{(4)}(x_{00'})(nn')^S\nonumber\\
 &\qquad\qquad+(-1)^S c_2(\nu,S) {\delta(\nu+\nu')\delta_{S,S'}} ((n \p_{x_0} )(n'\p_{x_{0'}})\ln x_{00'}^2)^S /(x_{00'}^2)^{2-2i\nu-S},
\end{align}
where the 4-dimensional coefficients $c_1$ and $c_2$ are given by
\begin{equation}\begin{split}\label{c1c2}
 &c_1= 
  \frac{ 2^{S-1}\, \pi ^7}{ (S+1)\nu ^2\left(4 \nu ^2
   +(S+1)^2\right)},
 \\
 &c_2=-\frac{i \pi ^5 (-1)^S\Gamma\! \left(\frac{S+\Delta_{\mathcal{O}_1}-\Delta_{\mathcal{O}_2}}{2}-i \nu +1\right)\!\Gamma\! \left(\frac{S-\Delta_{\mathcal{O}_1}+\Delta_{\mathcal{O}_2}}{2}-i \nu +1\right)  \!\Gamma (S+2 i \nu
   +1)}{\nu  (S+1) \Gamma\! \left(\frac{S+\Delta_{\mathcal{O}_1}-\Delta_{\mathcal{O}_2}}{2}+i \nu +1\right)\!\Gamma\! \left(\frac{S-\Delta_{\mathcal{O}_1}+\Delta_{\mathcal{O}_2}}{2}+i \nu +1\right)  \!\Gamma (S-2 i \nu
   +1)}.
\end{split}\end{equation}
The eigenfunction $\Phi_{\Delta,S,x_0}$ forms an orthonormal basis for $\nu\geq 0$ implying the following representation for the identity
\begin{equation}
\delta^{(4)}(x_{13})\delta^{(4)}(x_{24})=\!\sum_{S=0}^\infty\frac{(-1)^S}{(x_{34}^2)^{4-\Delta_{\mathcal{O}_1}-\Delta_{\mathcal{O}_2}}}\!\int_0^\infty \frac{d\nu}{c_1(\nu,S)}\int
d^4x_0 \Phi_{\Delta,S,x_0}(x_{1},x_{2}) \overline{ \Phi_{\Delta,S,x_{0}}}(x_{3},x_{4})\,,
\end{equation}
that, together with the definition \eqref{eigengeneral}, leads to the diagonalized representation
\begin{equation}\label{Hphiphi}
\mathcal{H}_i(x_1,x_2|x_3,x_4)=\!\sum_{S=0}^\infty\frac{(-1)^S}{(x_{34}^2)^{4-\Delta_{\mathcal{O}_1}-\Delta_{\mathcal{O}_2}}}\!\int_0^\infty \frac{d\nu \;{h_i}_{\Delta,S}}{c_1(\nu,S)}\int
d^4x_0 \Phi_{\Delta,S,x_0}(x_{1},x_{2}) \overline{ \Phi_{\Delta,S,x_{0}}}(x_{3},x_{4})\,,
\end{equation}
where $\mathcal{H}_i$ stands for the set of hamiltonians $\{\mathcal{V},\mathcal{H}_B,\mathcal{H}_F\}$ and ${h_i}_{\Delta,S}$ for the set of eigenvalues $\{{h_\mathcal{V}}_{\Delta,S},{h_B}_{\Delta,S},{h_F}_{\Delta,S}\}$ respectively. 

Plugging the representation \eqref{Hphiphi} for the graph-building operators $\mathcal{H}_i$ into \eqref{Ggeom}, we obtain the representation of the 4-point function in terms of their eigenvalues ${h_i}_{\Delta,S}$
\begin{equation}\begin{split}\label{Gphiphi}
G_{\mathcal{O}_1\mathcal{O}_2}&(x_1,x_2|x_3,x_4)=\!\sum_{S=0}^\infty\frac{(-1)^S}{(x_{34}^2)^{4-\Delta_{\mathcal{O}_1}-\Delta_{\mathcal{O}_2}}}\!\int_0^\infty \frac{d\nu}{c_1(\nu,S)}\times\\
&\!\!\!\!\times\frac{{h_B}_{\Delta,S}}{1-\chi_\mathcal{V}{h_\mathcal{V}}_{\Delta,S}-\chi_B{h_B}_{\Delta,S}-\chi_F{h_F}_{\Delta,S}}\int
d^4x_0 \Phi_{\Delta,S,x_0}(x_{1},x_{2}) \overline{ \Phi_{\Delta,S,x_{0}}}(x_{3},x_{4}).
\end{split}\end{equation}
The  integral over the auxiliary point $x_0$ can be expressed in terms of the four-dimensional conformal blocks $g_{\Delta,S}$ \cite{Dobrev:1977qv,Dolan:2000ut,Dolan:2011dv}
\begin{equation}\begin{split}\label{phiphi}
\int &d^4x_0 \Phi_{\Delta,S,x_0}(x_{1},x_{2}) \overline{ \Phi_{\Delta,S,x_{0}}}(x_{3},x_{4})\\
&\!\!\!=\!\left(\frac{1}{x_{12}^2x_{34}^2}\right)^{\!\!\frac{\Delta_{\mathcal{O}_1}\!+\Delta_{\mathcal{O}_2}}{2}}\!\!\!
\left(\frac{x_{24}^2}{x_{13}^2}\right)^{\!\!\frac{\Delta_{\mathcal{O}_1}\!-\Delta_{\mathcal{O}_2}}{2}}\!\!\!\left(\frac{c_1(\nu,S)}{c_2(\nu,S)}g_{\Delta,S}(u,v)+\frac{c_1(-\nu,S)}{c_2(-\nu,S)}g_{4-\Delta,S}(u,v)\right),
\end{split}\end{equation}
where the cross-ratios are $u=z\bar{z}=x_{12}^2 x_{34}^2/(x_{13}^2 x_{24}^2)$ and $v=(1-z)(1-\bar{z})=x_{14}^2x_{23}^2/(x_{13}^2 x_{24}^2)$ and we recall from~\cite{Dolan:2000ut} that
\begin{equation}
\begin{split}\label{defg}
&g_{\Delta,S}=(-1)^S \frac{z\bar  z}{
 z - \bar z} \left[ k(\Delta + S, z) k(\Delta - S - 2, \bar z) - 
   k(\Delta + S, \bar z) k(\Delta - S - 2, z)\right],\\
&\text{where} \quad k(\beta_, x) = x^{\beta/2}
  \,\,_2F_1\left(\frac{\beta - (\Delta_1 -\Delta_2)}{2}, \frac{\beta + (\Delta_3 - \Delta_4)}{2}, \beta, 
   x\right)\;.
\end{split}
\end{equation}
Inserting \eqref{phiphi} into \eqref{Gphiphi}, we obtain 
\begin{equation}\begin{split}\label{4pointgeneral}
G_{\mathcal{O}_1\mathcal{O}_2}&(x_1,x_2|x_3,x_4)=
\left(\frac{c_B^2}{x_{12}^2x_{34}^2}\right)^{\!\frac{\Delta_{\mathcal{O}_1}\!+\Delta_{\mathcal{O}_2}}{2}}\!
\left(\frac{x_{24}^2}{x_{13}^2}\right)^{\!\frac{\Delta_{\mathcal{O}_1}\!-\Delta_{\mathcal{O}_2}}{2}}\mathcal{G}_{\mathcal{O}_1\mathcal{O}_2}(u,v)\,,
\end{split}\end{equation}
where we defined
\begin{equation}\begin{split}\label{Gnu}
\mathcal{G}_{\mathcal{O}_1\mathcal{O}_2}(u,v)=\frac{1}{c_B^4}\sum_{S=0}^\infty
(-1)^S\!\!\!\int_{-\infty}^\infty \frac{d\nu}{c_2(\nu,S)}\frac{{h_B}_{\Delta,S}\;\,g_{\Delta,S}(u,v)}{1-\chi_\mathcal{V}{h_\mathcal{V}}_{\Delta,S}-\chi_B{h_B}_{\Delta,S}-\chi_F{h_F}_{\Delta,S}}\,.
\end{split}\end{equation}
Notice that we extended the integral over $\nu$ on the full real axis with the change of variable $\nu\rightarrow -\nu$ in the second term of \eqref{phiphi}. This is allowed by the symmetry of eigenvalues appearing in the spectral equation
\begin{equation}\label{hsymmetry}
{h_i}_{\,4-\Delta,S}={h_i}_{\,\Delta,S}\,,
\end{equation}
and can be interpreted as the fact that, for a given spin $S$, states with dimension $\Delta$ and $4-\Delta$ belong to a unitary equivalent representation of the conformal group. This symmetry is indeed satisfied for every  studied case, \eqref{hbS}, \eqref{hftildeS} and \eqref{hffinale}.

Before studying  the integral in \eqref{Gnu}, we want to focus on the role of the double-trace Hamiltonian and its eigenvalues in the perturbative and Bethe-Salpeter approaches. To find the correlation function \eqref{Ggeneral}, we have to sum up diagrams of the kind shown at the beginning of this section. These diagrams contain an involved scalar and fermionic structure generated by the operators $\hat{\mathcal{H}}_B$ and $\hat{\mathcal{H}}_F$ interspersed with the contributions of the double-trace vertices introduced in \eqref{doubletraces}. Since in general the integrals over the positions of the single-trace vertices develop ultraviolet (UV) divergencies at short distances, one needs the double-trace interactions to produce other UV divergent contributions which cancels against them. Therefore, the weak coupling expansion of the four-point correlation function remains UV finite at any order as expected for protected  $\mathcal{O}_1$ and $\mathcal{O}_2$. 

In the context of the Bethe-Salpeter equation the story is slightly different. Indeed consider the Hamiltonian operator $\hat{\mathcal{V}}$ associated to the double-trace kernel defined as follows
\begin{equation}\label{V}
\left[\hat{\mathcal{V}}\;\Phi\right]\!(x_1,x_2)=2 c_B^2\int \frac{d^4y_1 d^4y_2}{(x_1-y_1)^2(x_2-y_2)^2}\delta^{(4)}(y_{12}) \,\Phi(y_1,y_2)\,,
\end{equation}
where $\Phi(y_1,y_2)$ is a test function. We have to compute its spectrum by means of \eqref{eigen1} that, when applied to \eqref{Phi-def},  reads
\begin{equation}\label{eigenV}
 \left[\hat{\mathcal V}\; \Phi_{\Delta,S,x_0}\right]\!(x_1,x_2)  =  \frac{\delta^{(4)}(\nu)\delta_{S,0}}{(4\pi)^2}\, \Phi_{\Delta,S,x_0}(x_1,x_2)\qquad\Rightarrow\qquad {h_\mathcal{V}}_{\Delta,S}=\frac{\delta^{(4)}(\nu)\delta_{S,0}}{(4\pi)^2}.
\end{equation}
First of all, due to the form of the eigenvalue, the double-trace term can affect only the contribution to the sum in \eqref{Gnu} with spin $S=0$. Then we expect that the contribution to \eqref{Gnu} given by the Hamiltonian operators $\hat{\mathcal{H}}_B$ and $\hat{\mathcal{H}}_F$ are well-defined for $S\neq 0$ but in principle we have to take into account the double-trace term for $S=0$.

Since we want to write $\mathcal{G}_{\mathcal{O}_1\mathcal{O}_2}$ in the standard OPE form, we will consider the limit in which two of the external points are approaching, \textit{i.e.} $|x_{12}|\rightarrow 0$ (or $u\rightarrow 0$ and $v\rightarrow 1$). Since the conformal block scales as $u^{p}(1-v)^S$ decaying exponentially for $\text{Re}(i\nu)\rightarrow \infty$, one can close the contour in the integral over $\nu$ in lower-half plane and then compute it by residues. At short distances, the eigenstate \eqref{Phi-def} scales as $\Phi_{\Delta,S,x_0}(x_1,x_2)\sim (x_{12}^2)^{\tfrac{\Delta-\Delta_{\mathcal{O}_1}-\Delta_{\mathcal{O}_2}}{2}}$ and thus it vanishes in the lower-half plane (which is true in our case, since \(\Delta_{\mathcal{O}_1}=\Delta_{\mathcal{O}_2}=1 \) and \(\Re(\Delta)>2\)). In this case, the bosonic and fermionic operators do not develop UV divergencies (one can verify it in the two special cases that we  study in detail in Sec.\ref{sec:phi1phi1} and Sec.\ref{sector2}). Moreover, given the definition \eqref{V} and the formula \eqref{eigenV}, the double-trace operator $\hat{\mathcal{V}}$ annihilates $\Phi_{\Delta,S,x_0}$ with $\text{Im}(\nu)<0$ for any $S$ and therefore, it does not contribute. With this argument, we are able to neglect the double-trace contributions when we compute the four-point function $G_{\mathcal{O}_1\mathcal{O}_2}$ with the Bethe-Salpeter method. Then we can rewrite \eqref{Gnu} as follows
\begin{equation}\begin{split}\label{Gnu2}
\mathcal{G}_{\mathcal{O}_1\mathcal{O}_2}(u,v)=\frac{1}{c_B^4}\sum_{S=0}^\infty
(-1)^S\oint_{\mathcal{C}_-} \frac{d\nu}{c_2(\nu,S)}\frac{{h_B}_{\Delta,S}}{1-\chi_B{h_B}_{\Delta,S}-\chi_F{h_F}_{\Delta,S}}\;\,g_{\Delta,S}(u,v)\,,
\end{split}\end{equation}
where $\mathcal{C}_-$ is the close path in the lower-half plane.

In order to compute the integral over $\nu$ in \eqref{Gnu2} with residues, we have to identify the poles of the integrand. The \textit{physical poles} are given by the zeros of denominator under the integral, i.e. by  solutions of the equation
\begin{equation}\label{spectraleq}
{{h_B}_{\Delta,S}}^{-1}-\chi_F\,{h_F}_{\Delta,S}\,{{h_B}_{\Delta,S}}^{-1}=\chi_B\,  .
\end{equation}
We will refer to \eqref{spectraleq} as \textit{spectral equation}: indeed given the eigenvalues ${h_i}_{\Delta,S}$ and the constants $\chi_i$, solving the equation we will obtain the scaling dimensions $\Delta$ as  functions of the couplings $\xi_i$ with $i=1,2,3$ and the spin $S$.
In the integrand of \eqref{Gnu2}, two series of \textit{spurious poles} are generated by the measure $c_2$ and the conformal block $g_{\Delta,S}$. In App.\ref{appendix:sp} we will prove that the contribution of those poles cancel under the condition
\begin{align}
h_{i\,{3+S+k},S}-h_{i\,{3+S},S+k}=0\,\,\, \qquad k=0,2,4\dots\,\,,
\end{align} which happens to be satisfied.

Finally $\mathcal{G}_{\mathcal{O}_1\mathcal{O}_2}$ is given by the sum of only the residues at the physical pole \eqref{spectraleq}. Then we can rewrite the correlation function in the standard form of a conformal partial wave expansion as follows
\begin{equation}\label{OPE}
\mathcal{G}_{\mathcal{O}_1\mathcal{O}_2}=\sum_{\Delta,S\geq 0}C_{\Delta,S}\;g_{\Delta,S}(u,v)\,,
\end{equation}
with the OPE coefficients  $C_{\Delta,S}$ defined as
\begin{equation}
C_{\Delta,S}=\frac{(-1)^S}{c_B^4}\,4\pi\,\text{Res}_\Delta\left(\frac{1}{c_2(\nu,S)}\frac{{h_B}_{\Delta,S}}{1-\chi_B{h_B}_{\Delta,S}-\chi_F{h_F}_{\Delta,S}}\right)\,.
\end{equation}
The sum over $\Delta$ in \eqref{OPE} runs over the solutions of the spectral equation for scaling dimensions of exchanges operators with $\text{Re}(\Delta)>2$ \footnote{This condition in the OPE is equivalent to the restriction \(\text{Re}(i\nu)>0\) in the contour integral in \eqref{Gnu2}.}.

In the following sections, we will focus on the computation of the point-split four-point correlation functions of the operators introduced in \eqref{operators}, establishing the Hamiltonian operators $\hat{\mathcal{H}}_i$ and the constants $\chi_i$ appearing in \eqref{Ggeneral} from their Feynman diagram expansion. We closely follow in our analysis the logic of \cite{Gromov:2018hut}, but in contrast to this paper which treats the bi-scalar fishnet CFT, we have to introduce new types of diagrams into the Bethe-Salpeter procedure, reflecting a richer structure of the full three-coupling \(\chi\)CFT. To write the correlation function \eqref{Ggeneral}  in the standard OPE representation  requires, as the only dynamical input,  the knowledge of eigenvalues ${h_i}_{\Delta,S}$ of the Hamiltonian operators.  We will diagonalize $\hat{\mathcal{H}}_i$  to extract the conformal data, \textit{i.e.} the scaling dimensions of the exchanged operators and the OPE coefficients. In what follows we consider only single scalar fields as protected external operators and then we should set \(\Delta_{\mathcal{O}_1}=\Delta_{\mathcal{O}_2}=1\). Since the four-point correlator constructed from the second operator of \eqref{operators} is trivial (see Sec.\ref{sec:sector3}), in the following two sections we will analyze the remaining two.

\section{Exact four-point correlations function for $\mathcal{O}_1(x)=\mathcal{O}_2(x)=\phi_1(x)$}\label{sec:phi1phi1}

In this section we consider the four-point correlators associated to the first operator of \eqref{operators}, namely when $\mathcal{O}_1(x)=\mathcal{O}_2(x)=\phi_j(x)$ with $j=1,2,3$. 
Since the computation of the correlators is the same for any $j$, we will consider the case $j=1$ and then the four-point function we want to study takes the following form
\begin{equation}\label{phi1phi1}
G_{\phi_1\phi_{1}}(x_1,x_2|x_3,x_4)=\langle\text{Tr}[\phi_1(x_1)\phi_{1}(x_2)]\text{Tr}[\phi_1^\dagger(x_3)\phi_{1}^\dagger(x_4)]\rangle\,.
\end{equation}
This correlation function was extensively studied in \cite{Grabner:2017pgm} in the simplest case of the family of theories we are inspecting, \textit{i.e.} the bi-scalar theory \eqref{bi-scalarL}.

In the planar limit $N_c\rightarrow \infty$, once chosen $j=1$, the weak coupling expansion of \eqref{phi1phi1} in terms of Feynman diagrams is given by a combination of the following bosonic vertices
\begin{equation}\begin{split}\label{vvv}
(4\pi)^2\xi^2_{3}  &\text{Tr}[\phi_1^\dagger\phi_{2}^\dagger\phi_2\phi_{2}](x)\,, \qquad
(4\pi)^2\xi^2_{2}  \text{Tr}[\phi_{3}^\dagger\phi_{1}^\dagger\phi_{3}\phi_{1}](x)\,,\\
&\qquad\;(4\pi)^2\alpha_{1}^2 \, \text{Tr}[\phi_1\phi_1](x) \,\text{Tr}[\phi_1\phi_1]^\dagger (x)\,,
\end{split}\end{equation}
and the following Yukawa vertices
\begin{equation}\label{fff}
4\pi i\sqrt{\xi_{2} \xi_{3}} \,\text{Tr}[\overline{\psi}_{3} \phi_1^\dagger\, \overline{\psi}_{2} ](x)\,, \qquad 
4\pi i\sqrt{\xi_{2} \xi_{3}} \,\text{Tr}[\psi_{3} \phi_1 \,\psi_{2} ](x) \,.
\end{equation}
In the following we will study the correlation function \eqref{phi1phi1} with the Bethe-Salpeter method.

\subsection{The Bethe-Salpeter method for the correlator $G_{\phi_1\phi_{1}}$}

\begin{figure}[!t]
 \centering
   \includegraphics[width=\textwidth]{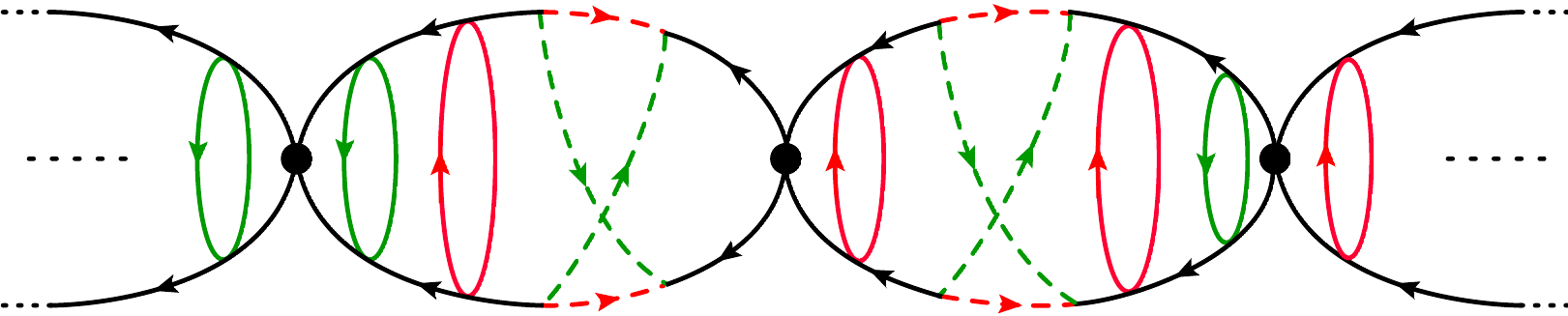}
\caption{ A Feynman diagram contributing to the perturbative expansion $G_{\phi_1\phi_{1}}^{(\ell)}$.  
The black dots stand for double-trace vertices and tick and dashed lines correspond to bosonic and fermionic propagators respectively. The colors represent different flavors $j$ of the particles $\phi_j$ and $\psi_j$: in particular black for $j=1$, red for $j=2$ and green for $j=3$. The propagators are not crossing and are curved to stress the fact that they have a cylindrical topology.}
  \label{fig:phi1phi1}
 \end{figure}

The perturbative expansion of \eqref{phi1phi1} can be written in the following form
\begin{equation}\label{Gpert1}
G_{\phi_1\phi_{1}}(x_1,x_2|x_3,x_4)=\sum_{\ell=0}^\infty(4\pi)^{4\ell}\,G_{\phi_1\phi_{1}}^{(\ell)}(x_1,x_2|x_3,x_4)\,,
\end{equation}
where $G_{\phi_1\phi_{1}}^{(\ell)}$ at any perturbative order $\ell$ contains contributions from the bosonic and fermionic integrals with different coupling dependencies. In Fig.\ref{fig:phi1phi1}, we present an example of an arbitrary Feynman diagram contributing to $G_{\phi_1\phi_{1}}^{(\ell)}$.
The black dots represents insertions of the double-trace vertex in the last line of \eqref{vvv} that in the Bethe-Salpeter picture are associated with the operator $\hat{\mathcal{V}}$ defined in $\eqref{V}$. Then it is straightforward to fix the normalization of its coupling constant in \eqref{Ggeom} as follows
\begin{equation}
\chi_\mathcal{V}=(4\pi)^2\,\alpha_1^2\,.
\end{equation}
In Sec.\ref{sec:bethe}, we discussed  the role of the double-trace terms in the computation of the four-point function, discovering that they are not contributing to the spectral equation. Then, similarly to observations of \cite{Gromov:2018hut,Grabner:2017pgm},   as far as we consider the perturbative expansion \eqref{Gpert1} in the point-splitting $x_1\neq x_2$ and $x_3\neq x_4$, we need only to sum over the single trace contributions, namely the diagrams inside the chain link of Fig.\ref{fig:phi1phi1}. In Sec.\ref{sec:feynman} we will present in detail how the relation between single- and double-trace terms is crucial for the setting of the fixed point \eqref{fixedpoint}.

\begin{figure}[!t]
 \centering
  \subfigure[$\xi_1^0\xi_2^0\xi_3^0$]
   {\includegraphics[width=3.2cm]{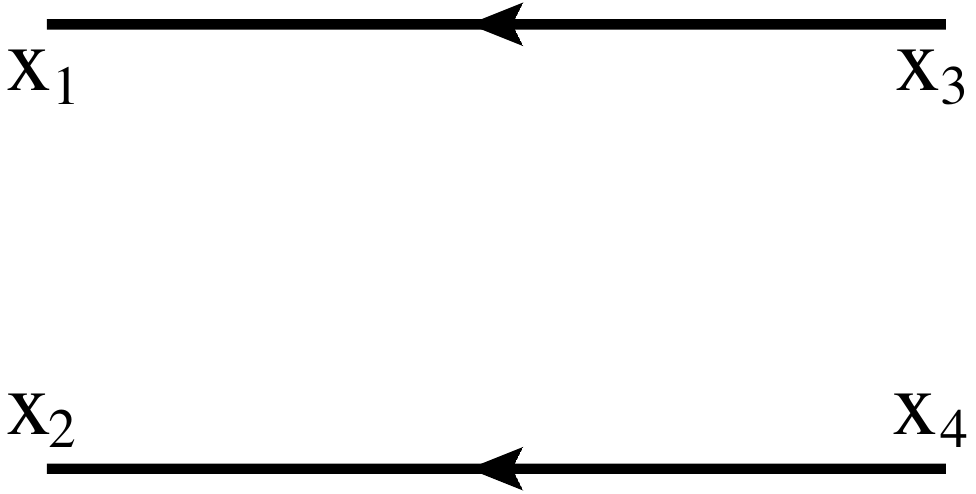}}
    \hspace{5mm}
    \subfigure[$\xi_3^4$]
   {\includegraphics[width=3.2cm]{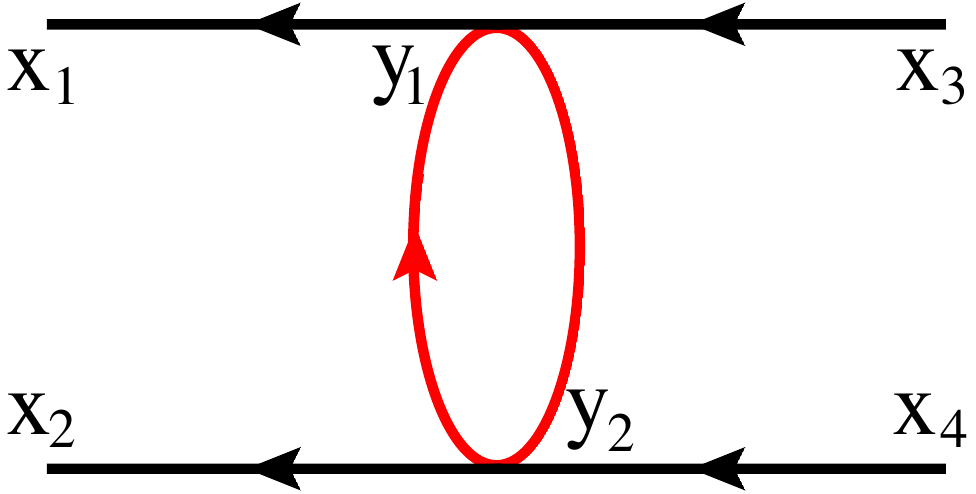}}
    \hspace{5mm}
    \subfigure[$\xi_2^4$]
   {\includegraphics[width=3.2cm]{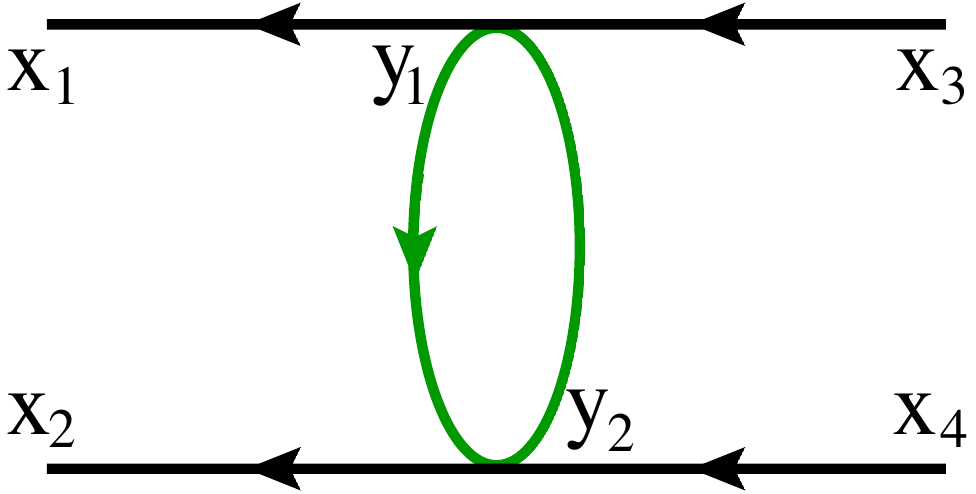}}
    \hspace{5mm}
    \subfigure[$\xi_2^2\xi_3^2$]
   {\includegraphics[width=3.2cm]{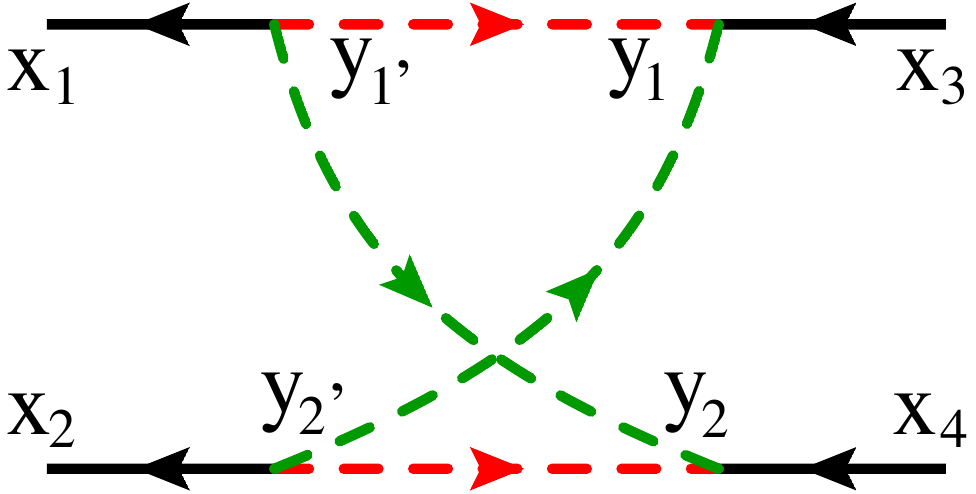}}
    \hspace{5mm}
 \caption{First contribution to the four-point functions $G_{\phi_1\phi_{1}}$.}
  \label{fig:pert1}
 \end{figure}

The first two orders of the perturbative expansion are given by the diagrams represented in Fig.\ref{fig:pert1} and they  can be written as follows
\begin{equation}\begin{split}\label{GGpert1}
G_{\phi_1\phi_{1}}^{(0)}\!&=\frac{c_B^2}{x_{13}^2x_{24}^2}\,,\\
G_{\phi_1\phi_{1}}^{(1)}\!&=c_B^6(\xi^4_{2}\!+\xi^4_{3}) \int \,\frac{d^4 y_1 d^4 y_2 }{(x_1-y_1)^2(x_2-y_2)^2(y_{12}^2)^2(y_1-x_3)^2(y_2-x_4)^2}\\
&\!\qquad-c_B^4c_F^4\xi_{2}^2\xi_{3}^2\!\!\int\!\!\frac{\prod_{i=1}^2 \! d^4 y_id^4 y_{i'} \;\text{Tr}\left[\sigma_\mu \overline{\sigma}_\rho \sigma_\eta \overline{\sigma}_\nu \right]\;
y_{22'}^\mu y_{2'1}^\rho y_{11'}^\eta y_{1'2}^\nu}{(x_1-y_{1'})^2 (x_2-y_{2'})^2 y_{22'}^4y_{2'1}^4y_{11'}^4y_{1'2}^4(y_1-x_3)^2 (y_2-x_4)^2 }, 
\end{split}\end{equation}
where each scalar propagator brings in the factor $c_B/x_{ij}^2$ and each fermionic propagator the factor $c_F \slashed{x}_{ij}/x_{ij}^4$, where $\slashed{x}$ can be $\sigma_\mu x^\mu$ or $\bar{\sigma}_\mu x^\mu$. Since  the fermionic propagator can also be written as $c_B \slashed{\partial}_{x_i} 1/x_{ij}^2$ we conclude that $c_F=-2c_B=-1/(2\pi^2)$. 
 These functions can be expressed in terms of a combination of the Hamiltonian graph-building operators $\hat{\mathcal{H}}_i$. 
  \begin{figure}[t]
 \centering
  \subfigure[$\mathcal{H}_B(x_1,x_2|x_3,x_4)$]
   {\includegraphics[width=3.2cm]{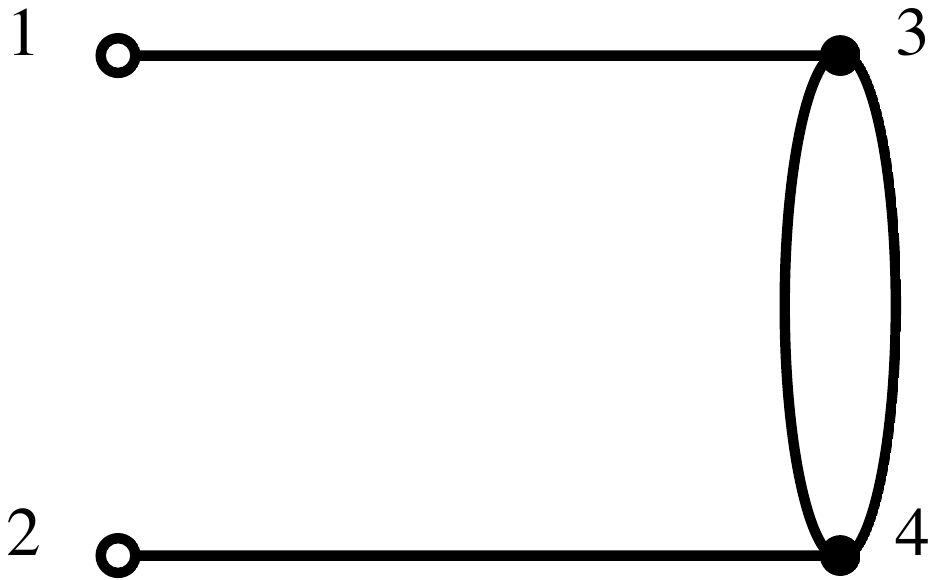}}
    \hspace{10mm}
    \subfigure[$\mathcal{H}_F(x_1,x_2|x_3,x_4)$]
   {\includegraphics[width=4.15cm]{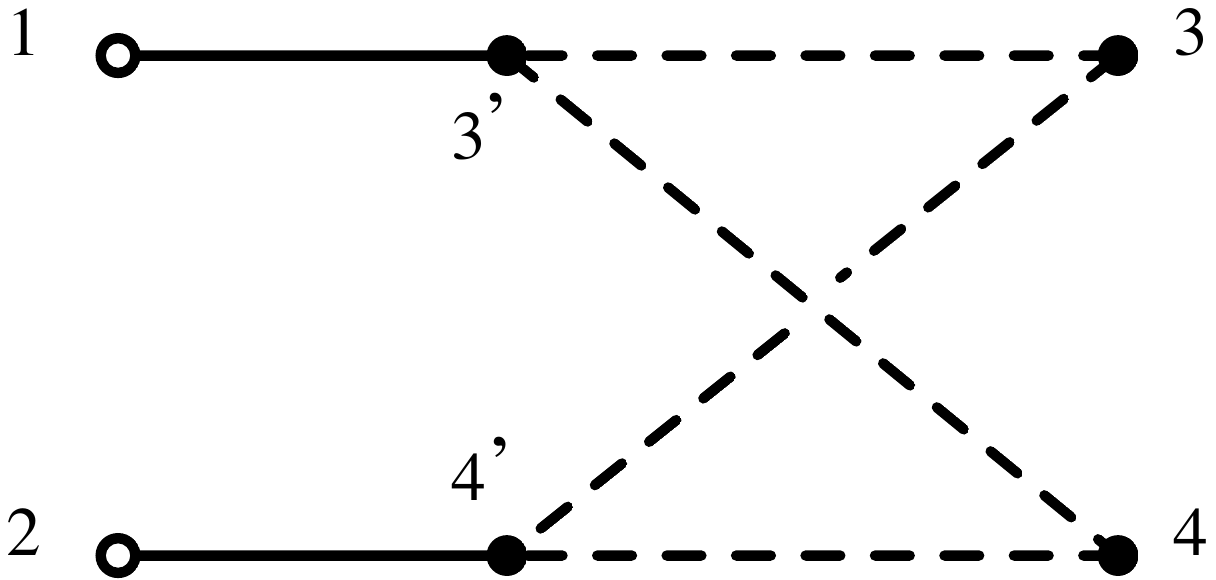}}    
 \caption{The kernels associated to the Hamiltonian graph-building operators $\hat{\mathcal{H}}_B$ and $\hat{\mathcal{H}}_F$ involved in the computation of the four-point function $G_{\phi_1\phi_{1}}$ with $j=1,2,3$. White dots represent external points and black dots integration over the full space \({R}^4\).}
  \label{fig:kernel1}
 \end{figure} 
Indeed defining the following kernels
\begin{equation}
\begin{split}\label{kernel1}
\mathcal{H}_B(x_1,x_2|x_3,x_4)=&\;\frac{c_B^4}{x_{13}^2x_{24}^2x_{34}^4}\,,\\
\mathcal{H}_F(x_1,x_2|x_3,x_4)=&-c_B^2\,c_F^4\int\frac{d^4x_{3'}d^4x_{4'}\,\text{Tr}\left[\sigma_\mu \overline{\sigma}_\rho \sigma_\eta \overline{\sigma}_\nu \right]\,x_{44'}^\mu x_{4'3}^\rho x_{33'}^\sigma x_{3'4}^\nu}{x_{13'}^2x_{24'}^2x_{44'}^4 x_{4'3}^4 x_{33'}^4 x_{3'4}^4}\,,
\end{split}
\end{equation}
represented in Fig.\ref{fig:kernel1}, we can rewrite \eqref{GGpert1} as follows
\begin{equation}\begin{split}\label{GpertHH}
G_{\phi_1\phi_{1}}^{(0)}\!&=\frac{x_{34}^4}{c_B^2}\mathcal{H}_B(x_1,x_2|x_3,x_4)\,,\\
G_{\phi_1\phi_{1}}^{(1)}\!&=\frac{x_{34}^4}{c_B^2}\int d^4y_1d^4y_2\biggl[(\xi^4_{2}\!+\xi^4_{3})\;\mathcal{H}_B(x_1,x_2|y_1,y_2)\;\mathcal{H}_B(y_1,y_2|x_3,x_4)\\
&\qquad\qquad\qquad\qquad\qquad\quad+\xi_{2}^2\xi_{3}^2\;\mathcal{H}_F(x_1,x_2|y_1,y_2)\;\mathcal{H}_B(y_1,y_2|x_3,x_4)\biggl]\,.
\end{split}\end{equation}
The kernels \eqref{kernel1} transform covariantly under conformal transformations\footnote{The easiest way to prove it is to apply the inversion operator to \eqref{kernel1}. For the fermionic Hamiltonian it is convenient to use its representation after the two integrations will be  performed later in \eqref{kernelF2}.}, then the corresponding Hamiltonian integral operators commute with the generators of the conformal group.

When carrying on  the perturbative expansion, it becomes clear that an arbitrary diagram at order $\ell$ is given by
\begin{equation}
\hat{G}_{\phi_1\phi_{1}}^{(\ell)}=\frac{x_{34}^4}{c_B^2}\;\hat{\mathcal{H}}_B\left[(\xi^4_{2}\!+\xi^4_{3})\hat{\mathcal{H}}_B+\xi_{2}^2\xi_{3}^2\hat{\mathcal{H}}_F\right]^\ell\,.
\end{equation}
Then the correlator \eqref{Gpert1} can be presented in the following operatorial form
\begin{equation}\label{Gresum}
\hat{G}_{\phi_1\phi_{1}}=\sum_{\ell=0}^\infty(4\pi)^{4\ell}\hat{G}_{\phi_1\phi_{1}}^{(\ell)}=\frac{x_{34}^4}{c_B^2}\frac{\hat{\mathcal{H}}_B}{1-(4\pi)^4(\xi^4_{2}\!+\xi^4_{3})\hat{\mathcal{H}}_B-(4\pi)^4\xi_{2}^2\xi_{3}^2\hat{\mathcal{H}}_F}\,.
\end{equation}
Comparing it with the definition \eqref{Ggeom} we fix the values of the constants $\chi_i$ (in this case $\hat{\mathcal{V}}$ is not contributing)
\begin{equation}\label{chi1}
\chi_B=(4\pi)^4(\xi^4_{2}\!+\xi^4_{3})\,,\qquad\chi_F=(4\pi)^4\xi_{2}^2\xi_{3}^2\,.
\end{equation}

\subsection{Eigenvalues of the Hamiltonian graph-building operators}
\label{sec:eigen1}

Writing the four-point correlation function in the standard OPE form, as presented in detail in Sec.\ref{sec:bethe}, involves the computation of the spectrum of the graph-building operators \eqref{kernel1}. The eigenstate that diagonalize the Hamiltonians is defined in \eqref{Phi-def} for $\Delta_{\mathcal{O}_1}=\Delta_{\mathcal{O}_2}=1$ and the eigenvalues are defined by means of equations \eqref{eigen2} and \eqref{eigen3}.
Substituting in the latter the kernels \eqref{kernel1} and using the definition \eqref{eigengeneral}, we will end up with a set of integrals that can be computed with the help of the star-triangle relations presented in App.\ref{sec:appendixC} (also known as \textit{uniqueness method}). The fact that all the integrals that we have to compute can be computed by means of the star-triangle relations is a consequence of the underlying conformal symmetry.

\paragraph{Bosonic eigenvalue:}
The bosonic eigenvalue ${h_B}_{\Delta,S}$ is defined in \eqref{eigen2}. Using the bosonic Hamiltonian \eqref{kernel1}, this relation can be written in the following integral form
\begin{equation}\label{inthb}
c_B^4\int\frac{d^4y_1d^4y_2}{(x_{1}-y_1)^2(x_{2}-y_2)^2y_{12}^4}\Phi_{\Delta,S,x_0}(y_1,y_2)={h_B}_{\Delta,S}\Phi_{\Delta,S,x_0}(x_1,x_2)\,.
\end{equation}
In the case of $S=0$, the function $\Phi_{\Delta,S,x_0}$ reduces to \eqref{eigenfunction} and the computation is straightforward. Indeed, one needs to apply the star-triangle relations two times as follows\footnote{\label{STR}It is convenient to perform this and other similar computations, together with the pictures,  with the \texttt{STR} package \cite{Preti:2018vog,Preti:2019rcq}.}
\begin{equation*}\label{STRbos}
\vcenter{\hbox{\includegraphics[trim={1.2cm 0 0 0},clip,width=2.5cm]{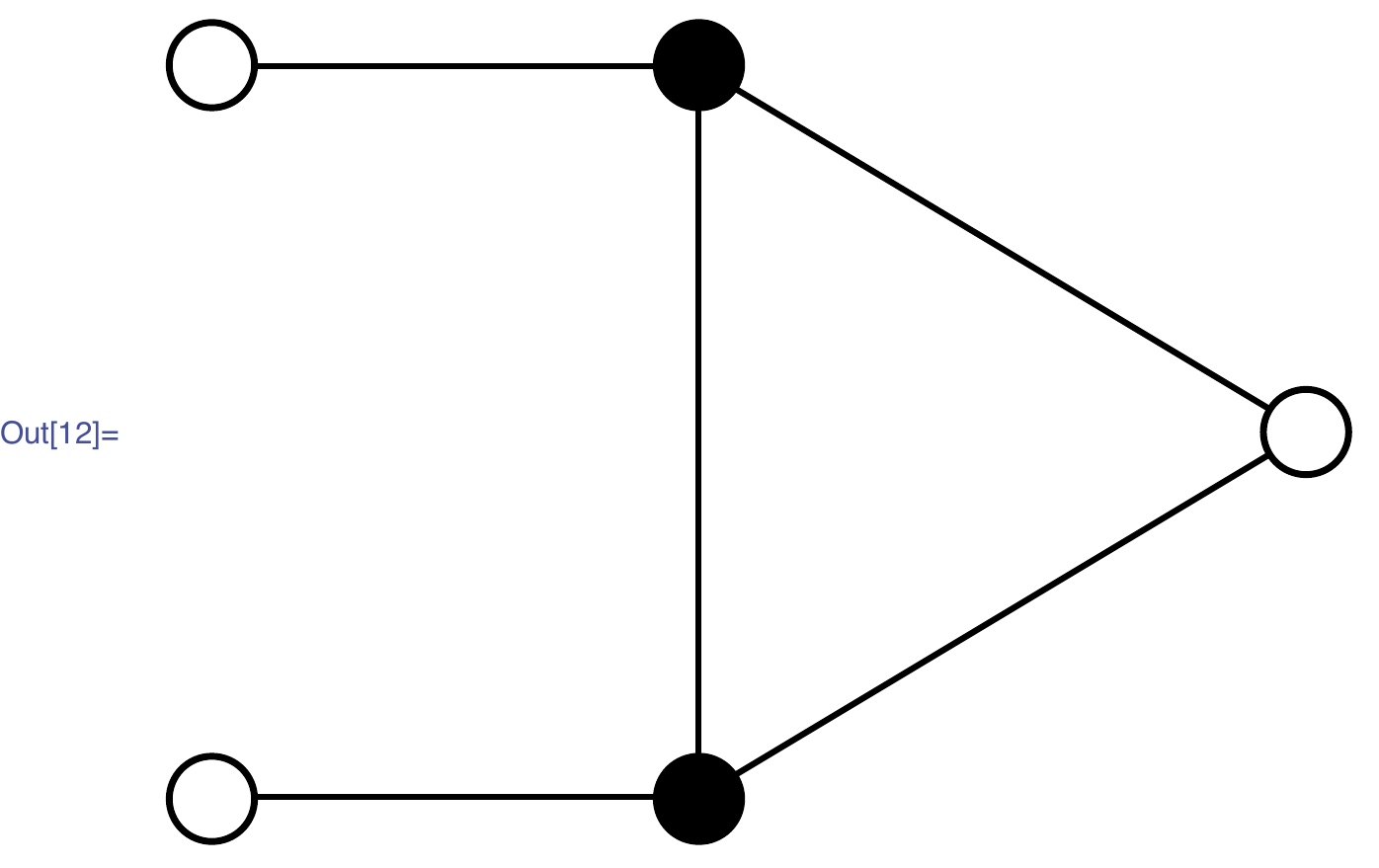}}}\quad\overset{\text{STR}}{\Longrightarrow}\quad
\vcenter{\hbox{\includegraphics[trim={1.2cm 0 0 0},clip,width=2.5cm]{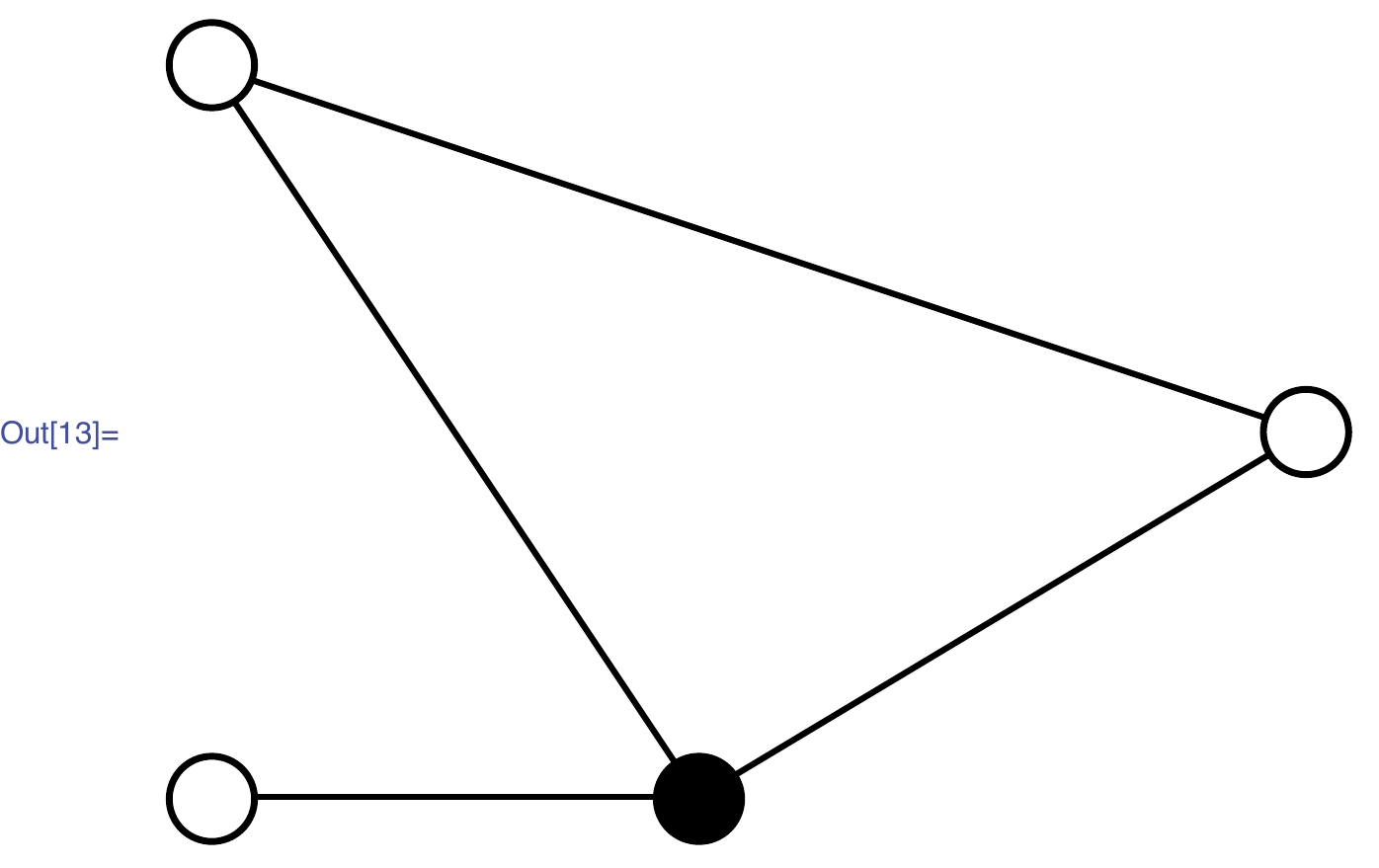}}}\quad\overset{\text{STR}}{\Longrightarrow}\quad
\vcenter{\hbox{\includegraphics[trim={1.2cm 0 0 0},clip,width=1.5cm]{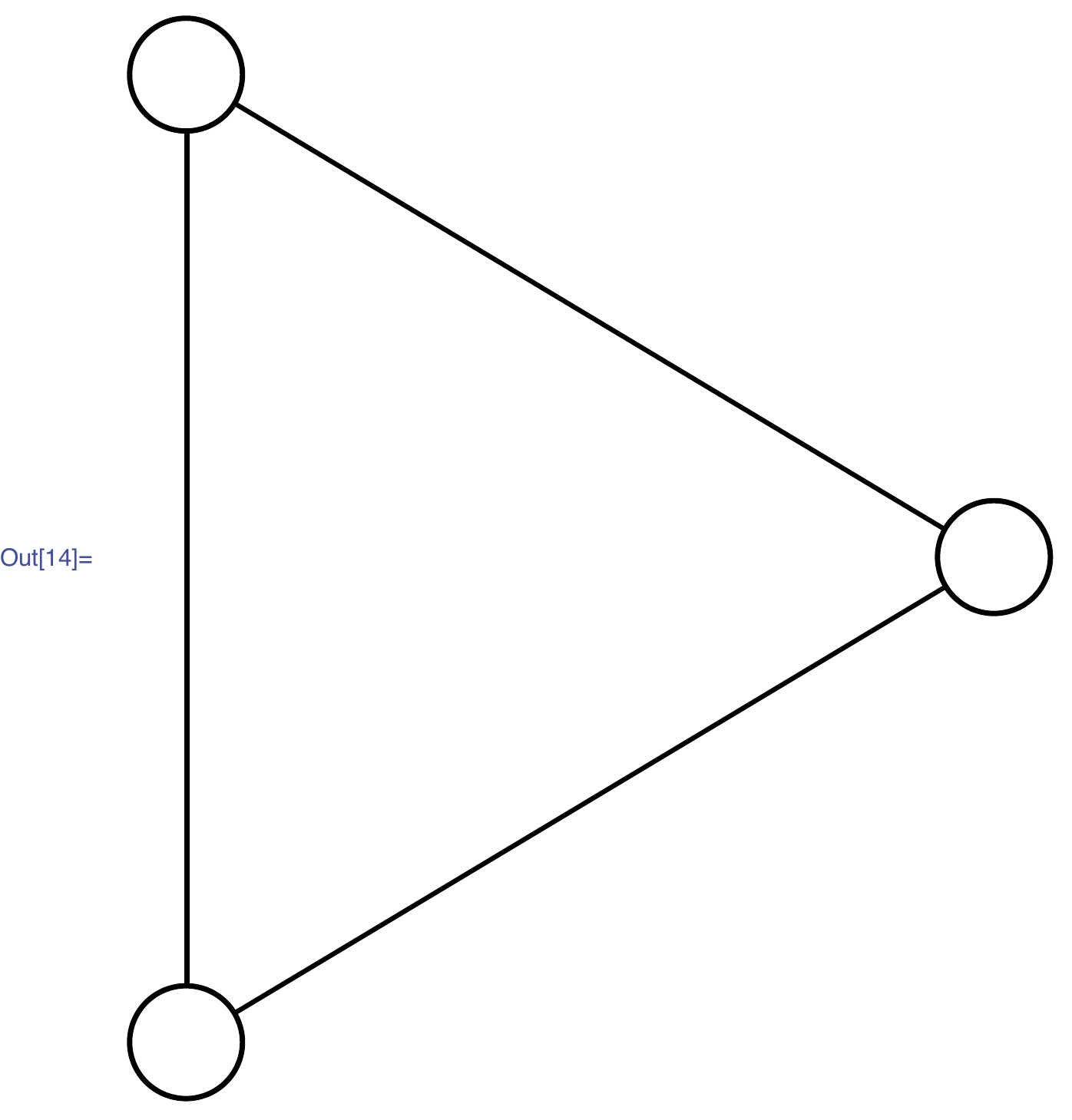}}}\,.
\end{equation*}
to obtain
\begin{equation}\label{hbS0}
{h_B}_{\Delta,0}=\frac{16\pi^4 c_B^4}{\Delta(\Delta-2)^2(\Delta-4)}\,.
\end{equation}
The eigenvalue at $S\neq 0$ can be computed in the same way, using the generalization of star-triangle relation to any-spin case \ref{uniqfermS} derived in \cite{Chicherin:2012yn}. The computation can be otherwise done in a more tedious and explicit way as presented in detail in \cite{Gromov:2018hut}. The result reads~\cite{Grabner:2017pgm}
\begin{equation}\label{hbS}
{h_B}_{\Delta,S}=\frac{16\pi^4 c_B^4}{(\Delta+S)(\Delta+S-2)(\Delta-S-2)(\Delta-S-4)}\,.
\end{equation}
The eigenvalue is invariant under $\Delta\rightarrow 4-\Delta$, as expected from \eqref{hsymmetry}.

\paragraph{Fermionic eigenvalue}
The fermionic eigenvalue ${h_F}_{\Delta,S}$ is defined in \eqref{eigen3}. This is a new object, absent in the similar correlator of  bi-scalar model treated in~\cite{Grabner:2017pgm}.  
First of all we can simplify the fermionic Hamiltonian in \eqref{kernel1} integrating the primed variables by means of the Yukawa star-triangle identity \eqref{uniqferm}
as follows (red lines are spin-1/2 fermionic propagators)\begin{equation*}\label{STRferm1}
\vcenter{\hbox{\includegraphics[trim={1.2cm 0 0 0},clip,width=3.2cm]{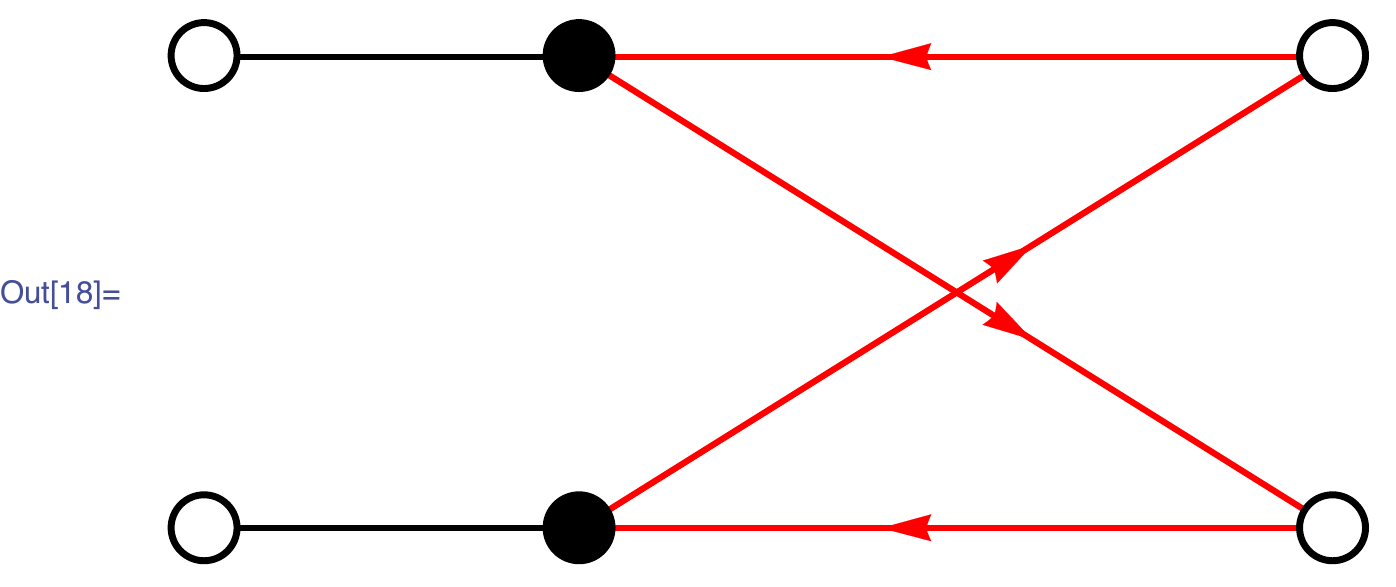}}}\quad\overset{\text{STR}}{\Longrightarrow}\quad
\vcenter{\hbox{\includegraphics[trim={1.2cm 0 0 0},clip,width=3.2cm]{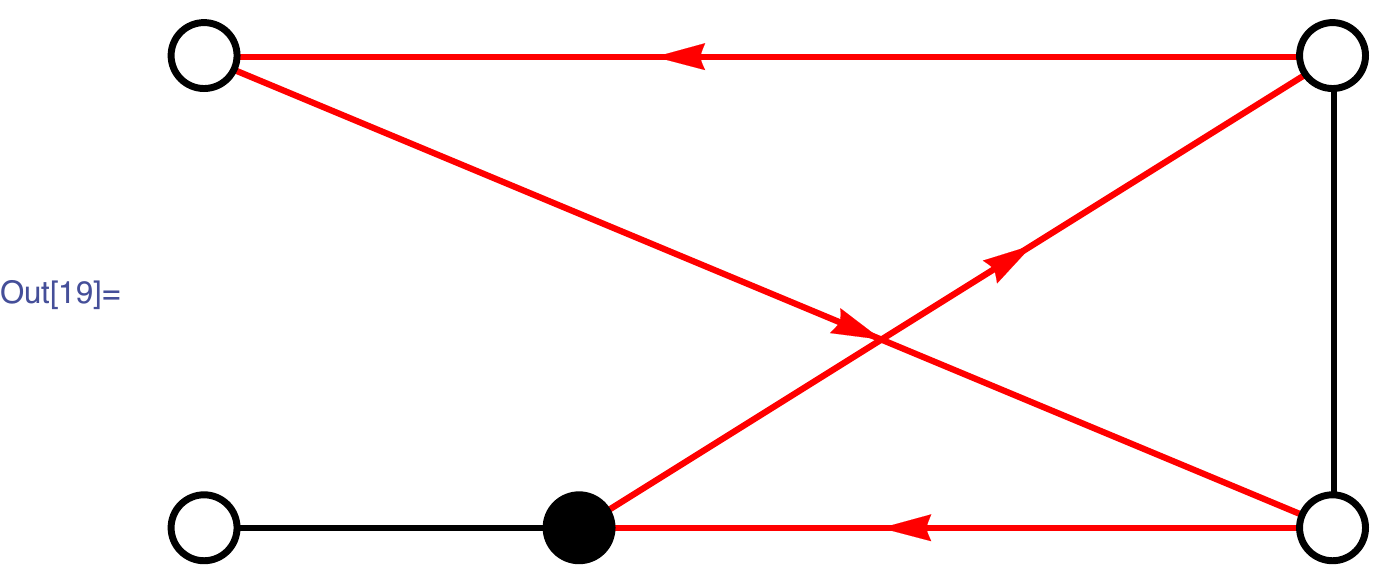}}}\quad\overset{\text{STR}}{\Longrightarrow}\quad
\vcenter{\hbox{\includegraphics[trim={1.2cm 0 0 0},clip,width=3.2cm]{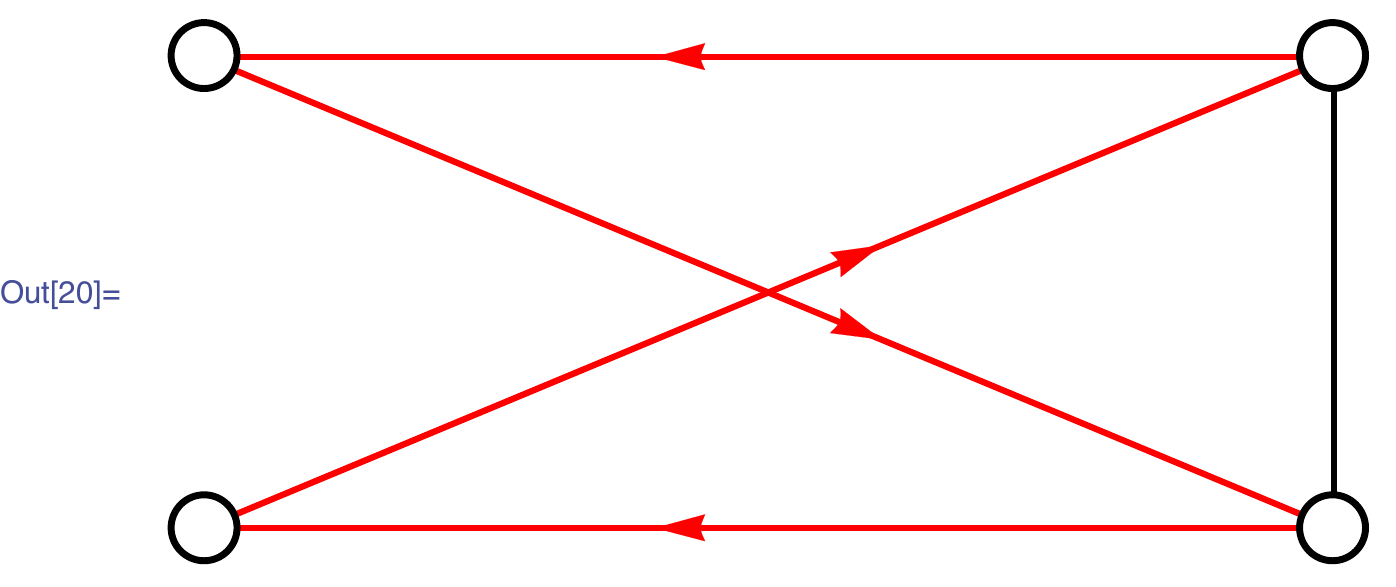}}}
\end{equation*}
where the computation and figures are made with the \texttt{STR} package (see footnote \ref{STR}). We obtain the following kernel
\begin{equation}
\mathcal{H}_F(x_1,x_2|x_3,x_4)=-\pi^4c_B^2\,c_F^4\frac{\text{Tr}\left[\sigma_\mu \overline{\sigma}_\rho \sigma_\eta \overline{\sigma}_\nu \right]\,x_{42}^\mu x_{23}^\rho x_{31}^\eta x_{14}^\nu}{x_{42}^2x_{23}^2x_{31}^2 x_{14}^2 x_{34}^4}\,.
\end{equation}
Using the formula for the trace of four \(\sigma\)-matrices \eqref{tracesigma} and simplifying the scalar products by means of \eqref{scalarproduct}, we can rewrite the fermionic hamiltonian in the following form
\begin{equation}\label{kernelF2}
\mathcal{H}_F(x_1,x_2|x_3,x_4)=\pi^4c_B^2\,c_F^4\tilde{\mathcal{H}}_F(x_1,x_2|x_3,x_4)-2\mathcal{H}_B(x_1,x_2|x_3,x_4)\,.
\end{equation}
where we used the symmetry $\mathcal{H}_B(x_1,x_2|x_3,x_4)=\mathcal{H}_B(x_2,x_1|x_3,x_4)$ of the bosonic hamiltonian studied in the previous paragraph, and  $\tilde{\mathcal{H}}_F$ is defined by
\begin{equation}\label{HFtilde1}
\tilde{\mathcal{H}}_F(x_1,x_2|x_3,x_4)\equiv \frac{x_{12}^2}{x_{42}^2x_{23}^2x_{31}^2 x_{14}^2 x_{34}^2}\,.
\end{equation}
Then the fermionic eigenvalue ${h_F}_{\Delta,S}$ consists of the bosonic eigenvalue \eqref{hbS} and the eigenvalue of $\tilde{\mathcal{H}}_F$ defined as follows
\begin{equation}\label{eigentilde}
\left[\hat{\tilde{\mathcal{H}}}_F\; \Phi_{\Delta,S,x_0}\right](x_1,x_2)  = {\tilde h_F}{}_{\Delta,S} \; \Phi_{\Delta,S,x_0}(x_1,x_2)\,,
\end{equation} 
such that
\begin{equation}\label{hf(p)}
{h_F}_{\Delta,S}=\pi^4c_B^2\,c_F^4\,{\tilde h_F}{}_{\Delta,S}-2\,{h_B}_{\Delta,S}\,.
\end{equation}
Let's focus on the relation \eqref{eigentilde}. It can be written in the following integral form
\begin{equation}\begin{split}\label{inthftilde}
\int\!\!\frac{d^4 y_1 d^4 y_{2}\;x_{12}^2}{(y_2\!-\!x_2)^2 (x_2\!-\!y_1)^2 (y_1\!-\!x_1)^2 (x_1\!-\!y_2)^2 y_{12}^2}\Phi_{\Delta,S,x_0}(y_1,y_2)={\tilde h_F}{}_{\Delta,S}\Phi_{\Delta,S,x_0}(x_1,x_2).
\end{split}\end{equation}
In order to simplify the computation, we consider the limit in which $x_0\rightarrow \infty$ on both sides of \eqref{inthftilde}. 
In this limit the eigenvalue ${\tilde h_F}{}_{\Delta,S}$ is given by the following integral
\begin{equation}\begin{split}\label{inthftilde2}
{\tilde{h}_F}{}_{\Delta,S}=\int\!\!\frac{d^4 y_1 d^4 y_{2}\;(n\,y_{12})^S}{(y_2\!-\!x_2)^2 (x_2\!-\!y_1)^2 (y_1\!-\!x_1)^2 (x_1\!-\!y_2)^2 (y_{12}^2)^{2-p}}\,,
\end{split}\end{equation}
where $p=\tfrac{\Delta-S}{2}$ and we put $x_{12}^2=(n x_{12})=1$ for convenience. Notice that the integrand is antisymmetric in the exchange $y_1\leftrightarrow y_2$ for odd S, then the eigenvalue ${\tilde h_F}{}_{\Delta,S}$ is non-zero only for even S.

In the $S=0$ case, the integral \eqref{inthftilde2} is known as a massless two-loop self-energy Feynman integral, or \textit{kite}. Its value is known for any power of the propagator $1/y_{12}^2$ in terms of an hypergeometric function \cite{Grozin:2012xi}, then  
\begin{equation}\begin{split}\label{I3sol}
{\tilde{h}_F}{}_{\Delta,0}
\!=\!-2\pi^4\Gamma(\tfrac{\Delta}{2}\!-\!1)\Gamma(1\!-\!\tfrac{\Delta}{2})\biggl[\frac{{}_3 F_2(1,2,\tfrac{\Delta}{2};\tfrac{\Delta}{2}+1,\tfrac{\Delta}{2}+1|1)}{\Delta/2\,\Gamma(\tfrac{\Delta}{2}+1)\Gamma(2-\tfrac{\Delta}{2})}\!+\pi \cot \pi(4\!-\!\tfrac{\Delta}{2})\biggl]\,,
\end{split}\end{equation}
where $\Delta=2+2i\nu$. Expanding \eqref{I3sol} around $\nu=0$, one can notice that the cotangent cancels all the odd terms of the hypergeometric functions. The analytic properties of \eqref{I3sol} are more clear when writing it in the following equivalent form
\begin{equation}\label{hftildeS0}
{\tilde h_F}{}_{\Delta,0} =\,\pi^4\frac{\psi^{(1)}\left(\tfrac{\Delta}{4}\right)-\psi^{(1)}\left(\tfrac{\Delta}{4}-\tfrac{1}{2}\right)}{2-\Delta}+(\Delta\rightarrow 4-\Delta)\,,
\end{equation}
where $\psi^{(1)}(x)=d\psi(x)/dx$ and $\psi(x)$ is the digamma function. 

When $S\neq 0$, we can appeal to a similar computation made in \cite{Gromov:2018hut}. In fact, the same integral of \eqref{inthftilde2} appears in the study of the spectrum of the graph-building operator associated to the 2-magnon correlation function. The 2-magnon Hamiltonian is $\mathcal{H}_{\text{2-magnon}}=x_{34}^2/x_{12}^2\tilde{\mathcal{H}}_F$ but, when applied to the eigenstate \(\Phi_{\Delta,S,x_0}\), that has in this case  \(\Delta_{\mathcal{O}_1}=\Delta_{\mathcal{O}_2}=2\), it leads to an eigenvalue with the same integral representation as \eqref{inthftilde2}. The strategy to compute the eigenvalue is to write the following recursion relation for the integrals
\begin{equation}
{\tilde{h}_F}{}_{\Delta,S}=\frac{1-S}{1+S}\,{\tilde{h}_F}{}_{\Delta,S-2}+\frac{64\pi^4 S}{(S+1)[S^2-(\Delta-2)^2]^2}\,.
\end{equation} 
Solving the recurrence with   the eigenvalue  \({\tilde{h}_F}{}_{\Delta,S=0}\),  given by \eqref{hftildeS0}, as initial condition, we obtain
\begin{equation}\label{hftildeS}
{\tilde h_F}{}_{\Delta,S} =\,\pi^4\frac{\psi^{(1)}\left(\tfrac{\Delta+S}{4}\right)-\psi^{(1)}\left(\tfrac{\Delta+S}{4}+\tfrac{1}{2}\right)}{(2-\Delta)(S+1)}+(\Delta\rightarrow 4-\Delta)\,.
\end{equation}
We can conclude that the eigenvalue \eqref{hf(p)} is manifestly invariant under $\Delta\rightarrow 4-\Delta$, as expected from \eqref{hsymmetry}.
\subsection{Spectrum of exchanged operators of $\mathcal{G}_{\phi_1\phi_1}(u,v)$}\label{sec:spectrumphi1phi1}

In this section we will use the eigenvalues \eqref{hbS} and \eqref{hf(p)} to compute the scaling dimensions of the operators contributing to the correlation function \eqref{OPE} for $\mathcal{O}_1=\mathcal{O}_2=\phi_1$. The spectrum of the exchanged operators is defined by the solutions of the equation for the physical poles \eqref{spectraleq}. Substituting in \eqref{spectraleq} the definition of bosonic and fermionic eigenvalues \eqref{hbS} and  \eqref{hf(p)} and the constants $\chi_i$ computed in \eqref{chi1}, we can rearrange the spectral equation in the following form
\begin{equation}\label{spec11}
{{h_B}_{\Delta,S}}^{-1}-(4\pi^2)^4\,c_B^2\,c_F^4\,\kappa^4\,{\tilde h_F}{}_{\Delta,S}\,{{h_B}_{\Delta,S}}^{-1}=(4\pi)^4\,\omega^4\,,
\end{equation}
where we defined the new couplings
\begin{equation}\label{coupdef1}
\kappa^4=\xi_{2}^2\xi_{3}^2\,,\qquad\text{and}\qquad
\omega^4=(\xi_{2}^2-\xi_{3}^2)^2\,.
\end{equation}
Plugging \eqref{hbS} and \eqref{hftildeS} into \eqref{spec11}, we obtain the  following spectral equation
\begin{align}\label{eq1}
\left(\tfrac{S^2}{4}\!+\!\nu^2\right)\!\!\!\left(\tfrac{(2+S)^2}{4}\!+\!\nu^2\right)\!\!\!\biggl[1+&\frac{i\kappa^4}{2\nu(S+1)}\biggl( \psi ^{(1)}\!\left(\frac{1}{2}\left(1+\tfrac{S}{2}-i \nu\right)\! \right)
\!-\!\psi ^{(1)}\!\left(\frac{1}{2}\left(2+\tfrac{S}{2}-i \nu\right)\! \right)+\nonumber\\
&\!\!\!\!\!\!\!\!\!\!\!\!+\psi ^{(1)}\!\left(\frac{1}{2}\left(2+\tfrac{S}{2}+i \nu\right)\! \right)
-\psi ^{(1)}\!\left(\frac{1}{2}\left(1+\tfrac{S}{2}+i \nu\right)\! \right)
\!\biggl)\biggl]=\omega^4,
\end{align}
with the additional constraint $\text{Im}\,\nu<0$. This equation can be studied perturbatively, for each individual anomalous dimension, expanding in $\nu$ around the value \(\nu_0\) corresponding to a bare dimension \(\Delta_0=2+2i\nu_0\)  at weak coupling, and in \(1/\nu\) at strong coupling.

\paragraph{Weak coupling expansion:}
The small coupling limit suggests that the equation has solutions with bare dimensions $2+S$ and $4+S$, in analogy with the same quantity in the bi-scalar theory \cite{Grabner:2017pgm,Gromov:2018hut}.  
There are six such solutions, but only half satisfies the physical requirement $\text{Re}\,\Delta \geq 2$: one of them corresponds to the scaling dimensions of exchanged operator with bare dimension $\Delta_0=2+S$ and two -- to the scaling dimensions of operators with bare dimensions $\Delta_0=4+S$. The remaining three solutions are related to the first ones b the transformation $\Delta\rightarrow 4-\Delta$ and describe \textit{shadow operators}, with $\text{Re}\,\Delta < 2$. In addition to that, there is an infinite series of physical solutions around the bare dimensions $\Delta_0=t+S$ with $t=6,8,...$  , due to the non algebraic eigenvalue ${h_F}_{\Delta,S}$, similarly to the two-magnon case studied in \cite{Gromov:2018hut}. For each value of the twist \(t\) there are two solutions; they describe the exchange of an infinite tower of local primary operators in the OPE of \eqref{fpoint1}.
Writing $\nu$ as a function of the two couplings \eqref{coupdef1} and expanding around the physical pole $\nu=-iS/2$ at weak coupling $\kappa$, $\omega \rightarrow 0$, we obtain the following expansion for the twist-two operator
\begin{equation}\begin{split}\label{D2S1}
\Delta^{(2)}=
2\!+\!S
\!-\!\frac{2\omega^4}{S(S+1)}&
\!+\!\frac{2\omega^4}{3S^3(S\!+\!1)^3}\biggl[3(S(S\!-\!1)\!-\!1)\omega^4\!-\!6 S(S\!+\!1)\kappa^4[2 H_S^{(2)}\!-\! H^{(2)}_{S/2}]\biggl]
+\dots
\end{split}\end{equation}
and, around the physical pole $\nu=-i(S+2)/2$, the twist-four operators. 
\begin{equation}\begin{split}\label{D4S1}
&\Delta^{(4)}=
4+S
+\frac{4\kappa^2}{\sqrt{(S+1)(S+2)}}
+\frac{(S+2)\omega^4-8\kappa^4}{(S+1)(S+2)^2}+\\
&\qquad\qquad\qquad+\frac{3\tfrac{\omega^8}{\kappa^2}-48\tfrac{(6\!+\!S(S\!+\!6))}{(S+1)(S+2)}\kappa^2\omega^4
-96\kappa^6\left[2 H_{S+2}^{(2)}\!-\!H_{{S}/{2}}^{(2)}\!-\!\frac{12}{(S+2)^2}\!\right]}{24(S+1)^{3/2}(S+2)^{3/2}}
+\dots\\
&\Delta^{(4')}=
4+S
-\frac{4\kappa^2}{\sqrt{(S+1)(S+2)}}
+\frac{(S+2)\omega^4-8\kappa^4}{(S+1)(S+2)^2}+\\
&\qquad\qquad\qquad-\frac{3\tfrac{\omega^8}{\kappa^2}-48\tfrac{(6\!+\!S(S\!+\!6))}{(S+1)(S+2)}\kappa^2\omega^4 
-96\kappa^6\left[2 H_{S+2}^{(2)}\!-\!H_{{S}/{2}}^{(2)}\!-\!\frac{12}{(S+2)^2}\!\right]}{24(S+1)^{3/2}(S+2)^{3/2}}
+\dots
\end{split}\end{equation}
where \(H^{(2)}_k\) are Harmonic numbers of order \(2\).
Remarkably, the expressions in square brackets in \eqref{D2S1}, \eqref{D4S1}~  are in fact  rational  numbers. In both cases,  we present only the first few terms since the following ones are cumbersome. We notice that the weak coupling expansions of \(\Delta^{(4)}\), \(\Delta^{(4')}\) are divergent but, as it will be  pointed out later in the analysis of Sec.\ref{fpoint1}, the sum of the two corresponding OPE contributions has a well defined expansion\footnote{ \label{foot:grisha} We are grateful to G. Korchemsky for the enlighting discussion about this point.}. 
Similar considerations can be made for the solutions at higher twist \(t=\Delta_0-S=6,8,\dots\)
\begin{equation}\begin{split}\label{DtS1}
\Delta^{(t)}_{\pm}\,=\,&
t+S
\pm\frac{4 i^{\frac{t}{2}}\kappa^2}{\sqrt{(S+1)(S+t-2)}}
-\frac{(-1)^{\frac{t}{2}} 8\kappa^4}{(S+1)(S+t-2)^2}+ \dots\, .
\end{split}
\end{equation}
The twist-2 solution corresponds to the operator
\begin{align}
\Tr[\phi_1 \,\partial^S \phi_1] + \text{permutations}\, ,
\end{align}
namely the traceless symmetric \(S\)-tensor obtained by insertion of light cone derivatives \(\partial = n_{\mu} \partial^{\mu}\), \(n^2=0\) into the operator \(\Tr[\phi_1^2]\). At twist-4 the matter content of the theory allows to find several \(S\)-tensor operators satisfying the condition \(\Delta_0-S=4\) and having the right \(U(1)^{\otimes 3}\) quantum numbers (e.g. for \(i=1,j=2\): \((2,0,0)\)). Twist-4 operators start mixing with each other. We perform an introductory analysis of this phenomenon for the simple scalar case \(S=0\) in Appendix~\ref{mix1}. At this stage the log-CFT effects \cite{Gurarie:1993xq} due to chiral interaction vertices in \eqref{fullL} show up. The analysis suggests the presence at twist-4 of only two non-protected physical operators, which should be identified with the two solutions \(\Delta^{(4)}\) and \(\Delta^{(4')}\) at \(S=0\), in contrast  to the bi-scalar fishnet CFT where only one type of twist-4 operators appears~\cite{Grabner:2017pgm}.Similar considerations apply to the higher twist operators \(\Delta_{\pm}^{(t)}\). Indeed also for value of twist \(t>4\) it is possible to find several \(S\)-tensor primary operators with the correct set of Cartan's charges. The detailed study of these operators and their mixing would be an interesting insight in the structure of operator algebra of \(\chi\)CFT. We will restrict from here on most of our analysis to solution of twist two and four, whose contribution to the OPE expansion appears to be enough for complete description of the first non-trivial order of the weak coupling  expansion, confirmed by direct computations in terms of Feynman diagrams.
 
Recalling the definition \eqref{coupdef1}, the weak coupling expansion
 \eqref{D2S1} for the twist-two operator goes in powers of $\xi^4$ of the original couplings which is exactly the expected behavior considering that the perturbative expansion in Fig.\ref{fig:pert1} alternates bosonic and fermionic \textit{wheels} attached to the diagrams with two quartic or four Yukawa single-trace vertices. On the contrary, the weak coupling expansions \eqref{D4S1} for the twist-four operators goes in power of $\xi^2$ of the original coupling. This fact can be understood looking at the expansion of \eqref{eq1} around the physical pole located at $\nu=-i(S+2)/2$. Indeed this expansion starts from $\kappa^4/(\nu+i(S+2)/2)^2$ and as a consequence the four-point correlation function \eqref{Gresum} is convergent when $\nu\rightarrow -i(S+2)/2$ if $\kappa$ is finite while it produces a divergence when we consider the weak coupling limit $\kappa,\,\omega\rightarrow 0$ such that $G_{\phi_1\phi_1}\sim\pm\kappa^2$. 

The zero-spin case presents some peculiar behaviours.
 Indeed, expanding \eqref{eq1} for $S=0$ around the physical poles $\nu=0,-i$ at weak coupling, we obtain the following expansions for the solutions of \eqref{spec11}
\begin{align}
&\Delta^{(2)}\big |_{S=0}=
2
-2i\omega
+i\omega^2[\omega^4\!\!-6\kappa^4\zeta_3]+\frac{i}{4}\omega^2[7\omega^8-12\omega^4\kappa^4(3\zeta_3+5\zeta_5)+108\kappa^8\zeta_3^2]+\dots\nonumber\\
&\Delta^{(4)}\big |_{S=0}=
4
+2\sqrt{2} \kappa^2
+\frac{1}{2}  [\omega^4-4 \kappa^4]
+\frac{ 16 \kappa^6-48\kappa^2 \omega^4+\tfrac{\omega^8}{\kappa^2}}{16 \sqrt{2}}+\dots\\
&\Delta^{(4')}\big |_{S=0}=
4
-2\sqrt{2} \kappa^2
+\frac{1}{2}  [\omega^4-4 \kappa^4]
-\frac{ 16 \kappa^6-48\kappa^2 \omega^4+\tfrac{\omega^8}{\kappa^2}}{16 \sqrt{2}}+\dots\\
&\Delta^{(t)}_{\pm}\big |_{S=0}=
t
\pm\frac{4 i^{\frac{t}{2}}\kappa^2}{\sqrt{(t-2)}}
-\frac{(-1)^{\frac{t}{2}} 8\kappa^4}{(t-2)^2}+ \dots\, \qquad t=6,4,8\dots .
\end{align}
where the one-loop order of the scaling dimension $\Delta^{(2)}\big |_{S=0}$ is in agreement with the prediction \eqref{predictionspectrum}. This twist-\(2\) solution is the scaling dimension of the operator \(\Tr[\phi_1 \phi_2^{\dagger}]\), while the two solutions of twist-\(4\) arise from the operatorial mixing in a similar way as to \(S=0\) case analysed in Appendix~\ref{mix1}.

Notice that the weak-coupling expansion of the twist-two operator is drastically different as compared to the $S\neq 0$ case, indeed it goes in powers of $\xi^2$. The same behavior was noticed in \cite{Gromov:2018hut} and the reason is similar to the one explained above . We observe that, expanding around the physical pole $\nu=0$, the spectral equation \eqref{eq1} goes as $\omega^4/\nu^2$. Then, when $\nu\rightarrow 0$, the correlation function \eqref{Gresum} is convergent if $\mu$ is finite, but it produces a square-root divergence when we expand at weak coupling, as in the previous case. The fact that the weak-coupling and $S\rightarrow 0$ limits are not commutative is related to this divergence.

The divergence in the expansion of the scaling dimension of the twist-two operator is not a surprise. In fact, as noticed also in some different contexts in \cite{Grabner:2017pgm}, in order to write the correlation function in the OPE form as in \eqref{OPE}, we assumed that in the integral \eqref{Gnu2} no physical poles are located on the real $\nu$-axes. However the poles that at weak coupling and when \(S\neq 0\) are situated at \(\nu=\mp iS/2\)  pinch the integration contour at the origin when \(S=0\), thus producing a divergence. Hence, the contribution of the double-traces is needed in this case to produce a non-vanishing term that cancels this divergence at weak coupling. 
Again, we stress that at finite couplings the solutions of \eqref{eq1} are well-defined even at zero spin.

\paragraph{Strong coupling expansion:}
At strong coupling, $\kappa\,,\;\omega\rightarrow \infty$, we consider the four solutions of eq.\eqref{eq1} of lowest twist. 
The solutions are related to the physical poles of the spectral equation located at $\nu=e^{i\pi \tfrac{k}{2}}\sqrt[4]{\omega^4+2\kappa^4}+\dots$ with $k=0,1,2,3$ but only two of them satisfy the condition $\text{Im}\,\nu<0$, the remaining solutions being associated to the shadow operators. However we stress that we are neglecting all the infinite non-algebraic solutions of higher twist, purely generated by ${h_F}_{\Delta,S}$.
Then we have
\begin{equation}\label{strong1}
\Delta_{\infty}=2e^{i\pi\tfrac{k}{2}}\sqrt[4]{\omega^4+2\kappa^4}+2+\frac{[S(S+2)+2]\omega^4+2[S(S+2)-2]\kappa^4}{4e^{i\pi\tfrac{k}{2}}\left[\sqrt[4]{\omega^4+2\kappa^4}\;\right]^5}+\dots
\end{equation}   
Notice that, if all couplings  scale as \(\xi_j\sim\xi\gg 1\),  the strong coupling expansion \eqref{strong1} is growing linearly with $\xi$. This becomes clear if one expands the eigenvalues appearing in  \eqref{spec11}. Indeed both of them decay at large $\nu$ as ${h_B}_{\Delta,S}\,,{\tilde{h}_F}{}_{\Delta,S}\sim~1/\nu^4$ then, since in the spectral equation the couplings appear in power of $\xi^4$, it is evident that the expansion will contain terms linear in $\xi$.  The $S\rightarrow 0$ limit is not singular at strong coupling and one can compute $\Delta_{\infty}\big|_{S=0}$ directly from \eqref{strong1}.

\paragraph{The spectrum of exchanged operators in reductions of \(\chi\)CFT}

In Sec.\ref{sec:section2}, we presented the $\gamma$-deformed $\mathcal{N}=4$ SYM theory in the double-scaling limit as a family of theories. In fact, playing with the three couplings $\xi_j$ with $j=1,2,3$ it is possible to describe different Lagrangians with different matter contents and symmetries. Thus we want to obtain the spectrum of exchanged operators for each theory of this family simply taking the limit on the couplings in the spectral equation \eqref{eq1} of the most general doubly-scaled theory. First of all, we recall the well-known result for the spectrum for the simplified Lagrangian \eqref{bi-scalarL} also known as \(4D\) \text{bi-scalar} fishnet CFT.
In this  theory the only non-trivial four-point correlation function is $G_{\phi_1\phi_1}$, and it can be written in the same OPE form as the one we are considering as \eqref{OPE}.
By the Bethe-Salpeter method it is possible to compute the correlator at all-loops, since its perturbative expansion is generated only by a bosonic graph-building operator $\mathcal{H}_B$ of \eqref{kernel1}, then we can extract the non-perturbative scaling dimension of the exchanged operators in the OPE s-channel. The corresponding spectral equation is the same of \eqref{eq1} with $\omega^4=\xi^4$ and $\kappa^4=0$ (indeed the bi-scalar theory has only one coupling $\xi^2$) and it has two solutions corresponding to the twist-two and -four operators with the following scaling dimensions
\begin{equation}\begin{split}\label{biscspe}
\Delta^{(2)}_{\text{bi}}(\xi^4)&=2+\sqrt{(S+1)^2+1-2\sqrt{(S+1)^2+4\xi^4}}\,,\\
\Delta^{(4)}_{\text{bi}}(\xi^4)&=2+\sqrt{(S+1)^2+1+2\sqrt{(S+1)^2+4\xi^4}}\,,
\end{split}\end{equation}  
together with two shadow solutions with $\Delta=4-\Delta$ for $\text{Re}\,\Delta<2$. The analytic properties of those solution and their weak- and strong- coupling expansions have been studied in detail in~\cite{Gromov:2018hut}.

The scaling dimensions of the exchanged operators in the correlation function $G_{\phi_1\phi_1}$ for theories defined as a reduction of $\chi$CFT as in Sec.\ref{sec:section2}, can be computed as solutions of the spectral equation \eqref{eq1} in which we are applying some limits on the couplings, or even directly on the weak- and strong-coupling expansions. In the
Tab.\ref{tab:family1} we present the summary of our results.
\begin{table}[!t]
\begin{center}
\renewcommand{\arraystretch}{1.5}
\begin{tabular}{l*{4}{c}r}
              & limit & $\Delta^{(2)}$ & $\Delta^{(4)}$ & $\Delta^{(4')}$& $\Delta^{(t)}_{\pm}$  \\
\hline
 \multirow{2}{*}{$\chi_0$CFT}& $\xi_1\rightarrow 0 $   &$\Delta^{(2)}(\kappa,\omega)$ &$\Delta^{(4)}(\kappa,\omega)$ &$\Delta^{(4')}(\kappa,\omega)$&$\Delta^{(t)}_{\pm}(\kappa,\omega)$\\
                     & $\xi_{2}\lor\xi_3\rightarrow 0 $ & $\Delta_{\text{bi}}^{(2)}(\xi_{3}^4\lor\xi_2^4)$ &  \multicolumn{2}{c}{$\Delta_{\text{bi}}^{(4)}(\xi_{3}^4\lor\xi_2^4)$}& $t+S$\\
                     \hline
 \multirow{2}{*}{bi-scalar}& $\xi_1\land(\xi_2\lor\xi_3)\rightarrow 0 \qquad\quad$ & $\Delta_{\text{bi}}^{(2)}(\xi_{3}^4\lor\xi_2^4)$ &  \multicolumn{2}{c}{$\Delta_{\text{bi}}^{(4)}(\xi_{3}^4\lor\xi_2^4)$}&$t+S$\\
                & $\xi_2\land\xi_3\rightarrow 0 $ & 2+S &  \multicolumn{2}{c}{4+S}&$t+S$\\             
                \hline
$\beta$-deformed & $\xi_1\!=\!\xi_2\!=\!\xi_3\!=\!\xi$ & 2+S & $\Delta^{(4)}(\xi^4,0)$ & $\Delta^{(4')}(\xi^4,0)$&$\Delta^{(t)}_{\pm}(\xi^4,0)$
\end{tabular}
\end{center}
\caption{In this table we summarize the operator and dimension content of exchange operators in three reductions of \(\chi\)CFT. }
\label{tab:family1}
\end{table}
\begin{itemize}
\item $\chi_0$CFT: since the spectrum of the exchanged operators for the four-point function $G_{\phi_1\phi_1}$ doesn't depend on $\xi_1$, the limit in which one of the couplings of the full $\chi$CFT is going to zero (reducing the theory to the $\chi_0$CFT) is not unique. Indeed if we set $\xi_1=0$, the scaling dimensions of the exchanged operators in the $\chi_0$CFT are the same of the full $\chi$CFT. On the contrary, if we set $\xi_2$ or $\xi_3$ to zero, the spectrum reduces to that of the bi-scalar theory \eqref{biscspe} depending on a single coupling. Notice that in this case the number of solution of twist-four operators reduces to a single one, while the higher-twist operators get protected.
\item bi-scalar theory: the reduction to bi-scalar theory corresponds to the limit in which two couplings of $\chi$CFT vanish. If one of the vanishing couplings is $\xi_1$, the  spectrum is the usual one of the bi-scalar theory \eqref{biscspe} while if $\xi_2=\xi_3=0$ the operators are protected because the only remaining interaction vertex is not  contributing.
\item $\beta$-deformed theory: when all the couplings are equal we reduce the full theory to its $\beta$-deformation. In this case, due of the restoration of one supersymmetry, the operator of twist-two is protected as pointed out in  \cite{Caetano:2016ydc} and confirmed by our computation (this reduction in terms of the new couplings $\kappa$ and $\omega$ corresponds to $\kappa\rightarrow\xi$ and $\omega\rightarrow 0$). The symmetry doesn't constrain the operators of twist-four to be protected, as well as for higher twist \(t>4\). Indeed, their spectrum can be easily read applying the limit, for example at weak coupling, to the expansions \eqref{D4S1},\eqref{DtS1}.
\end{itemize}

\subsection{The structure constant of the exchanged operators}\label{sec:OPEphi1phi1}

Once the spectrum of the exchanged operators is computed, in order to obtain the full set of conformal data for the four point function $G_{\phi_1\phi_1}$, one has to compute the OPE coefficients. From their definition \eqref{OPE}, we get
\begin{equation}\label{OPE2}
C_{\Delta,S}=\frac{\pi}{c_B^4}\frac{(-1)^{S+1}}{c_2(\nu,S)\,\mathcal{R}_{\Delta,S}}\,,
\end{equation}
where
\begin{equation}\label{OPE2der}
\mathcal{R}_{\Delta,S}\,=\,\frac{d}{d \Delta} \left({{h_B}_{\Delta,S}}^{-1}-\frac{\kappa^4}{\pi^4}\,{{\tilde{h}_F}{}_{\Delta,S}}\,{{h_B}_{\Delta,S}}^{-1}\right)\, .
\end{equation}
Here $c_2$ is given in \eqref{c1c2} and one puts $\Delta_{\mathcal{O}_1}=\Delta_{\mathcal{O}_2}=1$. The eigenvalues ${h_B}_{\Delta,S}$ and  ${{\tilde{h}_F}{}_{\Delta,S}}$ are presented in \eqref{hbS} and \eqref{hftildeS}, and the constants $c_F=-2 c_B=-1/(2\pi^2)$. Plugging these eigenvalues into \eqref{OPE2der} and performing the derivative, we obtain a rather cumbersome result that we will present in the next paragraph.

\paragraph{Weak coupling expansion}

Performing the derivative in \eqref{OPE2der} and substituting the weak coupling expansions of the scaling dimensions computed in \eqref{D2S1} and \eqref{D4S1}, we obtain the following expansions for the structure constants associated to the exchanged operators for $S\neq 0$
\begingroup\makeatletter\def\f@size{10.3}\check@mathfonts
\begin{equation}\begin{split}\label{structureS1}
C_{\Delta^{(2)},S}\!=&\frac{S!^2}{(2S)!}\!\!\left(1\!+\!\frac{2\kappa^4[2 H_S^{(2)}-\!H^{(2)}_{{S}/{2}}]
\!-\!2\omega^4[\tfrac{1}{S(S+1)}\!+\!H_{S-1}\!-\!H_{2S-2}]}{S(S+1)}\!+\!\dots\!\right)\\
C_{\Delta^{(4)},S}\!=&\frac{(S+1)!^2}{\!\sqrt{(S\!+\!1)(S\!+\!2)}(2S\!+\!2)!}\!\!\left(-\frac{\kappa^2}{2}\!-\!\frac{\omega^4
\!-8\kappa^4[\tfrac{9+S(11+3S)}{2(S+1)(S+2)}\!+\!H_{2S+2}\!-\!H_{S+1}]}{4\sqrt{(S+1)(S+2)}}\!+\!\dots\!\right)\\
C_{\Delta^{(4')},S}\!=&\frac{(S+1)!^2}{\!\sqrt{(S\!+\!1)(S\!+\!2)}(2S\!+\!2)!}\!\!\left(\frac{\kappa^2}{2}\!-\!\frac{\omega^4
\!-8\kappa^4[\tfrac{9+S(11+3S)}{2(S+1)(S+2)}\!+\!H_{2S+2}\!-\!H_{S+1}]}{4\sqrt{(S+1)(S+2)}}\!+\!\dots\!\right)
\\
C_{\Delta^{(t)}_{\pm},S}\!=&\frac{\pi\, i^t\, 2^{-2(t-4+S)} \Gamma\left( \frac{t}{2}-2 \right)\Gamma\left( \frac{t}{2}-1+S \right)}{(t-2)(t+2S)\,\Gamma\left(\frac{t-3}{2} \right) \Gamma\left(\frac{t-1}{2} +S\right)}\kappa^4 + \dots \, , 
\end{split}
\end{equation}
\endgroup
where \(t=6,8,10,\dots\) and \(H_k\), \(H_k^{(2)}\) are harmonic numbers. Again, the expressions in square brackets are in fact rational numbers. Similarly to the expansion of the scaling dimension, the OPE coefficient of the twist-two operator is singular for $S=0$. Indeed, as discussed in Sec.\ref{sec:spectrumphi1phi1}, due to the singularity arising at zero spin,  the weak coupling and  $S\rightarrow 0$ limits don't commute. In order to obtain the correct weak coupling expansion for the twist-two operator, one has to set $S=0$ in \eqref{OPE2} and then expand it in the coupling. The zero spin expansion the OPE coefficients of exchanged operators reads
\begin{equation}\begin{split}\label{structureS0}
C_{\Delta^{(2)},0}=&1+2i\omega^2-2[\omega^4\!\!-3\kappa^4\zeta_3]+i\omega^2[\omega^4(4\zeta_3-5)+18\kappa^4\zeta_3]+\dots\\
C_{\Delta^{(4)},0}=&-\frac{\kappa^2}{4\sqrt{2}}+\frac{22\kappa^4-\omega^4}{16}-\frac{3[\tfrac{\omega^8}{\kappa^2}-120\kappa^2\omega^4\!\!+912\kappa^6]}{256\sqrt{2}}+\dots\\
C_{\Delta^{(4')},0}=&\frac{\kappa^2}{4\sqrt{2}}+\frac{22\kappa^4-\omega^4}{16}+\frac{3[\tfrac{\omega^8}{\kappa^2}-120\kappa^2\omega^4\!\!+912\kappa^6]}{256\sqrt{2}}+\dots
\\
C_{\Delta^{(t)}_{\pm},0}\!=&\frac{\pi\, i^t\, 2^{(8-2t)} \Gamma\left( \frac{t}{2}-2 \right)\Gamma\left( \frac{t}{2}-1 \right)}{(t-2)t\,\Gamma\left(\frac{t-3}{2} \right) \Gamma\left(\frac{t-1}{2}\right)}\kappa^4 + \dots \,\qquad t=6,8,10,\dots\, .
\end{split}\end{equation}
In analogy with the spectrum analysis, the power counting shows that the twist-two operator goes in power of $\xi^4$ as expected if $S\neq 0$.  In the $S=0$ case it is going in powers of \(\xi^2\), suggesting that the weak coupling expansion is sensitive to the double trace counterterms. Moreover in both cases \eqref{structureS1} and \eqref{structureS0}, the twist-four OPE coefficients are suppressed by a factor of order $\xi^2$ as compared to those of the twist-2. 

\paragraph{Strong coupling expansion}

Since we know from the expansion at strong coupling of the scaling dimension \eqref{strong1}  that the scaling dimension becomes large, we can expand \eqref{OPE2} in the limit $\Delta\rightarrow\infty$ and  obtain
\begingroup\makeatletter\def\f@size{10.4}\check@mathfonts
\begin{equation}\label{OPEstrong1}
C_{\Delta,S}\!=\frac{2^{5-2\Delta}\,(S\!+\!1)}{\Delta}\tan\!\!\left(\!\pi\frac{\Delta\!-\!S}{2}\right)\!\!\left[
1\!+\!\frac{3}{2\Delta}\!+\!\frac{4(S\!+\!1)^2\!+\!25}{8\Delta^2}\!+\!\frac{36(S\!+\!1)^2\!+\!133}{16\Delta^3}\!+\!\mathcal{O}\!\!\left(\!\frac{1}{\Delta^4}\!\right)\!\right],
\end{equation}
\endgroup
where we have to substitute  $\Delta$ from the strong coupling spectrum $\Delta_\infty$ computed in \eqref{strong1} for low-twist operators. Naively, the expansion \eqref{OPEstrong1} looks the same as  the one of the structure constant of the bi-scalar model \cite{Gromov:2018hut}, but actually it is not. Indeed, one can notice from the definition \eqref{OPE2} that the OPE coefficient in our model depends explicitly on the coupling. Then in the expansion \eqref{OPEstrong1} some coefficients at higher order will start to depend on $\kappa^4$. The first contribution different from the bi-scalar expansion appear as $\kappa^4/\Delta^6$ which, after the substitution $\Delta_\infty$, contributes at order $\mathcal{O}(1/\xi^2)$ in the inverse coupling expansion. Hence, it is convenient to write \eqref{OPEstrong1} as follows
\begin{equation}\begin{split}\label{OPEstrong12}
C_{\Delta_\infty,S}
=&2^{5}\frac{S\!+\!1}{2^{2\Delta_\infty}\Delta_\infty}\tan\!\!\left(\!\pi\frac{\Delta_\infty\!-\!S}{2}\right)\!
\left[1+\frac{3}{4e^{i\pi k/2}(\omega^4+2\kappa^4)^{1/4}}+\right.\\
&\left.+\left(\frac{4(S+1)^2+1}{32e^{i\pi k}(\omega^4+2\kappa^4)^{1/2}}
-\frac{2(S+5)(2S+5)\kappa^4}{e^{i\pi k}(S+1)(\omega^4+2\kappa^4)^{3/2}}\right)
+\dots\right],
\end{split}\end{equation}
where $k=0,1,2,3$ labels the four solutions of the spectral equation \eqref{eq1} and dots stand for higher orders in  $1/\kappa$ and $1/\omega$.
Thus, given the scaling dimension $\Delta_\infty$ \eqref{strong1},  the OPE coefficient is exponentially small  at strong coupling due to the factor \(\frac{1}{2^{2\Delta_\infty}}\). 
The $S\rightarrow 0$ limit is not singular at strong coupling and one can compute $C_{\Delta_\infty,0}$ directly from \eqref{OPEstrong12}.

\subsection{The four-point correlation function}
\label{fpoint1}
Once the conformal data in Secc.\ref{sec:spectrumphi1phi1} and \ref{sec:OPEphi1phi1} is computed, one can determine the four-point function \eqref{phi1phi1} by means of \eqref{4pointgeneral}. In the case $\mathcal{O}_1=\mathcal{O}_2=\phi_1$ we obtain 
\begin{equation}
G_{\phi_1\phi_1}(x_1,x_2|x_3,x_4)=\frac{c_B^2}{x_{12}^2x_{34}^2}\,\mathcal{G}_{\phi_1\phi_1}(u,v)\,,
\end{equation} 
with the cross-ratios defined as $u=x_{12}^2 x_{34}^2/(x_{13}^2 x_{24}^2)$ and $v=x_{14}^2x_{23}^2/(x_{13}^2 x_{24}^2)$ and $\Delta_{\phi_1}=1$. 
The function $\mathcal{G}_{\phi_1\phi_1}(u,v)$ can be written in terms of the OPE representation \eqref{OPE} as a sum over the non-negative integer Lorentz spin $S$ and the states with scaling dimensions $\Delta$. From the study of the spectrum of exchanged operators in Sec.\ref{sec:spectrumphi1phi1} it turns out that 
infinitely many operators are exchanged in the OPE channel.
Then we have 
\begin{equation}\label{GGfinaleOPE}
\begin{split}
\mathcal{G}_{\phi_1\phi_1}(u,v)=&\sum_{S=0}^\infty \;\left[\,C_{\Delta^{(2)},S}\;g_{\Delta^{(2)},S} +C_{\Delta^{(4)},S}\;g_{\Delta^{(4)},S}+C_{\Delta^{(4')},S}\;g_{\Delta^{(4')},S}\right]+\\+&\sum_{t=6,8,\dots}\sum_{S=0}^\infty \; \left[ C_{\Delta^{(t)}_+,S}\;g_{\Delta^{(t)}_+,S}+C_{\Delta^{(t)}_{-},S}\;g_{\Delta^{(t)}_{-},S}\right]\, ,
\end{split}
\end{equation}
where the scaling dimensions $\Delta^{(i)}$ are defined by the spectral equation \eqref{eq1} and computed at weak coupling in \eqref{D2S1}, \eqref{D4S1} and\eqref{DtS1}, and for low twist \(t=2,4\) at strong coupling in~\eqref{strong1}. The structure constants $C_{\Delta^{(i)},S}$ associated to the exchanged operators are defined by \eqref{OPE2} are computed at weak coupling in \eqref{structureS1} and for low twist and strong coupling in~\eqref{OPEstrong12}. The four-dimensional conformal blocks $g_{\Delta,S}$ are defined in \eqref{defg}.

The proper definition of the four-point correlation function $G_{\phi_1\phi_1}$ takes into account the symmetrization $x_3\leftrightarrow x_4$. Under this symmetry, the cross-ratios transform as $u\rightarrow u/v$ and $v\rightarrow 1/v$.  Correspondingly, from the definition  \eqref{defg} the conformal blocks obey the symmetry $g_{\Delta,S}(u/v,1/v)=(-1)^Sg_{\Delta,S}(u,v)$.
Combining together this relation with \eqref{GGfinaleOPE}, it's easy to see that, imposing the symmetry $x_3\leftrightarrow x_4$,  the terms in \eqref{GGfinaleOPE} with odd $S$ cancel out whereas those with even $S$ get doubled.

Despite of the presence of singularities in the weak-coupling expansions of scaling dimensions and OPE coefficients, their combination in \eqref{GGfinaleOPE} is well-defined. Indeed, plugging the conformal data into \eqref{GGfinaleOPE}, we obtain an expansion in powers of the couplings that is compatible with the interpretation of the correlation function as a sum of Feynman diagrams in perturbation theory (see Sec.\ref{sec:feynman11} for an explicit example). In particular, since the first non-trivial order is fixed by the $S=0$ conformal data,  it is easy to write the very first contributions to $\mathcal{G}_{\phi_1\phi_1}$ in terms of the known functions, as follows
\begin{equation}\label{4pointfinal1}
\mathcal{G}_{\phi_1\phi_1}(u,v)=u-i\kappa^2\,u\,\Phi^{(1)}(u,v)+\dots\,.
 \end{equation}
where  $\Phi^{(L)}$ is the ladder three-point function \cite{Usyukina:1992jd} that in the case $L=1$ is given by
the Bloch-Wigner dilogarithm  function
\begin{equation}\label{phi1dav}
\Phi^{(1)}(u,v)=\frac{1}{\theta}\left[2(\text{Li}_2(-\rho u)+\text{Li}_2(-\rho v))+\log\frac{v}{u}\log\frac{1+\rho v}{1+\rho u}+\log{\rho u}\log{\rho v}+\frac{\pi^2}{3}\right]\, ,
\end{equation}
with
\begin{equation}
\theta(u,v)\equiv\sqrt{(1-u-v)^2-4 u v}\qquad\text{and}\qquad \rho(u,v)\equiv \frac{2}{1-u-v+\theta}\,.
\end{equation}

\section{Exact four-point correlations function for $\mathcal{O}_1(x)\!=\!\phi_1(x)$ and $\mathcal{O}_2(x)\!=\!\phi_{2}^\dagger(x)$}\label{sector2}

In this section we consider the four-point correlators associated to the last operator of \eqref{operators}, namely when $\mathcal{O}_1(x)=\phi_j(x)$ and $\mathcal{O}_2(x)=\phi_{k\neq j}^\dagger(x)$ with $j,k=1,2,3$. 
Since the computation of the correlators is the same for any $j$ and $k$, we will consider the case $j=1$ and $k=2$ and then the four-point function we want to study takes the following form
\begin{equation}\label{phi1phi2dagger}
G_{\phi_1\phi_{2}^\dagger}(x_1,x_2|x_3,x_4)=\langle\text{Tr}[\phi_1(x_1)\phi_{2}^\dagger(x_2)]\text{Tr}[\phi_1^\dagger(x_3)\phi_{2}(x_4)]\rangle\,.
\end{equation}
This correlation function was trivial in the bi-scalar model \cite{Grabner:2017pgm} but in the general double-scaled theory \eqref{fullL} it has a rich diagrammatic structure. Indeed a generic Feynman diagram in the weak coupling expansion of \eqref{phi1phi2dagger} in the planar limit $N_c\rightarrow \infty$ is given by a combination of all the single-trace vertices in \eqref{fullL} and the following double-trace vertex
\begin{equation}\label{vv1}
(4\pi)^2\alpha_{2}^2 \, \text{Tr}[\phi_1\phi_{2}^\dagger](x) \,\text{Tr}[\phi_1^\dagger\phi_{2}] (x)\,,
\end{equation}
coming from the counterterm Lagrangian \eqref{doubletraces}. In the following we will compute the conformal data of \eqref{phi1phi2dagger} with the Bethe-Salpeter method.

\subsection{The Bethe-Salpeter method for the correlator $G_{\phi_1\phi_{2}^\dagger}$}
\begin{figure}[!t]
 \centering
   \includegraphics[width=\textwidth]{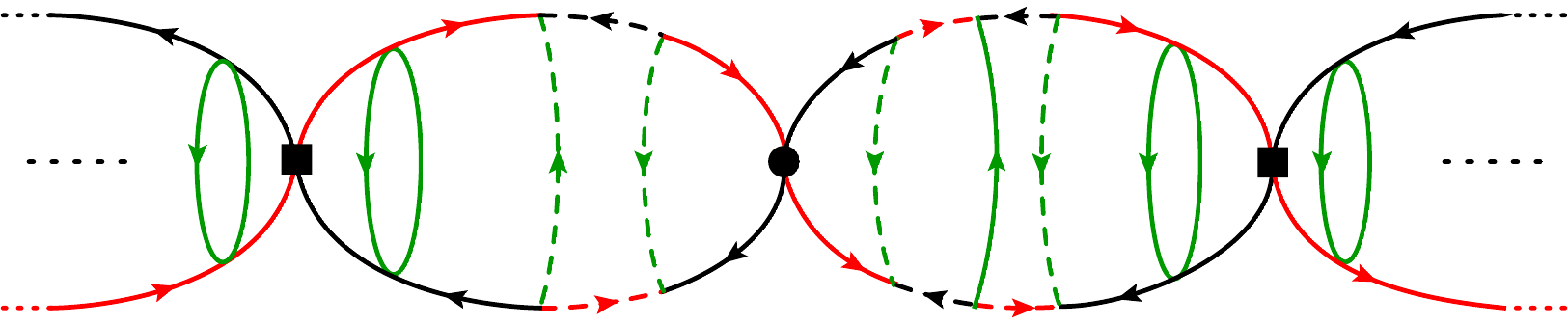}
\caption{ A Feynman diagram contributing to the perturbative expansion $G_{\phi_1\phi_{2}^\dagger}^{(\ell)}$.  
The black squares and dots stand for single- and double-trace vertices respectively. Tick lines are bosonic propagators and dashed lines fermionic ones. The colors represent different flavors $j$ of the particles $\phi_j$ and $\psi_j$: in particular black for $j=1$, red for $j=2$ and green for $j=3$. The propagators are not crossing and are curved to stress the fact that they have a cylindrical topology.}
  \label{fig:phi1phi2dagger}
 \end{figure}

The perturbative expansion of \eqref{phi1phi2dagger} can be written in the following form
\begin{equation}\label{Gpert2}
G_{\phi_1\phi_{2}^\dagger}(x_1,x_2|x_3,x_4)=\sum_{\ell=0}^\infty\,G_{\phi_1\phi_{2}^\dagger}^{(\ell)}(x_1,x_2|x_3,x_4)\,,
\end{equation}
where $G_{\phi_1\phi_{2}^\dagger}^{(\ell)}$ at any perturbative order $\ell$ contains contributions from the bosonic and fermionic integrals, with different  dependence on couplings. 
In Fig.\ref{fig:phi1phi2dagger}, we present an example of a  generic  Feynman diagram contributing to \eqref{Gpert2}.
As in the previous case, the Feynman diagrams of this four-point correlation function have  a cylindric topology and, at arbitrary order $\ell$, they take an iterative form allowing us to write the full expansion as an infinite geometric sum of the primitive divergencies, as in \eqref{Ggeom}. In contrast to  the previous case, the nodes of the chain diagrams in the expansion of $G_{\phi_1\phi_{2}^\dagger}$  are not only insertions of double-trace vertices but also  of the  single-trace vertex
\begin{equation}\label{vv2}
(4\pi)^2\,\xi_{3}^2\,\text{Tr}[\phi_1^\dagger\phi_{2}^\dagger\phi_1\phi_{2}](x)\,.
\end{equation}
In the Bethe-Salpeter procedure, both vertices enter only as  insertions of the operator $\hat{\mathcal{V}}$ defined in \eqref{V}. Then it's easy to conclude, as it was done in~\cite{Gromov:2018hut} for the biscalar fishnet model, that the coefficient of this operator in \eqref{Ggeom} is
\begin{equation}\label{alphatildedef}
\chi_\mathcal{V}=(4\pi)^2\tilde{\alpha}_2^2\qquad\text{where}\quad\tilde{\alpha}_2^2=\alpha_2^2+\xi_{3}^2\,.
\end{equation}
In Sec.\ref{sec:bethe}, we discussed the role of the operator $\hat{\mathcal{V}}$ in the computation of the four-point function, arguing that it is not contributing to the spectral equation for finite coupling or \(S\). Then, as far as we consider the perturbative expansion \eqref{Gpert2} in the point-splitting $x_1\neq x_2$ and $x_3\neq x_4$, we need only to resum the single trace contributions appearing inside the chain links of Fig.\ref{fig:phi1phi2dagger}. The contribution given by vertices \eqref{vv1} and \eqref{vv2} is crucial to calculate the fixed point \eqref{fixedpoint}. In Sec.\ref{sec:feynman} we will present this computation in detail.

\begin{figure}[!t]
 \centering
  \subfigure[$\xi_1^0\xi_2^0\xi_3^0$]
   {\includegraphics[width=3.2cm]{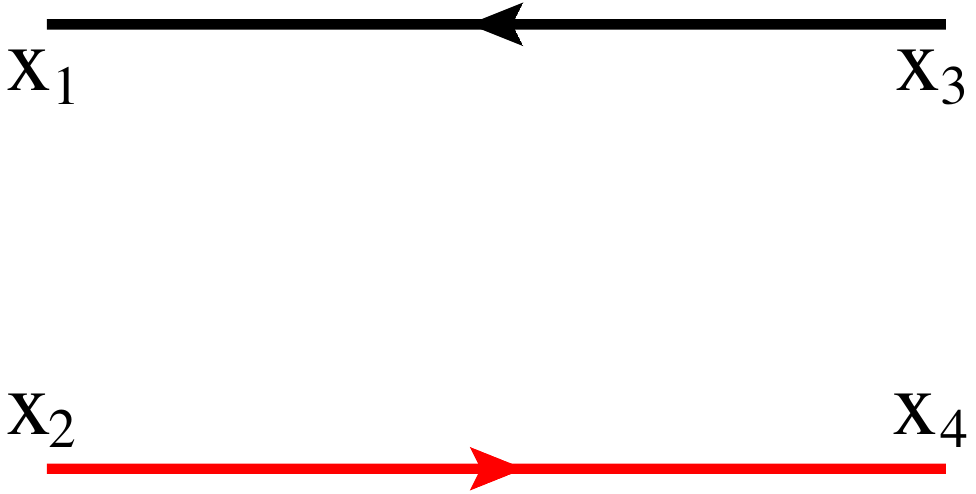}}
    \hspace{5mm}
    \subfigure[$\xi_1^2\xi_2^2$]
   {\includegraphics[width=3.2cm]{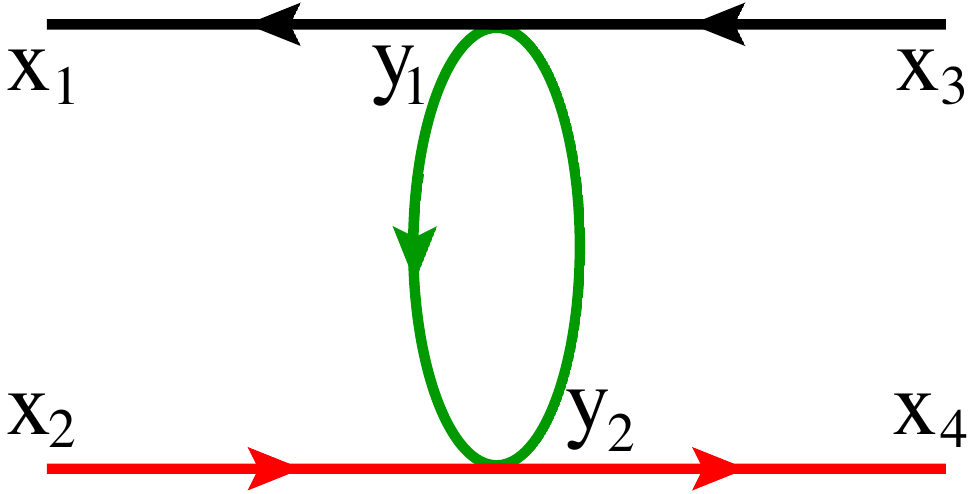}}
    \hspace{5mm}
    \subfigure[$\xi_1\xi_2\xi_3^2$]
   {\includegraphics[width=3.2cm]{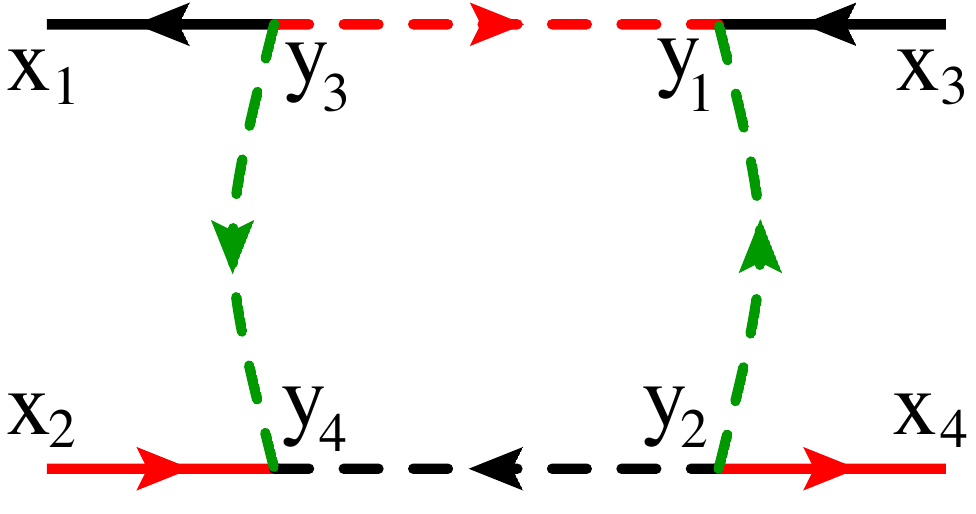}}
    \hspace{5mm}
    \subfigure[$\xi_1^2\xi_2^2\xi_3^2$]
   {\includegraphics[width=3.2cm]{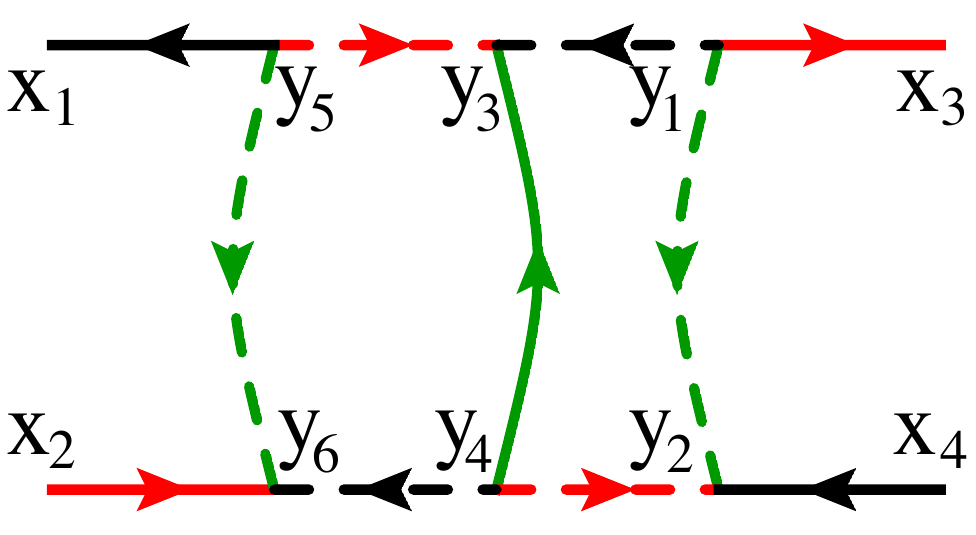}}
    \hspace{5mm}
 \caption{First contributions to the four-point functions $G_{\phi_1\phi_{2}^\dagger}$.}
  \label{fig:pert2}
 \end{figure}

The first few orders of the perturbative expansion are given by the diagrams represented in Fig.\ref{fig:pert2}. They can be written as follows
\begin{equation}\begin{split}\label{GGpert12}
G_{\phi_1\phi_{2}^\dagger}^{(0)}&\!\!=\frac{c_B^2}{x_{13}^2x_{24}^2},\\
G_{\phi_1\phi_{2}^\dagger}^{(1)}&\!\!=c_B^6(4\pi)^4\xi_{1}^2\xi_{2}^2 \!\int \,\frac{d^4 y_1 d^4 y_2 }{(x_1-y_1)^2(x_2-y_2)^2(y_{12}^2)^2(y_1-x_3)^2(y_2-x_4)^2}\\
&\!\!\!-c_B^4c_F^4(4\pi)^4\xi_1\xi_{2}\xi_{3}^2\!\!\int\!\!\frac{\prod_{i=1}^4 \! d^4 y_i\;\text{tr}\left[\sigma_\mu \overline{\sigma}_\rho \sigma_\eta \overline{\sigma}_\nu \right]\;y_{34}^\mu y_{42}^\rho y_{21}^\eta y_{13}^\nu}{(x_1-y_{3})^2 (x_2-y_{4})^2 y_{34}^4y_{42}^4y_{21}^4y_{13}^4(y_1-x_3)^2 (y_2-x_4)^2 }, \\
G_{\phi_1\phi_{2}^\dagger}^{(2)}&\!\!=\!-c_B^5c_F^6(4\pi)^6\xi_1^2\xi_{2}^2\xi^2_{3}\!\!\!\int\!\!\frac{(4\pi)^6\prod_{i=1}^6 \! d^4 y_i\;\text{tr}\left[\overline{\sigma}_\mu\sigma_\rho \overline{\sigma}_\eta \sigma_\nu \overline{\sigma}_\lambda \sigma_\sigma\right]\!y_{56}^\mu y_{64}^\rho y_{42}^\eta y_{21}^\nu y_{13}^\lambda y_{35}^\sigma}{(x_1\!-\!y_{5})^2 (x_2\!-\!y_{6})^2 y_{56}^4y_{64}^4y_{42}^4y_{21}^4y_{13}^4y_{35}^4(y_1\!-\!x_3)^2 (y_2\!-\!x_4)^2 },
\end{split}\end{equation}
where each scalar propagator brings in the factor $c_B/x_{ij}^2$ and each fermionic propagator -- the factor $c_F \slashed{x}_{ij}/x_{ij}^4$, where $\slashed{x}$ can be $\sigma_\mu x^\mu$ or $\bar{\sigma}_\mu x^\mu$ and $c_F=-2c_B=-1/(2\pi^2)$. 
 
\begin{figure}[!t]
 \centering
 \subfigure[$(\mathcal{H}_b)^{\dot{\alpha}\alpha}$]
   {\includegraphics[width=4cm]{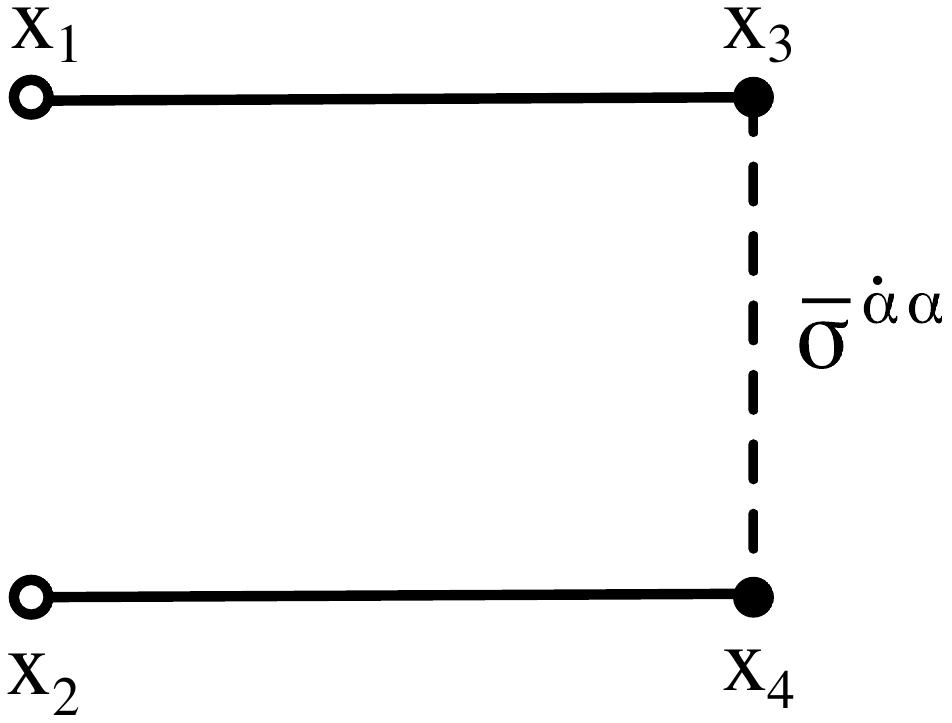}}
    \hspace{16mm}
    \subfigure[$(\mathcal{H}_r)_{\alpha\dot{\alpha};\beta\dot{\beta}}$]
   {\includegraphics[width=3.35cm]{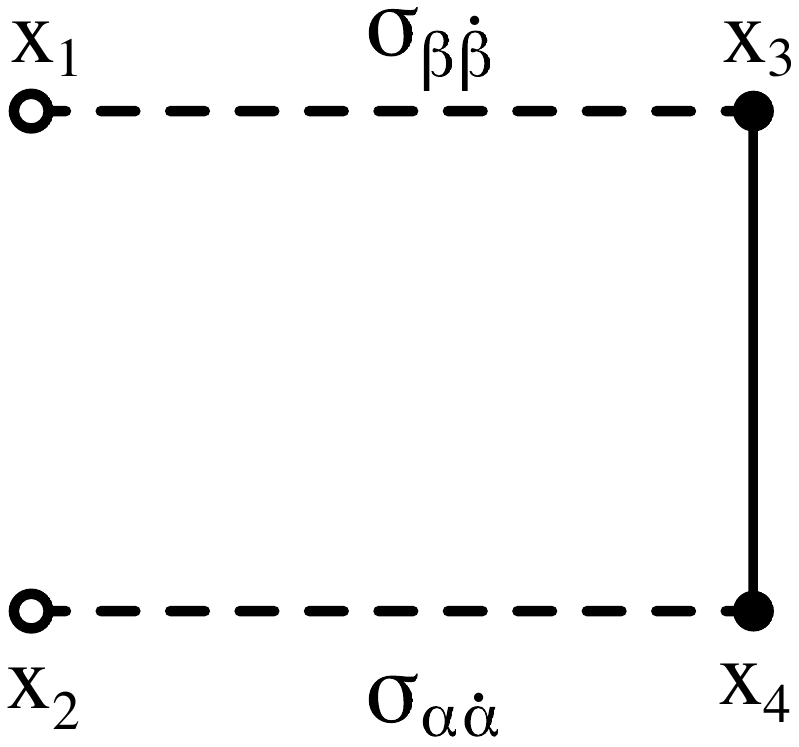}}
    \hspace{16mm}
 \subfigure[$(\mathcal{H}_t)^{\dot{\alpha}\alpha}$]
   {\includegraphics[width=4cm]{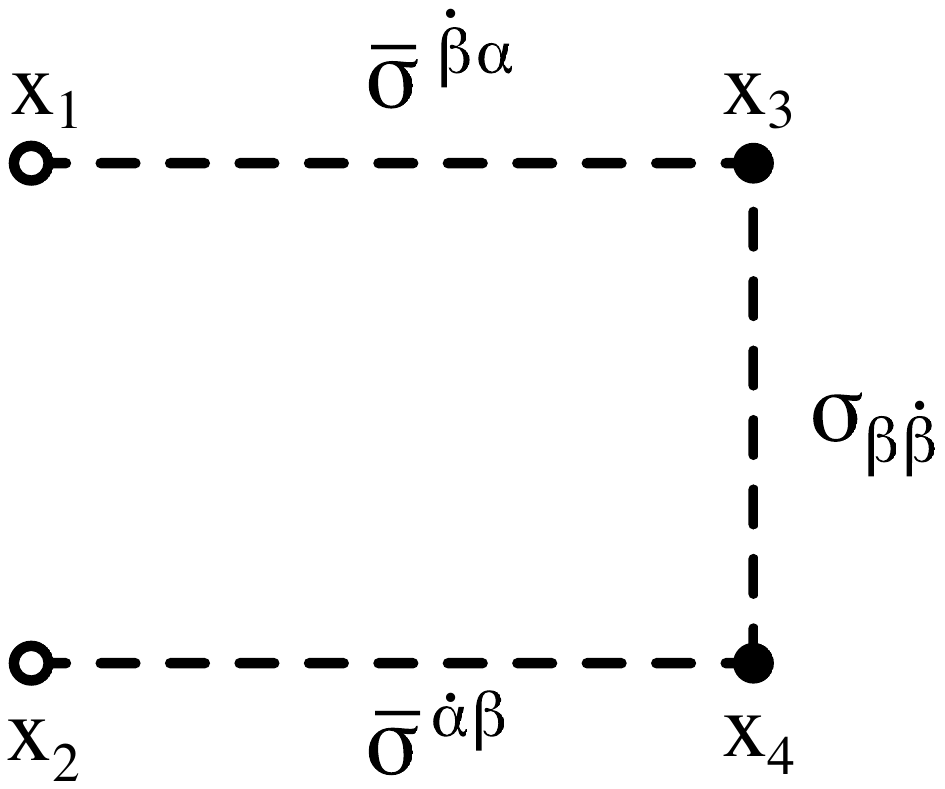}}
 \caption{The kernels associated to the Hamiltonian graph-building operators $\hat{\mathcal{H}}_F$ involved in the computation of the four-point function $G_{\phi_1\phi_{2}^\dagger}$ with $j=1,2,3$. White dots represent external points and black dots -- integration over the full space \({\mathbb{R}}^4\).}
  \label{fig:kernelferm2}
 \end{figure}

 These diagrams can be expressed in terms of a combination of the Hamiltonian graph-building operators $\hat{\mathcal{H}}_i$. 
Indeed, considering the bosonic part of \eqref{GGpert12}, the bosonic kernel is
\begin{equation}
\begin{split}\label{kernelbos2}
\mathcal{H}_B(x_1,x_2|x_3,x_4)=&\;\frac{c_B^4}{x_{13}^2x_{24}^2x_{34}^4}\,,
\end{split}
\end{equation}
that is clearly the same as studied in the previous case (see Sec.\ref{sec:phi1phi1}). The fermionic kernel is more involved. Considering the diagrams in Fig.\ref{fig:pert2}(c) and \ref{fig:pert2}(d) and their integral representation \eqref{GGpert12}, it is clear that they are not generated by the same repeated Hamiltonian operator. In fact, going on with the perturbative expansion of $G_{\phi_1\phi_{2}^\dagger}$, one can notice that at any order $\ell$ for $\ell>1$,   at least one fermionic diagram with the following ladder topology appears    
\begin{center}
\includegraphics[scale=0.6]{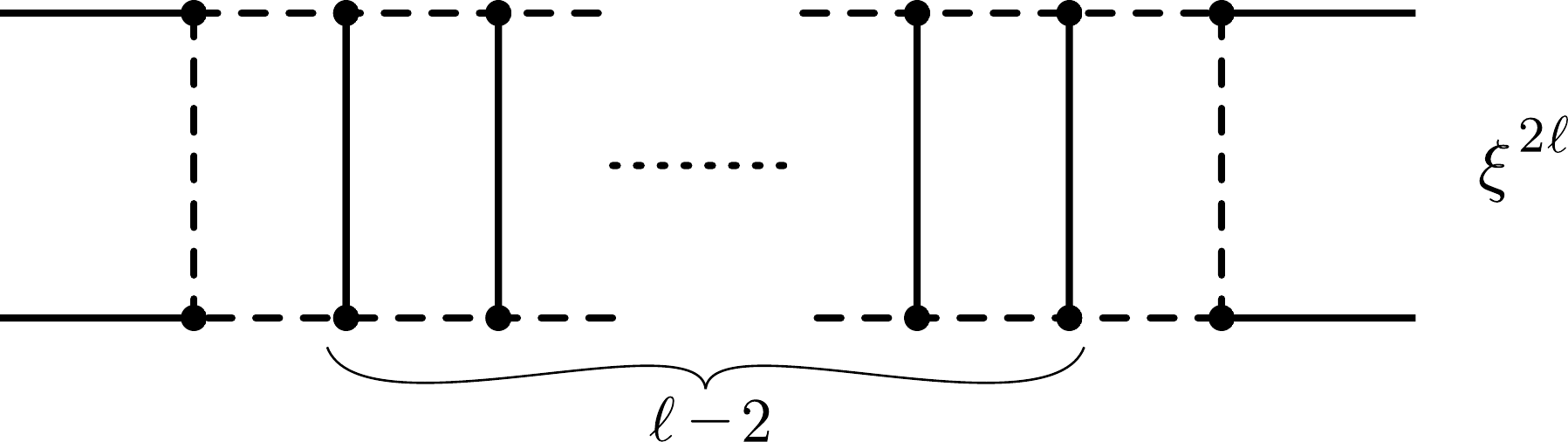}
\end{center}
Since any of them carries a $\xi^2$ power of the coupling, the maximum number of bosonic rungs in the ladder depends on the perturbative order, in particular is $\ell-2$. However for $\ell>3$, also the superpositions of ladders with less rungs contribute. For these reasons it is convenient to write the fermionic Hamiltonian as a product of \textit{sub-kernels} associated to the top and bottom parts of the ladder interspersed by $n$ copies of a rungs-building Hamiltonian
\begin{equation}\begin{split}\label{HF2}
\mathcal{H}_F^{(n)}(x_1,x_2|x_3,x_4)=\int\prod_{i=1}^{2n+2} &d^4y_i\;\mathcal{H}_b(x_1,x_2|y_{2n+1},y_{2n+2})\times\\
&\times\prod_{j=1}^n\mathcal{H}_r(y_{2j+1},y_{2j+2}|y_{2j-1},y_{2j})\mathcal{H}_t(y_1,y_2|x_3,x_4)\,,
\end{split}\end{equation}
where the fermionic sub-kernels $\mathcal{H}_b$, $\mathcal{H}_r$ and $\mathcal{H}_t$ are contracted in the spin indices in order to recompose the trace of $\sigma$-matrices.
In our convention, when $n=0$ the rung-building operator $\mathcal{H}_r$ is not contributing to $\mathcal{H}_F$. These Hamiltonians, graphically represented in Fig.\ref{fig:kernelferm2}, are defined as follows 
\begin{equation}\begin{split}\label{kernel2}
(\mathcal{H}_b)^{\dot{\alpha}\alpha}(x_1,x_2|x_3,x_4)&=-c_B^2c_F\frac{(\bar{\sigma}_\mu)^{\dot{\alpha}\alpha}\,x_{43}^\mu}{x_{24}^2x_{43}^4x_{31}^2}\,,\\
(\mathcal{H}_r)_{\alpha\dot{\alpha};\beta\dot{\beta}}(x_1,x_2|x_3,x_4)&=c_Bc_F^2\frac{(\sigma_\mu)_{\alpha\dot{\alpha}}(\sigma_\nu)_{\beta\dot{\beta}}\,x_{24}^\mu x_{31}^\nu}{x_{24}^4x_{43}^2x_{31}^4}\,,\\
(\mathcal{H}_t)^{\dot{\alpha}\alpha}(x_1,x_2|x_3,x_4)&=c_F^3\frac{(\bar{\sigma}_\mu \sigma_\nu \bar{\sigma}_\rho)^{\dot{\alpha}\alpha}\,x_{24}^\mu x_{43}^\nu x_{31}^\rho}{x_{24}^4x_{43}^4x_{31}^4}\,,
\end{split}\end{equation}
 and they can be used to rewrite the expansion \eqref{GGpert12} obtaining
 \begin{equation}\begin{split}\label{GGpert12H}
G_{\phi_1\phi_{2}^\dagger}^{(0)}\!&\!\!=\frac{x_{34}^4}{c_B^2}\;\mathcal{H}_B(x_1,x_2|x_3,x_4)\,,\\
G_{\phi_1\phi_{2}^\dagger}^{(1)}\!&\!\!=\frac{x_{34}^4}{c_B^2}(4\pi)^4\int d^4y_1d^4y_2\biggl[\xi_1^2\xi_{2}^2\;\mathcal{H}_B(x_1,x_2|y_1,y_2)\;\mathcal{H}_B(y_1,y_2|x_3,x_4)+\\
&\qquad\qquad\qquad\qquad\qquad+\xi_1\xi_{2}\xi_{3}^2\;\mathcal{H}_F^{(0)}(x_1,x_2|y_1,y_2)\;\mathcal{H}_B(y_1,y_2|x_3,x_4)\biggl]\,, \\
G_{\phi_1\phi_{2}^\dagger}^{(2)}\!&\!\!=\frac{x_{34}^4}{c_B^2}(4\pi)^6\xi_1^2\xi_{2}^2\xi_{3}^2\int d^4y_1d^4y_2 \;\mathcal{H}_F^{(1)}(x_1,x_2|y_1,y_2)\;\mathcal{H}_B(y_1,y_2|x_3,x_4)
\,,
\end{split}\end{equation}
where, using the definition \eqref{HF2}, we have
\begin{align}
\mathcal{H}_F^{(0)}(x_1,x_2|x_3,x_4)\!=&\!\!\int d^4y_1 d^4y_2\;{\mathcal{H}_b}_{\alpha\dot{\alpha}}(x_1,x_2|y_1,y_2)\mathcal{H}_t^{\;\dot{\alpha}\alpha}(y_1,y_2|x_3,x_4)\,,\\
\mathcal{H}_F^{(1)}(x_1,x_2|x_3,x_4)\!=&\!\!\int \!\!\prod_{i=1}^4\!d^4y_i\mathcal{H}_b^{\;\dot{\beta}\beta}(x_1,x_2|y_3,y_4){\mathcal{H}_r}_{\beta\dot{\alpha};\alpha\dot{\beta}}(y_3,y_4|y_1,y_2)\mathcal{H}_t^{\;\dot{\alpha}\alpha}(y_1,y_2|x_3,x_4)\,.\nonumber
\end{align}

The kernels \eqref{kernelbos2} and \eqref{HF2} transform covariantly under conformal transformations, then the corresponding Hamiltonian integral operators commute with the generators of the conformal group. The fermionic sub-kernels \eqref{kernel2}  have spinorial indices carried by the $\sigma$-matrices and  transform as two-components spinors.
Following the conventions we are using for the raising and lowering of spin indices, as explained in App.\ref{app:appendA}, we have the following transformations
\begin{equation}\label{eigentransf}
({\mathcal{H}}_b)_{\beta\dot{\beta}}=\epsilon_{\beta\alpha}\epsilon_{\dot\beta\dot\alpha}({\mathcal{H}}_b)^{\dot{\alpha}\alpha}\,,\!\!\qquad({\mathcal{H}_r})^{\dot{\gamma}\gamma;\dot{\delta}\delta}=\epsilon^{\delta\beta}\epsilon^{\dot\delta\dot\beta}\epsilon^{\gamma\alpha}\epsilon^{\dot\gamma\dot\alpha}({\mathcal{H}_r})_{\alpha\dot{\alpha};\beta\dot{\beta}}\,,\!\!\qquad({\mathcal{H}}_t)_{\beta\dot{\beta}}=\epsilon_{\beta\alpha}\epsilon_{\dot\beta\dot\alpha}({\mathcal{H}}_t)^{\dot{\alpha}\alpha}
\end{equation}
which corresponds to the exchange $\sigma\leftrightarrow \bar\sigma$. 
The rung-building operator $\mathcal{H}_r$ contains a couple of un-contracted $\sigma$-matrices, then it appears with two pairs of indices.
In order to build the general fermionic diagram, one has to contract the fermionic sub-kernels with the only constraint to obtain the trace of all the $\sigma$-matrices around the fermionic loop alternating $\sigma$'s with $\bar{\sigma}$'s. Once chosen if the top sub-kernel $\mathcal{H}_t$ contains the combination $\sigma\bar{\sigma}\sigma$ or $\bar{\sigma}\sigma\bar{\sigma}$, the first rung sub-kernel $\mathcal{H}_r$ has to have the right combination of indices to be contracted, in particular $\bar{\sigma}\bar{\sigma}$ and $\sigma\sigma$ respectively. Then the other $\mathcal{H}_r$ kernels have to alternate upper and lower indices.
Depending on parity of the number of rungs of the ladder $n$, the bottom sub-kernel ${\mathcal{H}}_b$ can carry $\sigma$ or $\bar\sigma$. Indeed, we can distinguish two different index structures, for  odd or even number, of repeated applications of $\mathcal{H}_r$ as follows
\begin{equation}\label{evenoddhf}
\mathcal{H}_r^n=\begin{cases}
(\mathcal{H}_r^{2\ell})^{\dot{\beta}\,\quad\beta}_{\;\;\dot{\alpha};\alpha}=(\mathcal{H}_r)^{\dot{\beta}\gamma_\ell;\dot{\delta}_\ell\beta}(\mathcal{H}_r)_{\gamma_\ell\dot{\gamma}_{\ell-1};\delta_{\ell-1}\dot{\delta}_\ell}\;...\;(\mathcal{H}_r)_{\gamma_2\dot{\gamma}_1;\delta_1\dot{\delta}_2}(\mathcal{H}_r)^{\dot{\gamma}_1\gamma_1;\dot{\delta}_1\delta_1}(\mathcal{H}_r)_{\gamma_1\dot{\alpha};\alpha\dot{\delta}_1}\\
(\mathcal{H}_r^{2\ell+1})_{\beta\dot{\alpha};\alpha\dot{\beta}}=(\mathcal{H}_r)_{\beta\dot{\gamma}_{\ell};\delta_{\ell}\dot{\beta}} (\mathcal{H}^{2\ell}_r)^{\dot{\gamma}_\ell\,\quad\delta_\ell}_{\;\;\;\,\dot{\alpha};\alpha}
\end{cases}
\end{equation}
where $\ell=0,1,...,\infty$. 

Carrying on in the perturbative expansion, one can find that for example the perturbative order $\ell=3$ is given by the sum of same combinations of kernels appearing at order $\ell=1$, namely  $\mathcal{H}_B^3$, ${\mathcal{H}_F^{(0)}}^2\mathcal{H}_B$ and $\mathcal{H}_F^{(0)}\mathcal{H}_B^2$, plus the new kernel $\mathcal{H}_F^{(2)}$ and so on. For this motivation, the $\ell$-th perturbative order $G_{\phi_1\phi_{2}^\dagger}^{(\ell)}$ cannot be written as the contribution for $\ell=1$ to the power $\ell$ as in the case studied in Sec.\ref{sec:phi1phi1}, but its sum takes the following form
\begin{equation}\label{Gell}
\sum_{\ell=0}^\infty \hat{G}_{\phi_1\phi_{2}^\dagger}^{(\ell)}=\frac{x_{34}^4}{c_B^2}\sum_{k=0}^\infty\left[(4\pi)^4\xi_{1}^2\xi_{2}^2\hat{\mathcal{H}}_B+
(4\pi)^4\xi_1\xi_{2}\xi_{3}^2\sum_{n=0}^\infty(4\pi)^{2n}\xi_{1}^n\xi_{2}^n\hat{\mathcal{H}}_F^{(n)}\right]^k\hat{\mathcal{H}}_B.
\end{equation}
Since the operatorial form of the fermionic Hamiltonian \eqref{HF2} in terms of the sub-kernels \eqref{kernel2} is 
\begin{equation}\label{hfop}
\hat{\mathcal{H}}_F^{(n)}=\hat{\mathcal{H}}_b\;{\hat{\mathcal{H}}_r}^n\;\hat{\mathcal{H}}_t\,,
\end{equation}
one can sum the two geometric series in \eqref{Gell}. Then the correlator \eqref{Gpert2} can be written as follows
\begin{equation}\label{Gresum2}
\hat{G}_{\phi_1\phi_{2}^\dagger}=\sum_{\ell=0}^\infty\hat{G}_{\phi_1\phi_{2}^\dagger}^{(\ell)}=
\frac{x_{34}^4}{c_B^2}\frac{1}{1-(4\pi)^4\xi^2_{1}\xi^2_{2}\hat{\mathcal{H}}_B-(4\pi)^4\xi_{1}\xi_{2}\xi_{3}^2\hat{\mathcal{H}}_F}\hat{\mathcal{H}}_B,
\end{equation}
where
\begin{equation}\label{hfsum}
\hat{\mathcal{H}}_F=\sum_{n=0}^\infty(4\pi)^{2n}\xi_{1}^n\xi_{2}^n\hat{\mathcal{H}}_F^{(n)}=\hat{\mathcal{H}}_b\;\frac{1}{1-(4\pi)^2\xi_1\xi_{2}\hat{\mathcal{H}}_r}\;\hat{\mathcal{H}}_t\,,
\end{equation}
Finally, comparing \eqref{Gresum} with the definition \eqref{Ggeom}, we can fix the value of the remaining constants $\chi_i$
\begin{equation}\label{chi2}
\chi_B=(4\pi)^4\xi^2_{1}\xi^2_{2}\,,\qquad\chi_F=(4\pi)^4\xi_{1}\xi_{2}\xi_{3}^2\,.
\end{equation}

\subsection{Eigenvalues of the Hamiltonian graph-building operators}\label{sec:eigen2}

In order to compute the four-point correlation function with the operator method presented in Sec.\ref{sec:bethe}, one has to obtain  the spectrum of the graph-building operators \eqref{kernelbos2} and \eqref{HF2} as was done in Sec.\ref{sec:eigen1} for another  four-point correlator. The eigenstate that diagonalizes all these Hamiltonians is defined in \eqref{Phi-def} for $\Delta_{\mathcal{O}_1}=\Delta_{\mathcal{O}_2}=1$ and the eigenvalues are defined by means of equations \eqref{eigen2} and \eqref{eigen3}.
Substituting in the latter the kernels \eqref{kernelbos2} and \eqref{kernel2} and using the definition \eqref{eigengeneral}, we will end up with a set of integrals that can be computed with the help of the star-triangle relations \eqref{uniqferm}. The fact that all the integrals that we have to compute can be solved by means of the star-triangle relations is a strong evidence of the underlying conformal symmetry.

\paragraph{Bosonic eigenvalue:} Since the bosonic Hamiltonian \eqref{kernelbos2} is the same as studied in the previous case, its eigenvalue is already computed: it is given by \eqref{hbS} for any \(S\) and by \eqref{hbS0} in the case of $S=0$. 

\paragraph{Fermionic eigenvalue:}

The fermionic eigenvalue is defined in \eqref{eigen3}. In the case we are studying,  the Hamiltonian operator \eqref{HF2} depends on the number of rungs $n$, thus  its eigenvalue will be also a function of \(n\), as follows
\begin{equation}\label{eigenF2}
\left[\hat{\mathcal{H}}_F^{(n)}\; \Phi_{\Delta,S,x_0}\right](x_1,x_2)  = {h_F}{}_{\Delta,S}^{(n)} \; \Phi_{\Delta,S,x_0}(x_1,x_2)\,.
\end{equation} 
The computation of the fermionic spectrum is simpler if we consider the Hamiltonian $\mathcal{H}_F^{(n)}$ in terms of the sub-kernels $\mathcal{H}_b$, $\mathcal{H}_r$ and $\mathcal{H}_t$, as in \eqref{hfop}. While the full fermionic Hamiltonian is diagonalized by the state $\Phi_{\Delta,S,x_0}$, the sub-kernels are not. Indeed, due to their fermionic structure,  individually they will turn the scalar conformal eigenfunction into a fermionic state. However,  compatibly with \eqref{eigenF2}, their combination \eqref{HF2} will leave the state $\Phi_{\Delta,S,x_0}$ unchanged. For instance, at \(S=0\) when the state is given by the conformal triangle \eqref{eigenfunction}, we have
\begin{equation}\begin{split}\label{hfsector2}
\underbrace{\hat{\mathcal{H}}_b\hat{\mathcal{H}}_r^n\hat{\mathcal{H}}_t}_{\hat{\mathcal{H}}_F^{(n)}}&\!\!\vcenter{\hbox{\includegraphics[width=1.7cm]{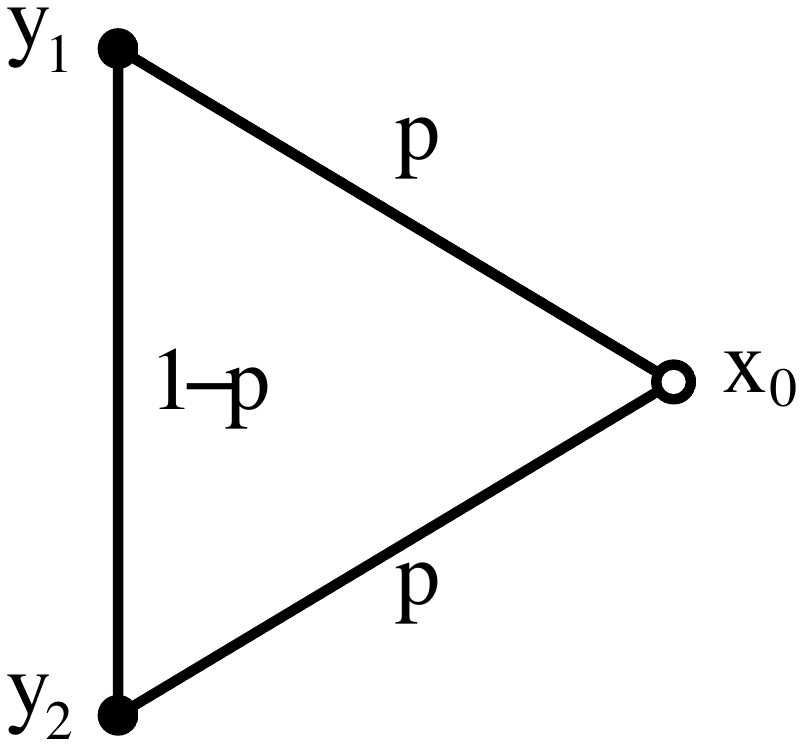}}}=\!\!
\begin{cases}
{\hat{\mathcal{H}}_b}{}_{\beta\dot{\beta}}(\hat{\mathcal{H}}_r^{2\ell})^{\dot{\beta}\,\quad\beta}_{\;\;\dot{\alpha};\alpha}\hat{\mathcal{H}}_t^{\dot{\alpha}\alpha}\!\!\vcenter{\hbox{\includegraphics[width=1.7cm]{triang1}}}=h_t\,\hat{\mathcal{H}}_b{}_{\beta\dot{\beta}}(\hat{\mathcal{H}}_r^{2\ell})^{\dot{\beta}\,\quad\beta}_{\;\;\dot{\alpha};\alpha}\!\!\vcenter{\hbox{\includegraphics[width=1.7cm]{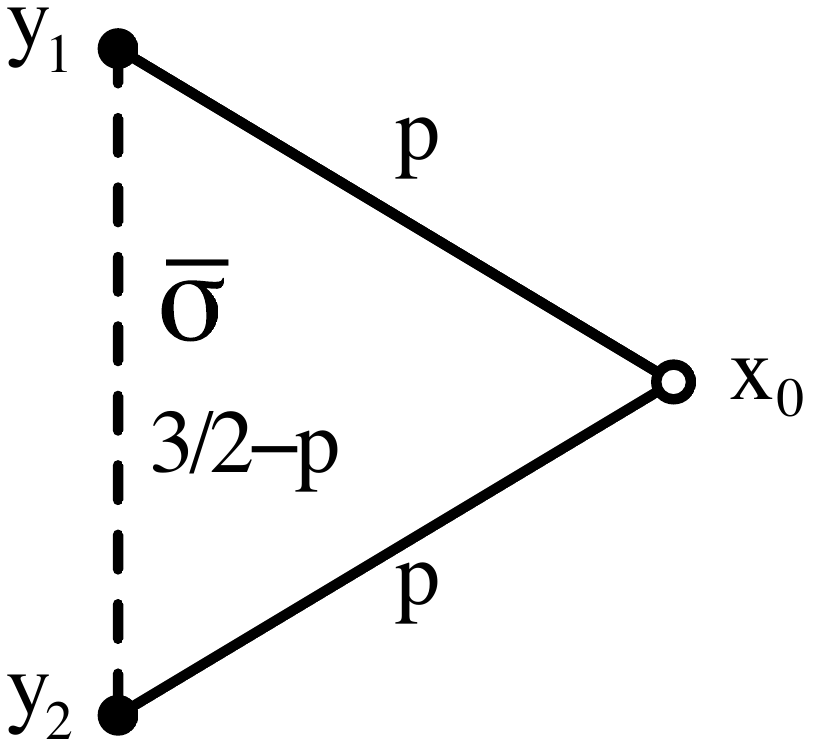}}}\\
\hat{\mathcal{H}}_b^{\dot{\beta}\beta}(\hat{\mathcal{H}}_r^{2\ell+1})_{\beta\dot{\alpha};\alpha\dot{\beta}}\hat{\mathcal{H}}_t^{\dot{\alpha}\alpha}\!\!\vcenter{\hbox{\includegraphics[width=1.7cm]{triang1}}}=h_t\,\hat{\mathcal{H}}_b^{\dot{\beta}\beta}(\hat{\mathcal{H}}_r^{(2\ell+1)})_{\beta\dot{\alpha};\alpha\dot{\beta}}\!\!\vcenter{\hbox{\includegraphics[width=1.7cm]{triangferm1y}}}
\end{cases}\\
&\begin{rcases}
&=h_r^{2\ell}\,h_t\,\hat{\mathcal{H}}_b{}_{\beta\dot{\beta}}\vcenter{\hbox{\includegraphics[width=1.7cm]{triangferm1y}}}=
h_b\,h_r^{2\ell}\,h_t\,\vcenter{\hbox{\includegraphics[width=1.7cm]{triang1}}}\\
&=h_r^{2\ell+1}\,h_t\,\hat{\mathcal{H}}_b^{\dot{\beta}\beta}\vcenter{\hbox{\includegraphics[width=1.7cm]{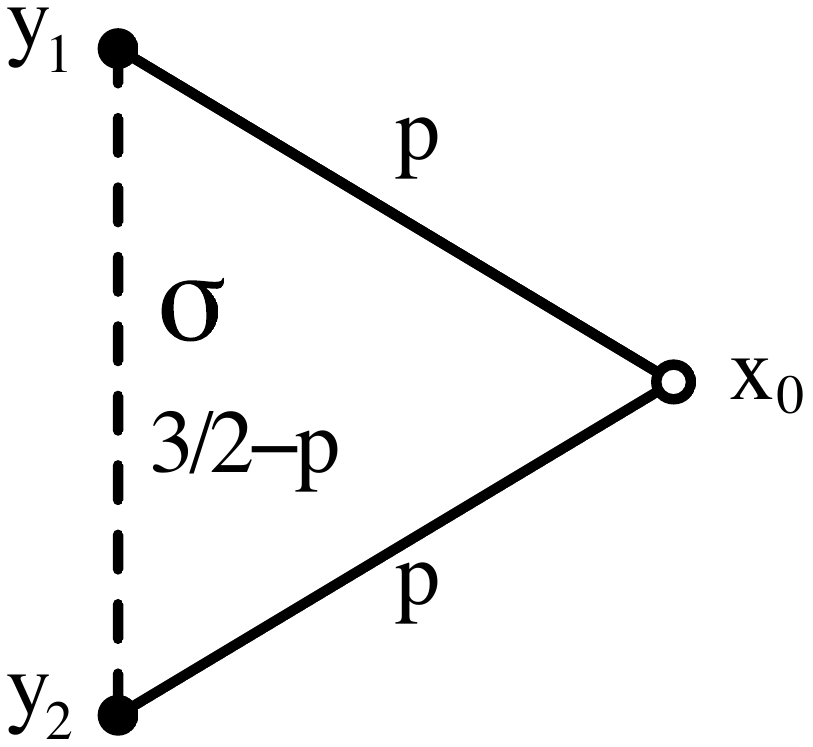}}}=h_b\,h_r^{2\ell+1}\,h_t\,\vcenter{\hbox{\includegraphics[width=1.7cm]{triang1}}}
\end{rcases}
=\underbrace{h_b\,h_r^n\,h_t}_{ {h_F}{}_{\Delta,S}^{(n)}}\vcenter{\hbox{\includegraphics[width=1.7cm]{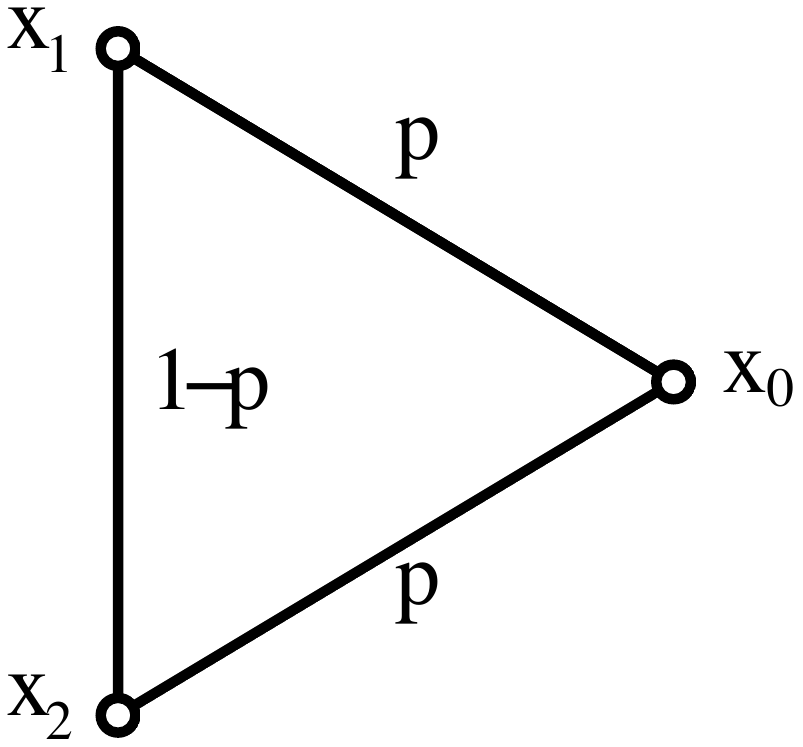}}}
\end{split}\end{equation}
where $\ell$ is a non-negative integer, $p=\Delta/2$, black and white dots are positions with and without integrations over \(\mathbb{R}^4\), and we defined the state
\begin{equation}\label{eigen}
\vcenter{\hbox{\includegraphics[width=2cm]{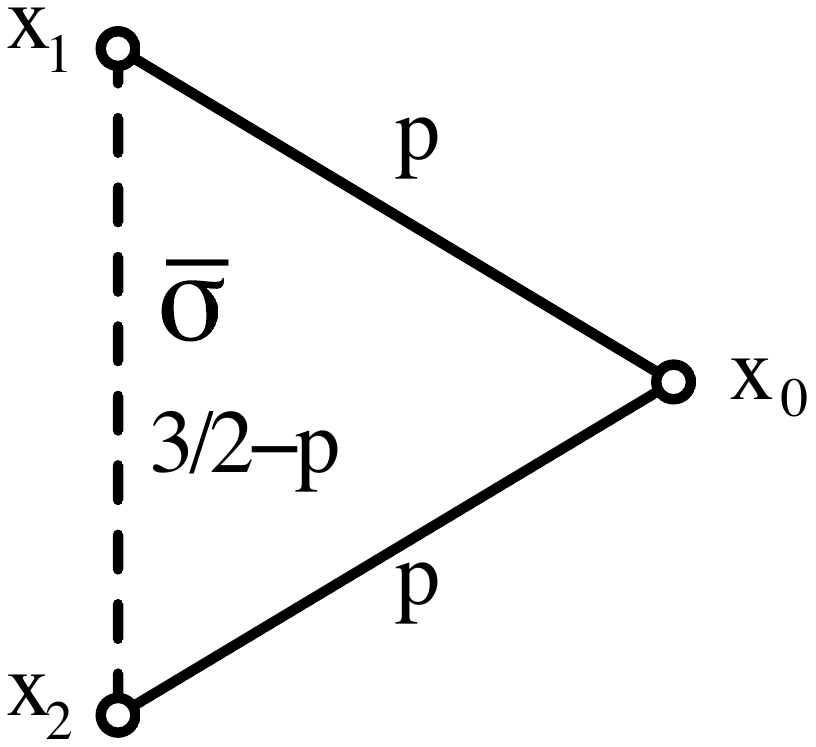}}}=\!\frac{(\bar{\sigma}_\mu)^{\dot{\alpha}\alpha}x_{21}^\mu}{x_{12}^2}\Phi_{\Delta,0,x_0}(x_1,x_2),\quad
\!\vcenter{\hbox{\includegraphics[width=2cm]{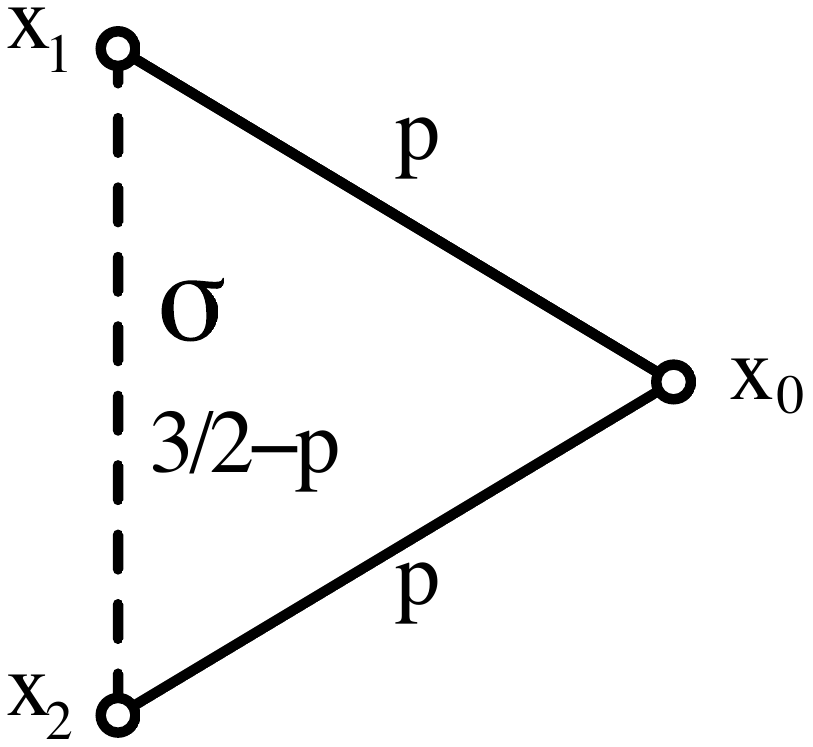}}}=\!\frac{(\sigma_\mu)_{\alpha\dot{\alpha}}x_{21}^\mu}{x_{12}^2}\Phi_{\Delta,0,x_0}(x_1,x_2)
\end{equation}
In \eqref{hfsector2} we consider the action of an even and odd number of rung-building operators separately. Indeed, while the top sub-kernel is changing the conformal triangle into a fermionic object and the bottom sub-kernel is turning it back to the original state, the $n$ copies of the operator $\hat{\mathcal{H}}_r$ are exchanging $\sigma\leftrightarrow\bar{\sigma}$ in the states \eqref{eigen} depending on the parity of $n$, according to \eqref{evenoddhf}.

These arguments hold also in the case of $S\neq 0$, then we have to solve the following equations
\begin{equation}\begin{split}\label{subeigen}
\left[\hat{\mathcal{H}}_t^{\;\dot{\alpha}\alpha}\; \Phi_{\Delta,S,x_0}\right](x_1,x_2)  &= {h_t}{}_{\Delta,S} \;\frac{(\bar{\sigma}_\mu)^{\dot{\alpha}\alpha}\,x_{21}^\mu}{x_{12}^2}\; \Phi_{\Delta,S,x_0}(x_1,x_2)\,,\\
\left[{\hat{\mathcal{H}}_r}^{\dot{\beta}{\alpha};\dot{\alpha}{\beta}}\;\frac{({\sigma}_\mu)_{\alpha\dot{\alpha}}\,y_{21}^\mu}{y_{12}^2}\; \Phi_{\Delta,S,x_0}\right](x_1,x_2)  &= {h_r}{}_{\Delta,S} \;\frac{(\bar{\sigma}_\mu)^{\dot{\beta}{\beta}}\,x_{21}^\mu}{x_{12}^2}\; \Phi_{\Delta,S,x_0}(x_1,x_2)\,,\\
\left[\hat{\mathcal{H}}_b^{\;\dot{\beta}\beta}\; \frac{({\sigma}_\mu)_{\beta\dot{\beta}}\,y_{21}^\mu}{y_{12}^2}\;\Phi_{\Delta,S,x_0}\right](x_1,x_2)  &= {h_b}{}_{\Delta,S} \; \Phi_{\Delta,S,x_0}(x_1,x_2)\,,
\end{split}\end{equation} 
and the same for the transformed kernels by means of \eqref{eigentransf}. Considering the first two equations of \eqref{subeigen} and the definitions \eqref{kernel2}, we have
\begin{align}\label{intht}
\left[\hat{\mathcal{H}}_t^{\;\dot{\alpha}\alpha}\, \Phi_{\Delta,S,x_0}\right]\!(x_1,x_2)=&
c_F^3\!\int\! d^4 y_1d^4 y_2\frac{(\bar{\sigma}_\mu\sigma_\nu\bar{\sigma}_\rho)^{\dot{\alpha}\alpha}(x_2-y_2)^\mu y_{21}^\nu (y_1-x_1)^\rho}{(x_2-y_2)^4 y_{12}^4 (y_1-x_1)^4}\Phi_{\Delta,S,x_0}(y_1,y_2)\nonumber\\
=&\frac{c_F}{c_B}\left[{\hat{\mathcal{H}}_r}^{\dot{\beta}{\alpha};\dot{\alpha}{\beta}}\;\frac{({\sigma}_\mu)_{\alpha\dot{\alpha}}y_{21}^\mu}{y_{12}^2}\; \Phi_{\Delta,S,x_0}\right](x_1,x_2)\,,
\end{align}
then, looking at \eqref{subeigen}, we can conclude that
\begin{equation}\label{Hr}
{h_r}{}_{\Delta,S}=\frac{c_B}{c_F}\,{h_t}{}_{\Delta,S}\,.
\end{equation}
Moreover, focusing on the last equation of \eqref{subeigen} and the definition of bottom sub-kernel given in \eqref{kernel2}, we have
\begin{equation}\label{hbottom}
\left[\hat{\mathcal{H}}_b^{\;\dot{\beta}\beta}\; \frac{({\sigma}_\mu)_{\beta\dot{\beta}}\,y_{21}^\mu}{y_{12}^2}\;\Phi_{\Delta,S,x_0}\right](x_1,x_2) =
-2c_B^2c_F\!\int\! d^4 y_1d^4 y_2\frac{\Phi_{\Delta,S,x_0}(y_1,y_2)}{(x_2-y_2)^2 y_{12}^4 (y_1-x_1)^2}\, ,
\end{equation} 
where we used equation \eqref{symferm} to simplify the fermionic structure.
The integral in the right-hand side of the equation is the same as appearing in the computation of the bosonic eigenvalue related to the operator $\hat{\mathcal{H}}_B$ given by \eqref{inthb}. Then eq.\eqref{hbottom} together with \eqref{subeigen} and \eqref{inthb} leads to
\begin{equation}\label{Hb}
{h_b}{}_{\Delta,S}=-2\frac{c_F}{c_B^2}\,{h_B}{}_{\Delta,S}\, ,
\end{equation}
where ${h_B}{}_{\Delta,S}$ corresponds to \eqref{hbS} for any $S$ and \eqref{hbS0} in the $S=0$ case. With these arguments it is clear that, in order to compute the fermionic eigenvalue of the operator \eqref{hfop}, we need only to compute the eigenvalue ${h_t}{}_{\Delta,S}$ defined by the first equation in \eqref{subeigen}.

In the $S=0$ case the computation of ${h_t}{}_{\Delta,S}$ can be performed  using only star-triangle relation \eqref{uniqferm}. Indeed starting from the integral \eqref{intht} together with the first equation of \eqref{subeigen} and performing two fermionic star-triangle integrations, as follows (see footnote \ref{STR})
\begin{equation*}\label{STRferm12}
\vcenter{\hbox{\includegraphics[trim={1.2cm 0 0 0},clip,width=2.8cm]{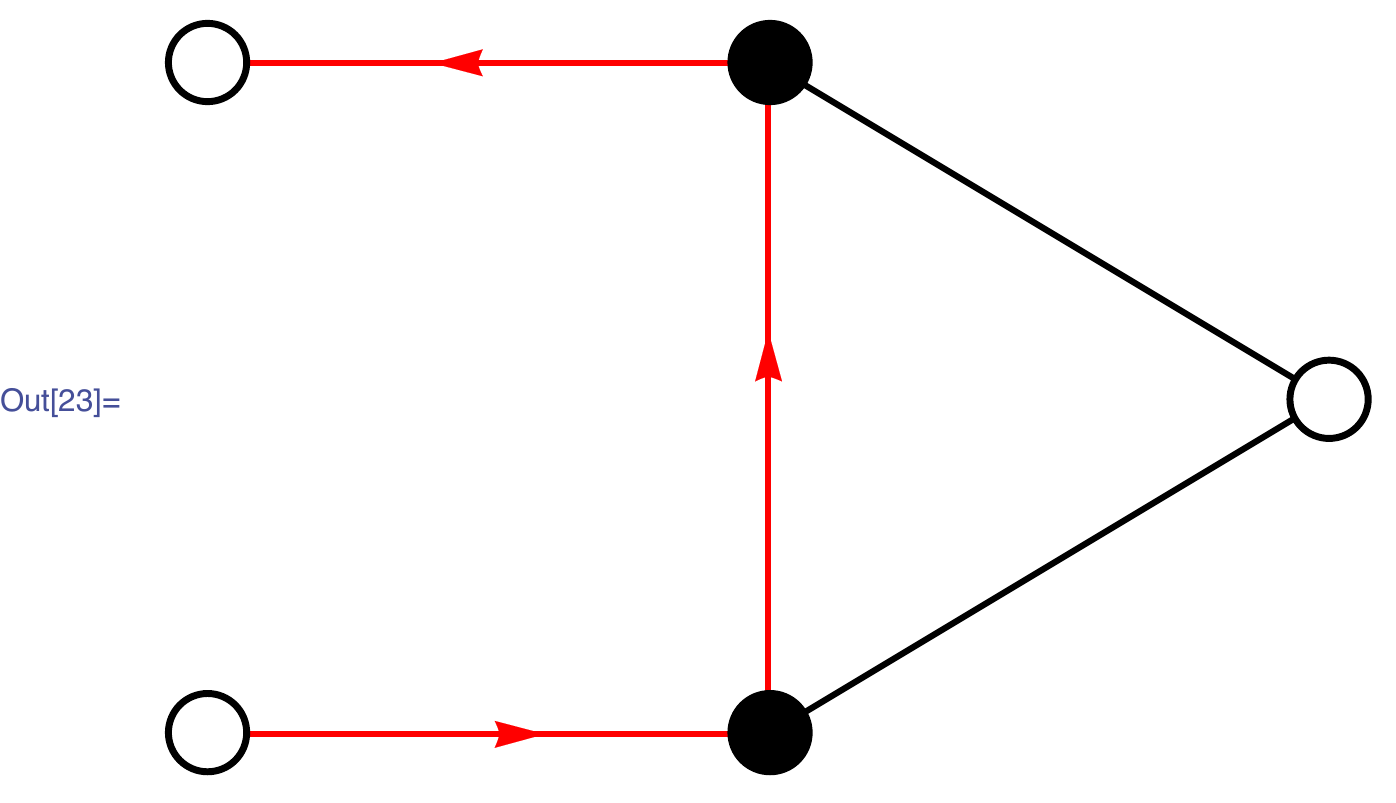}}}\quad\overset{\text{STR}}{\Longrightarrow}\quad
\vcenter{\hbox{\includegraphics[trim={1.2cm 0 0 0},clip,width=2.8cm]{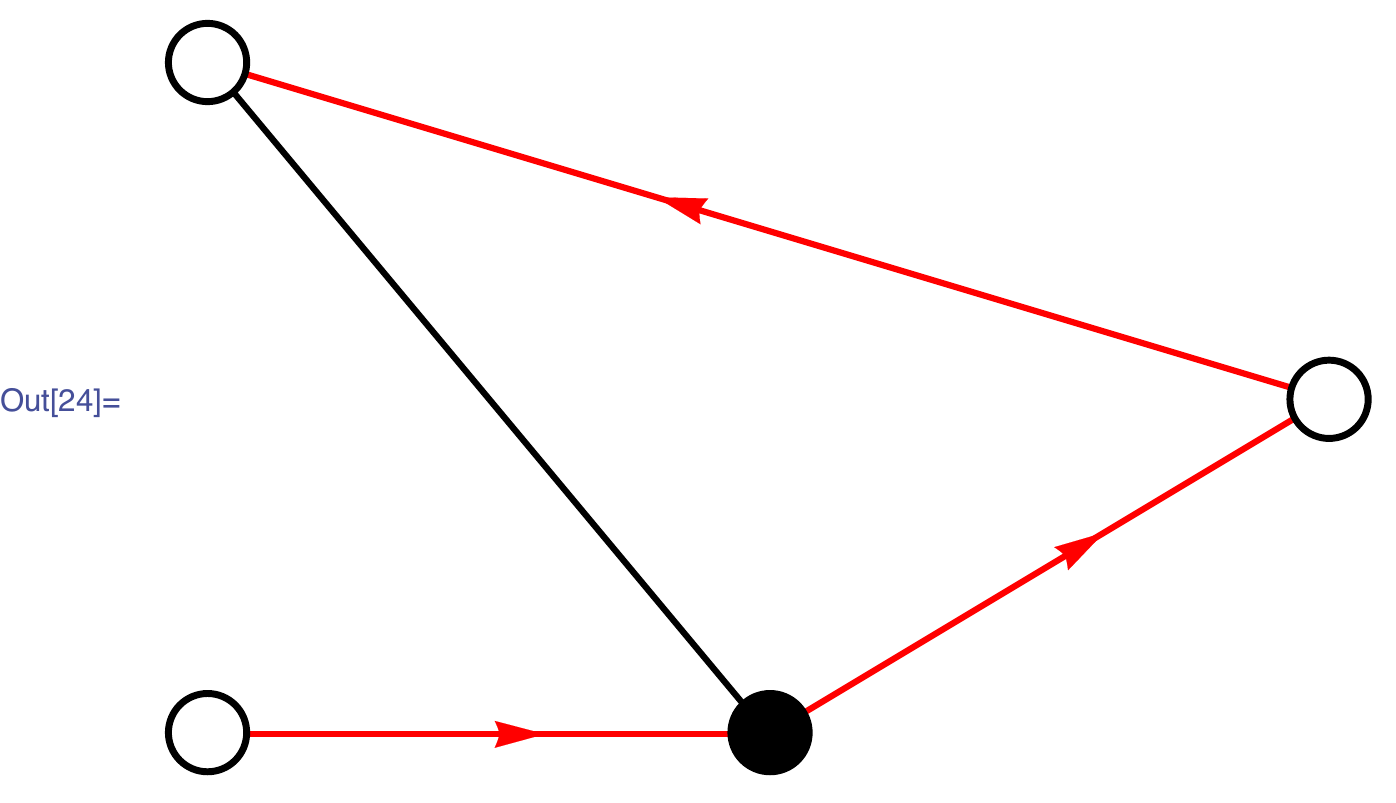}}}\quad\overset{\text{STR}}{\Longrightarrow}\quad
\vcenter{\hbox{\includegraphics[trim={1.2cm 0 0 0},clip,width=1.8cm]{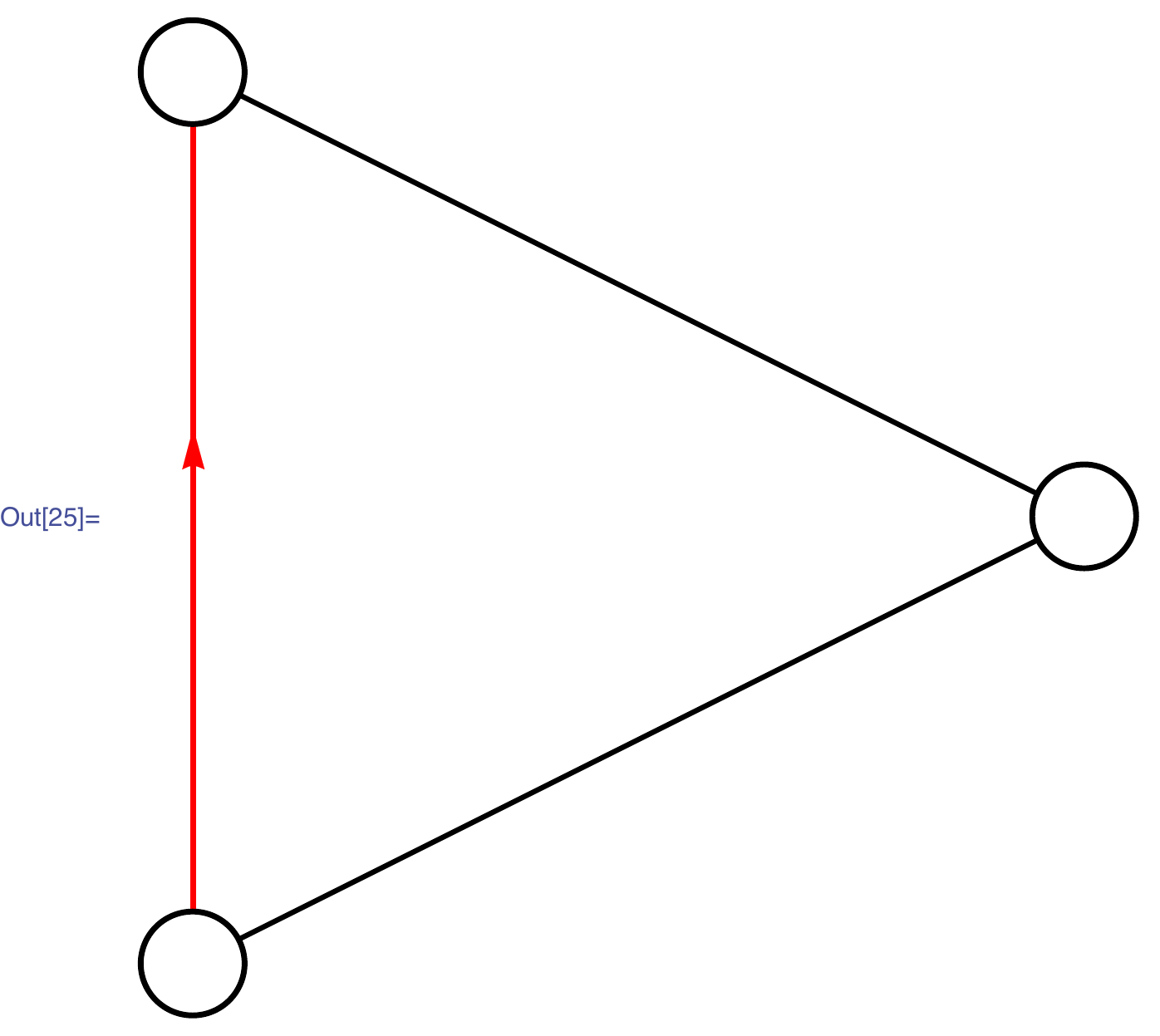}}}\,,
\end{equation*}
we obtain 
\begin{equation}\label{Ht}
{h_t}{}_{\Delta,0}=\frac{4\,c_F^3\,\pi^4}{\Delta(\Delta-4)}\, .
\end{equation}
Plugging \eqref{Hb}, \eqref{Hr} and \eqref{Ht} in the definition \eqref{eigenF2} we have
\begin{equation}\label{HfnS0}
{h_F}^{(n)}_{\Delta,0}={h_b}{}_{\Delta,0}\,({h_r}{}_{\Delta,0})^{n}\,{h_t}{}_{\Delta,0}
=-8\frac{(4\pi^4c_B c_F^2)^{n+2}}{\Delta^{n+2}(\Delta-2)^2(\Delta-4)^{n+2}}\, .
\end{equation}

The computation of the eigenvalue ${h_t}{}_{\Delta,S}$ for $S\neq 0$ is more involved. Without any loss of generality, we can consider the limit $x_0\rightarrow 0$ in both sides of the first equation of \eqref{subeigen} in order to simplify the spin structure. In this limit we are able to compute the resulting integral going to momentum space. We leave the details of the computation in App.\ref{appendix:ht}. Going through the calculation we obtain \eqref{htopS}. Notice that also this eigenvalue can be written in terms of the bosonic one as follows
\begin{equation}\label{Htt}
{h_t}{}_{\Delta,S}=\frac{c_F^3}{c_B^4}\frac{(\Delta-S)^2+S(S+2)}{4}\,{h_B}{}_{\Delta,S}\, ,
\end{equation}
where ${h_B}_{\Delta,S}$ is given in \eqref{hbS}.
Substituting \eqref{Hb}, \eqref{Hr} and \eqref{Htt} into the definition \eqref{eigenF2} we have
\begin{equation}\begin{split}\label{HfnS}
{h_F}^{(n)}_{\Delta,S}=&{h_b}{}_{\Delta,S}\,({h_r}{}_{\Delta,S})^{n}\,{h_t}{}_{\Delta,S}
=-8\left(\frac{c_F^2}{4 c_B^3}\right)^{n+2} \,[(\Delta-2)^2+S(S+2)]^{n+1}\,{h_B}_{\Delta,S}\\
=&-8(4\pi^4c_B c_F^2)^{n+2}\frac{[(\Delta-2)^2+S(S+2)]^{n+1}}{[(\Delta+S)(\Delta+S-2)(\Delta-S-2)(\Delta-S-4)]^{n+2}}\, ,
\end{split}\end{equation}
where in the last line we used the definition  \eqref{hbS}. 
Notice that setting $S=0$ we obtain \eqref{HfnS0} as expected. 
Finally, using the definition \eqref{hfsum}, we can resum the fermionic eigenvalue obtaining
\begin{equation}\label{hffinale}
{h_F}_{\Delta,S}=-\frac{c_F^4[(\Delta-2)^2+S(S+2)]{{h_B}_{\Delta,S}}^2}{2c_B^6-8c_B^3c_F^2\pi^2[(\Delta-2)^2+S(S+2)]\xi_{j}\xi_{j_+}{h_B}_{\Delta,S}}\, .
\end{equation}
We can conclude that the eigenvalue \eqref{hffinale} is manifestly invariant under $\Delta\rightarrow 4-\Delta$, as expected from \eqref{hsymmetry}.

\subsection{Spectrum of exchanged operators of $\mathcal{G}_{\phi_1\phi_{2}^\dagger}(u,v)$}\label{sec:spectrumphi1phi2}

In this section we will use the eigenvalues \eqref{hbS} and \eqref{hffinale} to compute the scaling dimensions of the operators contributing to the correlation function \eqref{OPE} for $\mathcal{O}_1=\phi_1$ and $\mathcal{O}_2=\phi_{2}^\dagger$. The spectrum of the exchanged operators is defined by the solutions of the equation for the physical poles \eqref{spectraleq}. Substituting in \eqref{spectraleq} the definition of bosonic and fermionic eigenvalues \eqref{hbS} and  \eqref{hffinale} and the constants $\chi_i$ computed in \eqref{chi2}, we can rearrange the spectral equation in the following form
\begin{equation}\label{spec22}
{{h_B}_{\Delta,S}}^{-1}+\frac{(4\pi)^4c_F^4[(\Delta-2)^2+S(S+2)]\lambda^2 \mu^2{h_B}_{\Delta,S}}{2c_B^6-8c_B^3c_F^2\pi^2[(\Delta-2)^2+S(S+2)]\lambda^2{h_B}_{\Delta,S}}
=(4\pi)^4\,\lambda^4\,,
\end{equation}
where we defined the new couplings
\begin{equation}\label{coupdef2}
\lambda^2=\xi_{1}\xi_{2}\qquad\mu^2=\xi_{3}^2\,.
\end{equation}
Plugging \eqref{hbS} into \eqref{spec22}, we obtain the  following equation
\begin{equation}\label{eq22}
[S^2+4\nu^2][(S+2)^2+4\nu^2]+\frac{128[S(S+2)-4\nu^2]\lambda^2\mu^2}{S(S+2)[S(S+2)-4\lambda^2]\!+\!8[2+S(S+2)+2\lambda^2]\nu^2\!+\!16\nu^4}=16\lambda^4
\end{equation}
with the additional constraint $\text{Im}\,\nu<0$ (i.e. $\text{Re}\,\Delta \geq 2$) for physical, exchange operators. 
Equation \eqref{eq22} has 8 solutions for $S\neq 0$.
Four of them correspond to the scaling dimensions of physical operators  satisfying this constraint.  Indeed, they are two couple of solutions with bare dimension $2+S$ and $4+S$.  The remaining four solutions are related to the first ones by the transformation $\Delta\rightarrow 4-\Delta$ and describe shadow operators with $\text{Re}\,\Delta < 2$. In the $S=0$ case \eqref{eq22} has 6 solutions. One corresponds to the operator with bare dimension $2$ and two -- to the one with bare dimension $4$. The remaining three solutions are their shadow operators.
This 4th order  equation in \(\nu^2\) can be solved exactly, but instead of these bulky formulas we prefer to present their perturbative expansions at weak and strong coupling.

The equation
\eqref{eq22} can be viewed as defining an algebraic curve of 8th degree in terms of \(\nu\), or of the corresponding dimension \(\Delta=2+2i\nu\) as a complex variable. In terms of \(\nu^2\) variable it is a 4th order algebraic curve whose branches -- the four sheets of the related Riemann surface -- describe directly the four physical dimensions as functions of couplings. Changing the couplings we can pass from one sheet to another, observing the transitions between various dimensions. The branch points correspond to the collisions of  physical dimensions. In contrast to this case, the spectral equation for the previous four-point function \eqref{eq1} is not algebraic and its Riemann surface contains infinitely many sheets.

\paragraph{Weak coupling expansion:}

Expanding around the physical pole $\nu=-iS/2$ at weak coupling $\lambda,\mu\rightarrow0$, we obtain the following expansions of dimensions for the two twist-two operators
\begin{equation}\begin{split}\label{D2S2}
\Delta^{(2)}=&
2+S
-\frac{\lambda}{2}\left[\lambda-\frac{\zeta}{\sqrt{S(S+1)}}\right]
+\frac{\lambda^2}{4S^2(S+1)^{2}\zeta}\biggl[S(S+1)\lambda^2[(S(S+1)-5)\zeta\\
&-\sqrt{S(S\!+\!1)}(S(S\!+\!1)\!+\!3)\lambda]\!+\!8\mu^2[\zeta(1\!-\!S^2)\!+\!\sqrt{S(S\!+\!1)}(S(2S\!+\!1)\!-\!2)\lambda]\biggr]
+\dots\\
\Delta^{(2')}=&
2+S
-\frac{\lambda}{2}\left[\lambda+\frac{\zeta}{\sqrt{S(S+1)}}\right]
+\frac{\lambda^2}{4S^2(S+1)^{2}\zeta}\biggl[S(S+1)\lambda^2[(S(S+1)-5)\zeta\\
&+\sqrt{S(S\!+\!1)}(S(S\!+\!1)\!+\!3)\lambda]\!+\!8\mu^2[\zeta(1\!-\!S^2)\!-\!\sqrt{S(S\!+\!1)}(S(2S\!+\!1)\!-\!2)\lambda]\biggr]
+\dots\,.
\end{split}\end{equation}
Expanding around the physical pole $\nu=-i(S+2)/2$, we get for the two remaining, twist-four operators
\begin{equation}\begin{split}\label{D4S2}
\Delta^{(4)}=&
4+S
+\frac{\lambda}{2}\left[\lambda+\frac{\tau}{\sqrt{(S\!+\!1)(S\!+\!2)}}\right]
-\frac{\lambda^2}{4(S+1)(S+2)^{2}}\biggl[(S+2)(S(S+3)-3)\lambda^2\\
&-8(S\!+\!3)\mu^2\!+\!\sqrt{\frac{S+2}{S+1}}\frac{\lambda}{\tau}[(S\!+\!1)(S\!+\!2)(S(S\!+\!3)\!+\!5)\lambda^2\!-\!8(S(2S\!+\!7)\!+\!4)\mu^2]\biggr]
+\dots\\
\Delta^{(4')}=&
4+S
+\frac{\lambda}{2}\left[\lambda-\frac{\tau}{\sqrt{(S\!+\!1)(S\!+\!2)}}\right]
+\frac{\lambda^2}{4(S+1)(S+2)^{2}}\biggl[(S+2)(S(S+3)-3)\lambda^2\\
&-8(S\!+\!3)\mu^2\!+\!\sqrt{\frac{S+2}{S+1}}\frac{\lambda}{\tau}[(S\!+\!1)(S\!+\!2)(S(S\!+\!3)\!+\!5)\lambda^2\!-\!8(S(2S\!+\!7)\!+\!4)\mu^2]\biggr]
+\dots
\end{split}\end{equation}
where we introduced the following short-hand notation
\begin{equation}\label{zetakappa}
\zeta=\sqrt{S(S+1)\lambda^2-16\mu^2}\qquad\text{and}\qquad
\tau=\sqrt{(S+1)(S+2)\lambda^2-16\mu^2}\,.
\end{equation}
In both cases we presented only the first few terms of the expansions since the following ones are quite cumbersome. Moreover, we notice that the weak coupling expansions of the solutions are divergent but, as we point out later, by the analysis of Sec.\ref{fpoint2} the sum of the corresponding OPE contributions has a well defined expansion (see footnote \ref{foot:grisha}).\\ Taking into account the quantum numbers of the external state \(\Tr[\phi_1\phi_2^\dagger]\) we can list the operators of twist-2 and -4 which mix, obtaining the expression for the exchanged operators in the OPE from the diagonalization of the mixing matrix. At \(\Delta_0-S=2\) we recognize that \(\Tr[\phi_1 \,\partial_+^S \phi_2^\dagger]\) and \(\Tr[\psi_1\, \partial_+^{S-1} (n\cdot \sigma) \bar \psi_2]\), where \(n\cdot n=0\) and \(\partial_+=(n\cdot \partial)\), operators can mix if \(S>0\). Interestingly, due to trace ciclicity, at twist-\(2\), the mixing transitions are symmetric and then there are no logarithmic operators despite chiral interactions of \eqref{fullL} (see App.\ref{appendix:sec2}).  On the other hand, at \(\Delta_0-S=4\) many more operators can realise the \(U(1)^{\otimes 3}\) quantum numbers \((1,-1,0)\), building up a large mixing matrix. As pointed out in App.\ref{appendix:sec2}, already for \(S=0\) the anomalous dimension matrix is not diagonalizable. We conjecture there that  its canonical Jordan form presents several logarithmic multiplets (Jordan blocks) of various ranks, together with two non-zero diagonal elements, corresponding to solutions \eqref{D4S2} (or  \eqref{D4S20} and \(S=0\)).

The zero-spin case presents some peculiar behaviors. Indeed, expanding \eqref{eq22} for $S=0$ around the physical poles $\nu=0$ at weak coupling, we obtain the following expansion for the twist-two operator
\begin{equation}\begin{split}\label{D2S20}
\Delta^{(2)}\big|_{S=0}=&
2
-2i\lambda\sqrt{\lambda^2+2\mu^2}
+\frac{2i\lambda^3\mu^2}{\sqrt{\lambda^2+2\mu^2}}
+\frac{i\lambda^3(\lambda^6+6\lambda^4\mu^2+17\lambda^2\mu^4+16\mu^6)}{(\lambda^2+2\mu^2)^{3/2}}
+\dots
\end{split}\end{equation}
and around the pole $\nu=-i$ --  the following twist-four operators
\begin{equation}\begin{split}\label{D4S20}
\Delta^{(4)}=&
4
+\frac{\lambda}{2}\left[\lambda+\sqrt{\lambda^2-8\mu^2}\right]
+\frac{\lambda^2}{8}\left[3\lambda^2+12\mu^2-\frac{5\lambda^3-16\lambda\mu^2}{\sqrt{\lambda^2-8\mu^2}}\right]
+\dots\\
\Delta^{(4')}=&
4
+\frac{\lambda}{2}\left[\lambda-\sqrt{\lambda^2-8\mu^2}\right]
+\frac{\lambda^2}{8}\left[3\lambda^2+12\mu^2+\frac{5\lambda^3-16\lambda\mu^2}{\sqrt{\lambda^2-8\mu^2}}\right]
+\dots\end{split}\end{equation}
 The fact that at $S=0$ there is only one solution corresponding to operators of length-two is not surprising: the only scalar operator which has twist-2 and the correct quantum numbers is indeed $\text{Tr}[\phi_1\phi_{2}^\dagger]$, that is the one exchanged in the OPE channel (see App.\ref{appendix:sec2}). Moreover, similarly to the case of the spectrum of the  length-two operator in the correlation function computed in Sec.\ref{sec:spectrumphi1phi1}, the limit $S\rightarrow 0$ and the weak-coupling limit are not commutative, thus the scaling dimension $\Delta^{(2)}$ for $S=0$ presents a different expansion in power of the coupling w.r.t. the case $S>0$. The explanation is the same of the previous case. Indeed, since the physical pole is situated at $\nu=-iS/2$, and the mirror pole at $\nu=+iS/2$, at weak coupling,   when $S\to 0$, they pinch the integration contour of \eqref{Gnu2} at the origin producing a divergence. Hence, the contribution of the double-traces is needed in this case to produce a non-vanishing term that cancels this divergence at weak coupling. Again we stress that at finite couplings the solutions \eqref{eq22} are well-defined even at zero spin.

\paragraph{Strong coupling expansion:}
Equation \eqref{eq22} has 6 solutions at strong coupling \textit{i.e.} for $\lambda,\mu\rightarrow\infty$. 
The explanation of the fact that we have a different number of solutions w.r.t the weak coupling case can be identified by the analysis of the exchanged operators in App.\ref{appendix:sec2}. In particular when $\lambda,\mu\rightarrow\infty$ the two operators $\mathcal{O}'_{\pm}$ defined in \eqref{Oprimepm} collapse to a single operator.
Then the resulting physical solutions are
\begingroup\makeatletter\def\f@size{10}\check@mathfonts
\begin{equation}\begin{split}\label{D2S2strong}
\Delta^{(2)}_\infty=&
\!-\!2i\lambda\!+\!2
\!+\!\frac{i[(2\!+\!S(S\!+\!2))\lambda^2\!-\!2\mu^2]}{4\lambda^3}
\!+\!\frac{i[(S^2\!-\!2)(2\!+\!S(S\!+\!4))\lambda^4\!+\!4S(S\!+\!2)\lambda^2\mu^2\!+\!20\mu^4]}{64\lambda^{7}}
\!+\!\dots\\
\Delta^{(4)}_\infty=&
2\lambda\!+\!2
\!+\!\frac{i\mu}{\lambda}
\!+\!\frac{(3+2S(S+2))\lambda^2+2\mu^2}{4\lambda^3}
\!-\!\frac{i[(S^2\!+\!1)^4\lambda^4\!+\!2(5S(S+2)\!+\!7)\lambda^2\mu^2\!+\!13\mu^4]}{32\lambda^{5}\mu}
\!+\!\dots\\
\Delta^{(4')}_\infty=&
2\lambda\!+\!2
\!-\!\frac{i\mu}{\lambda}
\!+\!\frac{(3+2S(S+2))\lambda^2+2\mu^2}{4\lambda^3}
\!+\!\frac{i[(S^2\!+\!1)^4\lambda^4\!+\!2(5S(S+2)\!+\!7)\lambda^2\mu^2\!+\!13\mu^4]}{32\lambda^{5}\mu}
\!+\!\dots
\end{split}\end{equation}\endgroup
and the remaining solutions are associated to the shadow operators. 
Notice that, doing simply the power counting in terms of the original couplings all the strong coupling expansions are growing linearly with $\xi$. The $S\rightarrow 0$ limit is not singular at strong coupling, then one can compute $\Delta_{{\infty}}\big|_{S=0}$ directly from \eqref{D2S2strong}. 
 
\paragraph{The spectrum of exchanged operators in reductions of \(\chi\)CFT}   

The scaling dimensions of the exchanged operators in the correlation function $G_{\phi_1\phi_{2}^\dagger}$ for the reductions of $\chi$CFT introduced in Sec.\ref{sec:section2}, can be computed as solutions of the spectral equation \eqref{eq22} in which we are applying the limits on the couplings, or even directly on the weak- and strong-coupling expansions. In Tab.\ref{tab:family2} we summarize our results.
\begin{table}[!t]
\begin{center}
{\renewcommand{\arraystretch}{1.5}
\begin{tabular}{l*{5}{c}r}
              & limit & $\Delta^{(2)}$ &$\Delta^{(2')}$& $\Delta^{(4)}$ & $\Delta^{(4')}$  \\
\hline
 \multirow{2}{*}{$\chi_0$CFT}& $\xi_1\lor\xi_{2}\rightarrow 0 $   &\multicolumn{2}{c}{$2+S$}&\multicolumn{2}{c}{$4+S$} \\
                     & $\xi_{3}\rightarrow 0 $    &\multicolumn{2}{c}{$\Delta_{\text{bi}}^{(2)}(\lambda^4)$}&\multicolumn{2}{c}{$\Delta_{\text{bi}}^{(4)}(\lambda^4)$}\\
                     \hline
bi-scalar & $\xi_i\land\xi_k\rightarrow 0 \;\;\forall i,k \qquad\quad$ &\multicolumn{2}{c}{$2+S$}&\multicolumn{2}{c}{$4+S$}\\   
                \hline
$\beta$-deform & $\xi_1\!=\!\xi_2\!=\!\xi_3\!=\!\xi$ & $\Delta^{(2)}(\xi)$ &$\Delta^{(2')}(\xi)$& $\Delta^{(4)}(\xi)$ & $\Delta^{(4')}(\xi)$  
\end{tabular}}
\end{center}
\caption{In this table we summarize the operator and dimension content of exchange operators in three reductions of \(\chi\)CFT for the correlator $G_{\phi_1\phi_2^\dagger}$. In our notation $\Delta^{(i)}(\xi)=\Delta^{(i)}(\lambda\rightarrow\xi,\mu\rightarrow \xi)$ and $\Delta_{\text{bi}}^{(i)}$ is defined in \eqref{biscspe}.  }
\label{tab:family2}
\end{table}
\begin{itemize}
\item $\chi_0$CFT: sending one of the couplings to zero we perform the reduction from the full $\chi$CFT to the $\chi_0$CFT. 
In this case, the reduction doesn't give a unique spectrum. Indeed, setting $\xi_1$ or $\xi_2$ to zero, or equivalently $\lambda\rightarrow 0$, we obtain two protected operators. This is clear when considering the spectral equation \eqref{eq22} in which only the first term on the left-hand side will contribute and clearly it describes two protected solutions (plus the shadow operators associated to them). If we consider  $\xi_3$ to vanish, or equivalently $\mu\rightarrow 0$, and again referring to the spectral equation \eqref{eq22}, it is clear that the second term in the left-hand side vanishes but, in contrast  to the previous case, in the right-hand side of the equation the dependence on the coupling is still present. The resulting equation is the same as describing the spectrum of exchanged operators for the four-point function $G_{\phi_1\phi_1}$ in the bi-scalar theory. Then we obtain 2 solutions with the spectrum given by \eqref{biscspe} where $\xi=\lambda$. 
\item bi-scalar theory: the reduction to bi-scalar theory consists of the limit in which two couplings of $\chi$CFT vanish.
For the correlator we are considering, any choice of the vanishing couplings leads to protected solutions, as expected. Indeed, in the bi-scalar theory the lack of interaction vertices doesn't develop an anomalous dimension for exchanged operators with external states given by $\phi_1$ and $\phi_2^\dagger$ as previously noticed in \cite{Grabner:2017pgm,Gromov:2018hut}.  
\item $\beta$-deformed theory (all three couplings are equal): In this case, the restoration of one supersymmetry is not sufficient to constrain any of the solutions. Then their spectrum can be easily read off applying the equal couplings limit, for example at weak coupling, to the expansions \eqref{D2S2} and \eqref{D4S2}.
\end{itemize}

\subsection{The structure constant of the exchanged operators}\label{sec:OPEphi1phi2dag}

Once the spectrum of the exchanged operators is computed, in order to obtain the full set of conformal data for the four point function $G_{\phi_1\phi_{2}^\dagger}$, one has to compute the OPE coefficients. Their definition is given by \eqref{OPE} or, equivalently, by \eqref{OPE2} where
\begin{equation}\label{OPE3}
\mathcal{R}_{\Delta,S}\,=\,\frac{d}{d \Delta} \left({{\frac{1}{{h_B}_{\Delta,S}}}}+\frac{(4\pi)^8\lambda^2 \mu^2\, [(\Delta-2)^2+S(S+2)]\,{h_B}_{\Delta,S}}{2- 8(2\pi)^4 \lambda^2\,[(\Delta-2)^2+S(S+2)]\,{h_B}_{\Delta,S}}\right)\, ,
\end{equation}
where the eigenvalues ${h_B}_{\Delta,S}$ is defined in \eqref{hbS} and the constants $c_F=-2 c_B=-1/(2\pi^2)$. Plugging the eigenvalues into \eqref{OPE3} and performing the derivative, we obtain explicit but rather cumbersome result that we will not present here. In the following paragraphs we will consider its weak- and strong- coupling expansions.

\paragraph{Weak coupling expansion}

  Substituting  the weak coupling expansions of the scaling dimensions computed in \eqref{D2S2} into  \eqref{OPE2},\eqref{OPE3}, we obtain the following weak coupling expansions for the structure constants associated to the twist-two operators for $S\neq 0$
\begin{equation}\begin{split}\label{structureS22}
C_{\Delta^{(2)},S}=&\frac{\sqrt{\pi}\Gamma(S+1)}{2^{S}\Gamma(S+1/2)}\frac{(\zeta+\sqrt{S(S+1)})^2}{[(\zeta+\sqrt{S(S+1)})^2-16\mu^2]}+\dots\\
C_{\Delta^{(2')},S}=&\frac{\sqrt{\pi}\Gamma(S+1)}{2^{S}\Gamma(S+1/2)}\frac{(\zeta-\sqrt{S(S+1)})^2}{[(\zeta-\sqrt{S(S+1)})^2-16\mu^2]}+\dots\, ,
\end{split}\end{equation}
and a similar substitution of  \eqref{D4S2} for the twist-four operators at any \(S\) gives
\begin{equation}\begin{split}\label{structureS42}
C_{\Delta^{(4)},S}=&\frac{\sqrt{\pi}\lambda\Gamma(S+2)(\tau-\sqrt{(S+1)(S+2)}\lambda)^2(\tau+\sqrt{(S+1)(S+2)}\lambda)}{2^{2S+5}\Gamma(S+3/2)\sqrt{(S+1)(S+2)}[(\tau-\sqrt{(S+1)(S+2)}\lambda)^2-16\mu^2]}+\dots\\
C_{\Delta^{(4')},S}=&-\frac{\sqrt{\pi}\lambda\Gamma(S+2)(\tau-\sqrt{(S+1)(S+2)}\lambda)(\tau+\sqrt{(S+1)(S+2)}\lambda)^2}{2^{2S+5}\Gamma(S+3/2)\sqrt{(S+1)(S+2)}[(\tau+\sqrt{(S+1)(S+2)}\lambda)^2-16\mu^2]}+\dots\, .
\end{split}\end{equation}
where the functions $\zeta$ and $\tau$ are defined in \eqref{zetakappa}. Similarly to the expansion of the scaling dimensions, the OPE coefficients of the twist-two operators are singular for $S=0$. Indeed, as discussed in Sec.\ref{sec:spectrumphi1phi2}, due to the singularity arising at zero spin,  the weak coupling and  $S\rightarrow 0$ limits don't commute. In order to obtain the correct weak coupling expansion for the twist-two operators, one has to set $S=0$ in \eqref{OPE3} and then expand in the couplings. 
In the $S=0$ case we found that the spectral equation \eqref{spec22} has 6 solutions instead of 8, as expected from the operator mixing analysis in App.\ref{appendix:sec2} for twist-\(2\) solutions. In particular the zero spin expansion of the structure constant related to the twist-two operator is
\begin{equation}\begin{split}\label{structure2S0}
C_{\Delta^{(2)},0}=&1+2i\lambda\sqrt{\lambda^2+2\mu^2}
-\lambda^2\left[2\lambda^2+6\mu^2+\frac{2i\lambda\mu^2}{\sqrt{\lambda^2+2\mu^2}}\right]+\dots\,.
\end{split}\end{equation}
Since the \(S\to 0\) singularity is not arising for the twist four operators, one can read off the spinless OPE coefficients for the twist-four operators directly setting $S=0$ in \eqref{structureS42}, thus obtaining
\begin{equation}\begin{split}\label{structure4S0}
C_{\Delta^{(4)},0}=&\frac{-\lambda\mu^2}{4\sqrt{\lambda^2-8\mu^2}}\!-\!\frac{\lambda^2[2\lambda^5\!-\!19\lambda^3\mu^2\!+\!72\lambda\mu^4\!-\!(2\lambda^4-5\lambda^2\mu^2-88\mu^4)\sqrt{\lambda^2-8\mu^2}]}{32(\lambda^2-8\mu^2)^{3/2}}+\dots\\
C_{\Delta^{(4')},0}=&\frac{\lambda\mu^2}{4\sqrt{\lambda^2-8\mu^2}}\!+\!\frac{\lambda^2[2\lambda^5\!-\!19\lambda^3\mu^2\!+\!72\lambda\mu^4\!+\!(2\lambda^4-5\lambda^2\mu^2-88\mu^4)\sqrt{\lambda^2-8\mu^2}]}{32(\lambda^2-8\mu^2)^{3/2}}+\dots\, .
\end{split}\end{equation}
In analogy with the  analysis of spectrum, the divergence arising at  $S=0$ for the twist-two operators suggests that the weak coupling expansion is sensitive to the double trace counterterms. Moreover, for any $S$ the twist-four OPE coefficients are suppressed by a factor of order $\xi^{3/2}$ as compared to the twist-two ones.

\paragraph{Strong coupling expansion}

In Sec.\ref{sec:spectrumphi1phi2} we studied the strong coupling spectrum of the exchanged operator for the four-point function $G_{\phi_1\phi_{2}^\dagger}$.  
Since the scaling dimensions are growing linearly with the coupling  $\lambda$,  the OPE coefficients can be computed expanding \eqref{OPE2} and \eqref{OPE3} in $\Delta\rightarrow \infty$ and then plugging the strong coupling asymptotics of scaling dimensions in the resulting expansion. Since in this limit the coefficient $c_2(\Delta,S)$ in \eqref{OPE2} is dominant w.r.t. the one of \eqref{OPE3}, the first terms of the expansion are exactly the same as in \eqref{OPEstrong1}. Obviously at higher orders in $1/\Delta$, \eqref{OPE3}  starts  contributing and then we obtain
\begin{equation}\begin{split}\label{OPEstrong24}
C_{\Delta^{(2)}_\infty,S}
=&\frac{2^{5}(S\!+\!1)}{2^{2\Delta^{(2)}_\infty}\Delta^{(2)}_\infty}\tan\!\!\left(\!\pi\frac{\Delta^{(2)}_\infty\!-\!S}{2}\right)\!
\left[1+\frac{3i}{4\lambda}-\frac{[5\!+\!4S(S\!+\!2)]\lambda^2+32\mu^2}{32\lambda^4}+\dots\right],\\
C_{\Delta^{(4)}_\infty,S}
=&\frac{2^{5}(S\!+\!1)}{2^{2\Delta^{(4)}_\infty}\Delta^{(4)}_\infty}\tan\!\!\left(\!\pi\frac{\Delta^{(4)}_\infty\!-\!S}{2}\right)\!
\left[1+\frac{3}{4\lambda}+\frac{[5\!+\!4S(S\!+\!2)]\lambda^2-12i \lambda\mu+32\mu^2}{32\lambda^4}+\dots\right],\\
C_{\Delta^{(4')}_\infty,S}
=&\frac{2^{5}(S\!+\!1)}{2^{2\Delta^{(4')}_\infty}\Delta^{(4')}_\infty}\tan\!\!\left(\!\pi\frac{\Delta^{(4')}_\infty\!-\!S}{2}\right)\!
\left[1+\frac{3}{4\lambda}+\frac{[5\!+\!4S(S\!+\!2)]\lambda^2+12i \lambda\mu+32\mu^2}{32\lambda^4}-\dots\right]\, .
\end{split}\end{equation}
Thus, given the linear growth of the scaling dimensions $\Delta_{\infty}$ with coupling,  the OPE coefficients are exponentially small at strong coupling. 
The $S\rightarrow 0$ limit is not singular at strong coupling, thus one can compute $C_{\Delta_{\infty},0}$ directly from expansion \eqref{OPEstrong24}, setting there $S=0$.

\subsection{The four-point correlation function}
\label{fpoint2}
Once we computed the conformal data in Sec.\ref{sec:spectrumphi1phi2} and Sec.\ref{sec:OPEphi1phi2dag}, we can determine the four-point function \eqref{phi1phi2dagger} by means of \eqref{4pointgeneral}. In the case $\mathcal{O}_1=\phi_1$ and $\mathcal{O}_2=\phi_{2}^\dagger$ we obtain 
\begin{equation}
G_{\phi_1\phi_{2}^\dagger}(x_1,x_2|x_3,x_4)=\frac{c_B^2}{x_{12}^2x_{34}^2}\,\mathcal{G}_{\phi_1\phi_{2}^\dagger}(u,v)\, .
\end{equation} 
The cross-ratios are defined as $u=x_{12}^2 x_{34}^2/(x_{13}^2 x_{24}^2)$ and $v=x_{14}^2x_{23}^2/(x_{13}^2 x_{24}^2)$ and $\Delta_{\phi_i}=1$. 
The function $\mathcal{G}_{\phi_1\phi_1}(u,v)$ can be written in terms of the OPE representation \eqref{OPE} as a sum over the non-negative integer Lorentz spin $S$ and the states with scaling dimensions $\Delta$. From the study of the spectrum of exchanged operators in Sec.\ref{sec:spectrumphi1phi2} it turns out that in the OPE channel 
 two operators of length-two for $S\neq 0$ are exchanged, but only one when $S=0$. Thus, in order to write the four point function in a compact form, we set \(C_{\Delta^{(2')},S}\) to vanish at \(S=0\), or equal to \eqref{structureS22} otherwise, then we have
\begin{equation}\label{GGfinaleOPE2}
\mathcal{G}_{\phi_1\phi_{2}^\dagger}(u,v)=
\sum_{S=0}^\infty[C_{\Delta^{(2)},S}\;g_{\Delta^{(2)},S}+C_{\Delta^{(2')},S}\;g_{\Delta^{(2')},S}+C_{\Delta^{(4)},S}\;g_{\Delta^{(4)},S}+C_{\Delta^{(4')},S}\;g_{\Delta^{(4')},S}]\, ,
\end{equation}
where the scaling dimensions are defined by the spectral equation \eqref{spec22}.  At weak coupling they are computed in \eqref{D2S2} and \eqref{D4S2}, and at strong coupling -- in \eqref{D2S2strong}. The structure constants are defined by \eqref{OPE2} and \eqref{OPE3}. They are computed at weak and strong coupling in \eqref{structureS22}, \eqref{structureS42} and \eqref{OPEstrong24} respectively. The four-dimensional conformal blocks $g_{\Delta,S}$   are defined \eqref{defg}.

As already pointed out for the four-point correlator \(G_{\phi_1\phi_1}\), the proper definition of the four-point correlation function $G_{\phi_1\phi_2^\dagger}$,  takes into account the symmetrization $x_3\leftrightarrow x_4$. Under this symmetry, the cross-ratios transform as $u\rightarrow u/v$ and $v\rightarrow 1/v$ and correspondingly, from the definition  \eqref{defg}, the conformal blocks $g_{\Delta,S}(u/v,1/v)=(-1)^Sg_{\Delta,S}(u,v)$.
Combining together this relation with \eqref{GGfinaleOPE2}, it's easy to see that including the symmetry $x_3\leftrightarrow x_4$ means that the terms in \eqref{GGfinaleOPE2} with odd $S$ cancel out whereas those with even $S$ get doubled.

Despite  the presence of singularities in the weak-coupling expansions of scaling dimensions and OPE coefficients, their sum in \eqref{GGfinaleOPE2} is well-defined and non-singular. Indeed plugging the conformal data into \eqref{GGfinaleOPE2}, we obtain an expansion in powers of the couplings that is compatible with the interpretation of the correlation function as a sum of Feynman diagrams in perturbation theory (see Sec.\ref{sec:feynman12} for an explicit example). In particular, since the first non-trivial order is fixed by the $S=0$ conformal data,  it is easy to write the two leading contributions to $\mathcal{G}_{\phi_1\phi_2^\dagger}$ in terms of known functions as follows
\begin{equation}\label{4pointfinal2}
\mathcal{G}_{\phi_1\phi_2^\dagger}(u,v)=u+i\lambda\sqrt{\lambda^2+2\mu^2}\,u\,\Phi^{(1)}(u,v)+\dots\,.
 \end{equation}
where  $\Phi^{(1)}$ is the Bloch-Wigner dilogarithm  function defined in \eqref{phi1dav}. 
Expanding \eqref{GGfinaleOPE2} at higher order in the couplings we obtain a cumbersome result. However, we notice that the maximum transcendental weight of the involved functions grows linearly with the order. This suggest, together with the presence of the Bloch-Wigner dilogarithm at the first non-trivial order, the possibility to interpret the four-point function \eqref{GGfinaleOPE2} as a combination of a class of special iterated integrals, the so-called harmonic polylogarithms \cite{Remiddi:1999ew}  following the same idea as of \cite{Gromov:2018hut}.

\section{Correlation functions at weak coupling from Feynman diagrams}\label{sec:feynman}

In Sec.\ref{sec:phi1phi1} and Sec.\ref{sector2}, we have analyzed two different four-point functions computing their conformal data by means of the Bethe-Salpeter method. With this procedure we were able to diagonalize the graph-building operators and write exact equations for the spectrum of exchanged operators even though we ignored on the way the contribution of the double-trace interactions \eqref{doubletraces}. The double-trace counterterms are necessary in the action to have a consistent description of the double-scaled theory \eqref{fullL} in the perturbative regime and in particular for the restoration of conformal symmetry.  

In this section, we will study  the weak coupling expansions of the four-point functions related to the operators \eqref{operators} and clarify the role of the double-trace terms for this expansion. As we already mentioned, bosonic and fermionic wrappings in the related Feynman graphs develop UV divergencies at short distances.  Adding the double-trace vertices in the perturbative expansion we will be able to determine the conformal fixed points \eqref{fixedpoint} canceling divergencies generated by the single trace terms. However, they will not affect the finite coupling solutions computed in the section above.

The double-trace counterterms are given by the Lagrangian \eqref{doubletraces}. In general, this action contains 9 terms but, due of the cylindric  topology of Feynman diagrams for the observables we are computing. For any four-point function only one double-trace term is contributing generating a new local four-scalar vertex (see for instance Fig.\ref{fig:phi1phi1}). This fact is crucial to ensure that conformal symmetry is restored. Indeed, in this case we know that the $\beta$-function can be written as \eqref{betageneral} and it admits two fixed points $\alpha^2_{j\star}$ as in \eqref{fixedpoint}.

If we focus on the Feynman diagram expansion of the four point-functions we have to deal with divergent integrals. Then we have to introduce dimensional regularization setting $D=4-2\epsilon$. One important observation is that the diagrams containing fermionic contributions produce the same divergence as the bosonic ones. In other words the fermionic kernels contains the divergent part of the bosonic one plus a remainder function that is finite in $D=4$ which therefore does not require regularization. Then the divergent operator can be written as
\begin{equation}\begin{split}\label{HVdimreg}
\left[\hat{\mathcal{V}}\;\Phi\right]\!(x_1,x_2)=&2 c_B^2\int \frac{d^{4-2\epsilon}y_1 d^{4-2\epsilon}y_2}{[(x_1-y_1)^2(x_2-y_2)^2]^{1-\epsilon}}\delta^{(4-2\epsilon)}(y_{12}) \,\Phi(y_1,y_2)\\
\left[\hat{\mathcal{H}}_B\;\Phi\right]\!(x_1,x_2)=&c_B^4\int \frac{d^{4-2\epsilon}y_1 d^{4-2\epsilon}y_2}{[(x_1-y_1)^2(x_2-y_2)^2y_{12}^4]^{1-\epsilon}}\,\Phi(y_1,y_2)\, ,
\end{split}\end{equation}
where $\Phi(x_1,x_2)$ is a test function.

Given the Hamiltonians \eqref{HVdimreg} and the definition \eqref{Ggeom}, the four-point correlation function is defined as follows
\begin{equation}\label{Gdimreg}
G(x_1,x_2|x_3,x_4)=\lim_{\epsilon\rightarrow 0}\left(\frac{c_B}{x^2_{34}}\right)^{\Delta_{\mathcal{O}_1}+\Delta_{\mathcal{O}_2}-D}\langle x_1,x_2|\frac{\hat{\mathcal{H}}_B}{1-\chi_\mathcal{V}^\star\hat{\mathcal{V}}-\chi_B\hat{\mathcal{H}}_B-\chi_F\hat{\mathcal{H}}_F}|x_3,x_4\rangle\, ,
\end{equation}
where the effective coupling  \(\chi_\mathcal{V}^\star=\chi_\mathcal{V}\vert_{\alpha_j^2=\alpha_{j\star}^2}\), i.e. it is taken at the fixed point $\alpha_j^2=\alpha_{j\star}^2$. 
It is clear that for $\epsilon\neq 0$ conformal symmetry is broken. However, expanding \eqref{Gdimreg} at weak-coupling in terms of Feynman diagrams, one can demonstrate order-by-order how conformal symmetry is restored. 
In the following sections, we present some examples of this mechanism for the four-point correlation functions associated to the operators \eqref{operators}.

\subsection{Four-point function of  $\mathcal{O}_1(x)=\phi_1(x)$ and $\mathcal{O}_2(x)=\phi_{2}(x)$}\label{sec:sector3}

Let's start from this simplistic example. Consider $\mathcal{O}_1(x)=\phi_j(x)$ and $\mathcal{O}_2(x)=\phi_{k\neq j}(x)$ with $j,k=1,2,3$. 
Since the correlators are the same for any $j$ and $k$, we choose $j=1$ and $k=2$.  Then the four-point function we want to study is as follows
\begin{equation}\label{phi1phi2}
G_{\phi_1\phi_{2}}(x_1,x_2|x_3,x_4)=\langle\text{Tr}[\phi_1(x_1)\phi_{2}(x_2)]\text{Tr}[\phi_1^\dagger(x_3)\phi_{2}^\dagger(x_4)]\rangle\,.
\end{equation}
In the planar limit $N_c\rightarrow \infty$, the weak coupling expansion of \eqref{phi1phi2} in terms of Feynman diagrams is given by a combination of the following vertices
\begin{equation}
 (4\pi)^2\alpha_{3}^2 \, \text{Tr}[\phi_1\phi_2](x) \,\text{Tr}[\phi_1\phi_2]^\dagger (x)\,,\qquad \quad  (4\pi)^2\xi^2_3 \, \text{Tr}[\phi_1^\dagger\phi_2^\dagger\phi_1\phi_2](x)\, ,
\end{equation}
 and it can be written as follows
\begin{equation}
G_{\phi_1\phi_{2}}(x_1,x_2|x_3,x_4)=\sum_{\ell=0}^\infty(4\pi)^{2\ell}(\alpha_3^2+\xi_3^2)^\ell\,G_{\phi_1\phi_{2}}^{(\ell)}(x_1,x_2|x_3,x_4)\, ,
\end{equation}
where $\ell$ represents the perturbative order. It is straightforward to see that Feynman diagrams at any order $\ell$ form a chain structure alternating groups of single- and double-trace vertices. These vertices insert into the graphs identical primitive divergencies. Then at the conformal critical point ${\alpha_{3}^2}{}_\star= -\xi^2_3$ the graphs cancel each other except for the $\ell=0$ term, namely
\begin{equation}
G_{\phi_1\phi_2}^{(0)}(x_1,x_2|x_3,x_4)=\frac{c_B^2}{x_{13}^2x_{24}^2}\, .
\end{equation}
Restoring the two point function in the limit $x_1\rightarrow x_2$  and $x_3\rightarrow x_4$ we notice that the spectrum of the operator $\text{Tr}[\phi_1\phi_{2}]$ is not affected by quantum corrections. Indeed, its scaling dimension is protected and equal to the bare one
\begin{equation}
\Delta^{(2)}=2\,.
\end{equation}

\subsection{Four-point function of  $\mathcal{O}_1(x)=\mathcal{O}_2(x)=\phi_1(x)$}\label{sec:feynman11} 

In Sec.\ref{sec:phi1phi1} we studied the contribution to the four-point function $G_{\phi_1\phi_1}$ of diagrams generated by the bosonic and fermionic Hamiltonians $\hat{\mathcal{H}}_B$ and $\hat{\mathcal{H}}_F$ but ignored the double-trace vertices, which works well for finite couplings. Since any diagram in Fig.\ref{fig:pert1}  (except for the trivial leading order diagram) is UV divergent,  we will re-introduce in this section the double-trace counterterms in the perturbative expansion in order to make the weak-coupling expansion UV finite and to restore conformal symmetry.
  \begin{figure}[!t]
 \centering
  \subfigure[$\alpha_1^2$]
   {\includegraphics[width=4cm]{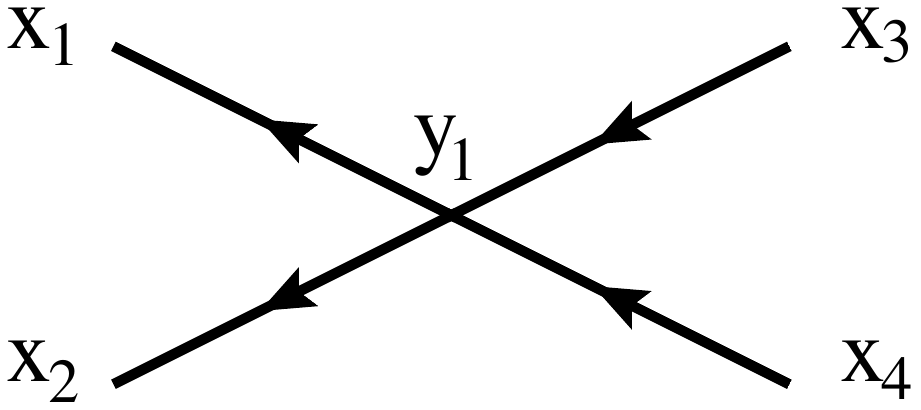}}
    \hspace{15mm}
    \subfigure[$\alpha_1^4$]
   {\includegraphics[width=6.9cm]{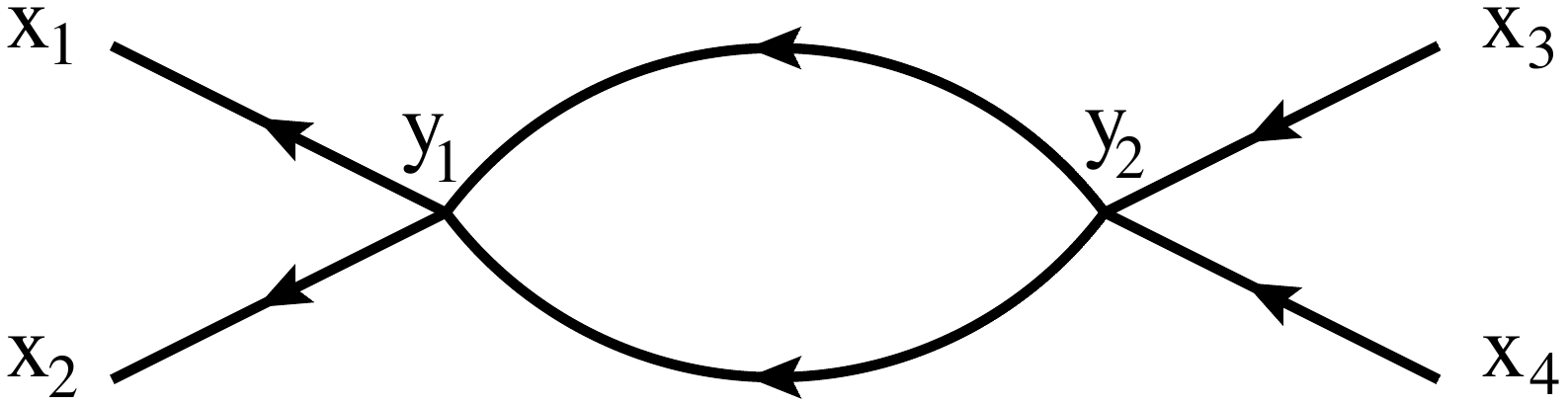}}
 \caption{First two contributions of the double trace vertex $\text{Tr}[\phi_1\phi_1]\text{Tr}[\phi_1^\dagger\phi_1^\dagger]$ to the four-point functions $G_{\phi_1\phi_{1}}$.}
  \label{fig:dtr1}
 \end{figure}
 
 Let's compute the first few orders of the weak coupling expansion of $G_{\phi_1\phi_1}$. In terms of Feynman diagrams, we have to compute the graphs given in Figg.\ref{fig:pert1} and \ref{fig:dtr1}. Defining a function $G^{(a,b,c)}_{\phi_1\phi_1}$ where $a$ counts the number of double-trace vertices, $b$ the number of bosonic vertices \eqref{vvv} and $c$ the number of fermionic vertices \eqref{fff}, we have
{\begin{equation}\label{expfixpoint1}
G_{\phi_1\phi_1}\!\!=\!G^{(0,0,0)}_{\phi_1\phi_1}\!\!+(4\pi)^2\!\alpha_1^2G^{(1,0,0)}_{\phi_1\phi_1}+(4\pi)^4[\alpha_1^4G^{(2,0,0)}_{\phi_1\phi_1}\!\!+(\xi_{2}^4\!\!+\xi_{3}^4)G^{(0,2,0)}_{\phi_1\phi_1}\!+\xi_{2}^2\xi_{3}^2\!G^{(0,0,4)}_{\phi_1\phi_1}]\!+\dots\, .
 \end{equation}} 
 The leading order $G^{(0,0,0)}_{\phi_1\phi_1}$ is already defined in \eqref{GGpert1}, thus $G^{(0,0,0)}_{\phi_1\phi_1}=G^{(0)}_{\phi_1\phi_1}$. The first correction is given by the diagram in Fig.\ref{fig:dtr1}(a). This contribution is finite and it can be written as follows
 \begin{equation}\label{1loopalpha}
 G^{(1,0,0)}_{\phi_1\phi_1}
 =\frac{2\pi^2c_B^4}{x_{12}^2x_{34}^2}u\Phi^{(1)}(u,v)\, ,
 \end{equation}
where  $\Phi^{(L)}$ is the ladder three-point function \cite{Usyukina:1992jd} that in the case $L=1$ is given by
the Bloch-Wigner dilogarithm  function\begin{equation}
\Phi^{(1)}(u,v)=\frac{1}{\theta}\left[2(\text{Li}_2(-\rho u)+\text{Li}_2(-\rho v))+\log\frac{v}{u}\log\frac{1+\rho v}{1+\rho u}+\log{\rho u}\log{\rho v}+\frac{\pi^2}{3}\right]\, ,
\end{equation}
with
\begin{equation}
\theta(u,v)\equiv\sqrt{(1-u-v)^2-4 u v}\qquad\text{and}\qquad \rho(u,v)\equiv \frac{2}{1-u-v+\theta}\,.
\end{equation}
The cross-ratios are $u=x_{12}^2 x_{34}^2/(x_{13}^2 x_{24}^2)$ and $v=x_{14}^2x_{23}^2/(x_{13}^2 x_{24}^2)$ and the constant $c_B=1/(4\pi^2)$.
 
The bosonic part of the two-loop correction given by $G^{(2,0,0)}_{\phi_1\phi_1}$ and $G^{(0,2,0)}_{\phi_1\phi_1}$ comes from the diagrams in Figg.\ref{fig:dtr1}(b), \ref{fig:pert1}(b) and \ref{fig:pert1}(c). The corresponding integrals are divergent and they need dimensional regularization, then we have
\begin{equation}\label{GI}
G^{(2,0,0)}_{\phi_1\phi_1}=4c_B^6\mathcal{I}(x_1,x_2|x_3,x_4)\,, \quad \text{and}\quad
G^{(0,2,0)}_{\phi_1\phi_1}=c_B^6\mathcal{I}(x_1,x_3|x_2,x_4)\, ,
\end{equation}
where we defined the short-hand notation
\begin{equation}\label{intI}
\mathcal{I}(x_1,x_2|x_3,x_4)=\int \frac{d^{4-2\epsilon}y_1 d^{4-2\epsilon}y_2}{[(x_1-y_1)^2(x_2-y_1)^2y_{12}^4(y_2-x_3)^2(y_2-x_4)^2]^{1-\epsilon}}\, .
\end{equation}
This integral is UV divergent at short distances $y_{12}^2\rightarrow 0$. Using the identity $1/(y_{12}^2)^{2-2\epsilon}=\pi^2\delta^{4-2\epsilon}(y_{12})/\epsilon+\mathcal{O}(\epsilon^0)$, one can compute the divergent part of the integral \eqref{intI} that is proportional to the same one-loop function found in \eqref{1loopalpha}, as follows
\begin{equation}\label{nnn}
\alpha_1^4G^{(2,0,0)}_{\phi_1\phi_1}\!\!+(\xi_{2}^4\!\!+\xi_{3}^4)G^{(0,2,0)}_{\phi_1\phi_1}=
\frac{\pi^4c_B^6}{x_{12}^2x_{34}^2}\!\left(\frac{4\alpha_1^4+\xi_{2}^4\!\!+\xi_{3}^4}{\epsilon}\right)\!u\Phi^{(1)}(u,v)+\text{finite}\, ,
\end{equation}
where for the purpose of this section we are not interested in the finite part.

Let us finally consider the fermionic contribution $G^{(0,0,4)}_{\phi_1\phi_1}$. This term corresponds to the Feynman diagram in Fig.\ref{fig:pert1}(d) and its integral representation in four dimensions is given in the last line of \eqref{GGpert1}. Since we are only interested in the computation of the UV divergent part of the diagram, we can avoid  computing the whole integral \eqref{GGpert1} in dimensional regularization and proceed in a more naive way. Indeed, representing the integral as in the last line of \eqref{GpertHH} and considering that the fermionic Hamiltonian can be written as a combination of the bosonic one and some finite reminder function as in \eqref{kernelF2}, we know that all the UV divergence is arising from $\mathcal{H}_B$. Then we can write   
\begin{equation}
G^{(0,0,4)}_{\phi_1\phi_1}=
-\frac{2\pi^4c_B^6}{x_{12}^2x_{34}^2}\,\frac{1}{\epsilon}\,u\,\Phi^{(1)}(u,v)+\text{finite}\, .
\end{equation}
Combining this result with \eqref{nnn} and \eqref{1loopalpha}, we obtain that the expansion \eqref{expfixpoint1} takes the expected form \eqref{4pointgeneral} with the function 
$\mathcal{G}_{\phi_1\phi_1}(u,v)$ given by
\begin{equation}\label{mathG1}
\mathcal{G}_{\phi_1\phi_1}(u,v)=u+2\alpha_1^2\,u\,\Phi^{(1)}(u,v)+\frac{4\alpha_1^4+\omega^4}{\epsilon}\,u\,\Phi^{(1)}(u,v)+\text{finite}(\kappa^4,\omega^4)+\dots\, ,
 \end{equation}
where the new couplings $\kappa$ and $\omega$ are defined in \eqref{coupdef1} and finite$(\kappa^4,\omega^4)$ stands for the finite part of $\mathcal{G}_{\phi_1\phi_1}$ at two-loop.
Finally, imposing the UV finiteness of the correlation function we obtain the first order of the fixed point as follows
\begin{equation}
\alpha^2_{1\star}=\pm\, i\, \frac{\omega^2}{2}+\dots\, ,
\end{equation}
and notice that it matches exactly the prediction given in \eqref{predictionspectrum}. 
Replacing the double-trace coupling in \eqref{mathG1} with its value\footnote{This choice is coherent with the sign convention used in Sec.\ref{sec:spectrumphi1phi1}.} \(\alpha^2_{1\star}=-\, i\, \frac{\omega^2}{2}\) we obtain the one-loop expansion of the correlation function as follows
\begin{equation}
\mathcal{G}_{\phi_1\phi_1}(u,v)=u-i\omega^2\,u\,\Phi^{(1)}(u,v)+\dots\,.
 \end{equation}
This perfectly matches the same quantity computed via OPE, with conformal data fixed by the Bethe-Salpeter method \eqref{4pointfinal1}.

\subsection{Four-point function of  $\mathcal{O}_1(x)=\phi_1(x)$ and $\mathcal{O}_2(x)=\phi_{2}^\dagger(x)$}\label{sec:feynman12}

In Sec.\ref{sector2}, we studied the four-point function $G_{\phi_1\phi_{2}^\dagger}$ by the Bethe-Salpeter method, considering the diagrams generated by bosonic and fermionic Hamiltonians \(\hat{\mathcal{H}}_B\) and \(\hat{\mathcal{H}}_F^{(n)}\), but ignoring the operator $\hat{\mathcal{V}}$, which is valid at any finite couplings. Since we want to analyze the weak coupling perturbative expansion of the correlator using Feynman diagrams, the graphs generated by this operator are needed in order to cancel the UV divergencies arising from the diagrams in Fig.\ref{fig:pert2}.
There is here a substantial difference w.r.t. the case of \(G_{\phi_1\phi_{1}}\) studied in Sec.\ref{sec:phi1phi1}: indeed the diagrams generated by $\hat{\mathcal{V}}$ are not only produced by the double-trace vertex $\text{Tr}[\phi_1\phi_{2}^\dagger]\text{Tr}[\phi_1^\dagger\phi_{2}]$ but also by the single-trace $\text{Tr}[\phi_1^\dagger\phi_{2}^\dagger\phi_1\phi_{2}]$. For this reason, we redefine the double-trace coupling as \eqref{alphatildedef} taking into account  both  contributions.
  \begin{figure}[!t]
 \centering
  \subfigure[$\alpha_1^2$]
   {\includegraphics[width=4cm]{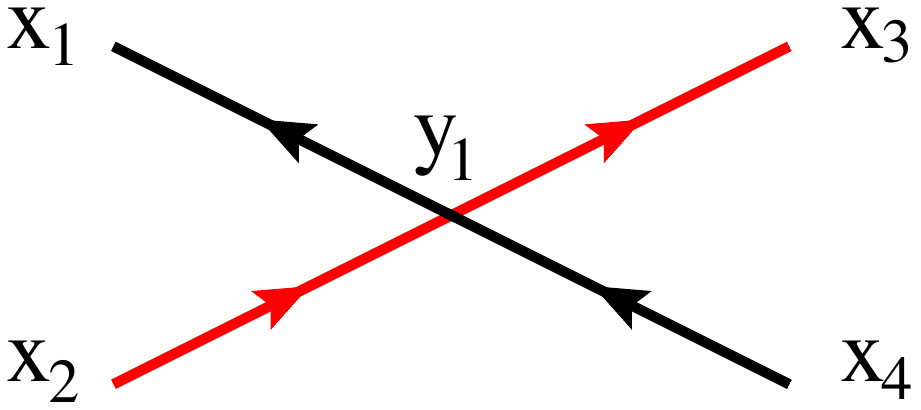}}
    \hspace{15mm}
    \subfigure[$\alpha_1^4$]
   {\includegraphics[width=6.9cm]{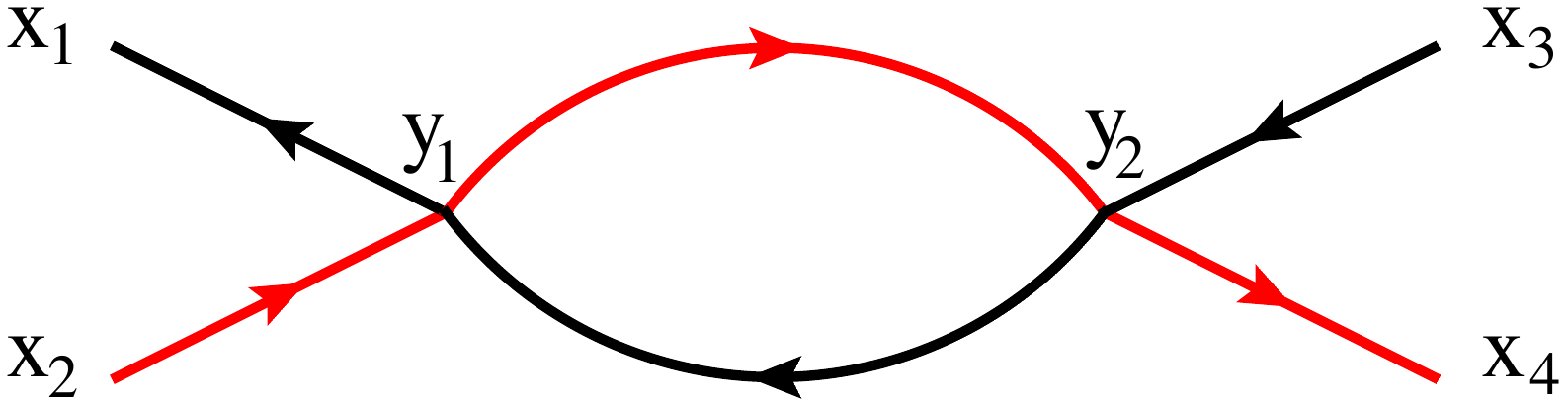}}
 \caption{First two contributions of the double- and single-trace vertices $\text{Tr}[\phi_1\phi_2^\dagger]\text{Tr}[\phi_1^\dagger\phi_2]$ and $\text{Tr}[\phi_1^\dagger\phi_2^\dagger\phi_1\phi_2]$ to the four-point functions $G_{\phi_1\phi_{2}^\dagger}$.}
  \label{fig:dtr2}
 \end{figure}
 
 Let's compute the first few orders of the weak coupling expansion of $G_{\phi_1\phi_{2}^\dagger}$. In terms of Feynman diagrams, we have to compute the first three graphs given in Fig.\ref{fig:pert2} and the ones in Fig.\ref{fig:dtr2}. 
Defining the function $G^{(a,b,c)}_{\phi_1\phi_{2}^\dagger}$ where $a$ counts the number of double- and single-trace vertices contributing with coupling $\tilde{\alpha}^2_2$ and  $b$ and $c$ the number of bosonic and fermionic vertices, respectively, we have
{\begin{equation}\label{expfixpoint2}
G_{\phi_1\phi_{2}^\dagger}=\!G^{(0,0,0)}_{\phi_1\phi_{2}^\dagger}\!\!+(4\pi)^2\tilde{\alpha}_2^2G^{(1,0,0)}_{\phi_1\phi_{2}^\dagger}+(4\pi)^4[\tilde{\alpha}_2^4G^{(2,0,0)}_{\phi_1\phi_{2}^\dagger}\!\!+\lambda^4G^{(0,2,0)}_{\phi_1\phi_{2}^\dagger}\!+\lambda^2\mu^2G^{(0,0,4)}_{\phi_1\phi_{2}^\dagger}]\!+\dots
 \end{equation}}
 where the new couplings $\lambda$ and $\mu$ are defined by \eqref{coupdef2} and $\tilde{\alpha}^2_2=\alpha^2_2+\mu^2$. 

The leading order $G^{(0,0,0)}_{\phi_1\phi_{2}^\dagger}$ is already defined in \eqref{GGpert12} then $G^{(0,0,0)}_{\phi_1\phi_{2}^\dagger}=G^{(0)}_{\phi_1\phi_{2}^\dagger}$. The first correction is given by the diagram in Fig.\ref{fig:dtr2}(a) that is a half\footnote{The two diagrams have a different symmetry factor given by the Wick contractions.} of the one computed in the previous section in \eqref{1loopalpha}.  
The bosonic part of the two-loop correction given by $G^{(2,0,0)}_{\phi_1\phi_{2}^\dagger}$ and $G^{(0,2,0)}_{\phi_1\phi_{2}^\dagger}$ comes from the diagrams in Fig.\ref{fig:dtr2}(b) and Fig.\ref{fig:pert2}(b) and they are divergent. Their integral representation is exactly the same as in the previous case, namely \eqref{GI}, and then we can write their sum as follows
\begin{equation}\label{nnn2}
\tilde{\alpha}_2^4G^{(2,0,0)}_{\phi_1\phi_{2}^\dagger}+\lambda^4G^{(0,2,0)}_{\phi_1\phi_{2}^\dagger}=
\frac{\pi^4c_B^6}{x_{12}^2x_{34}^2}\!\left(\frac{\tilde{\alpha}_2^4+\lambda^4}{\epsilon}\right)\!u\Phi^{(1)}(u,v)+\text{finite\,.}
\end{equation}

Let's finally consider the fermionic contribution $G^{(0,0,4)}_{\phi_1\phi_{2}^\dagger}$. This term corresponds to the Feynman diagram in Fig.\ref{fig:pert2}(c) and its integral representation in four dimensions is given in the next-to-the-last line of \eqref{GGpert12}. In this case, our goal is also to identify the UV divergent part of the diagram. Thus we will not compute the whole integral in \eqref{GGpert12} in dimensional regularization but we will rather proceed following the method of the previous section. 
Performing two integrations by means of the star-triangle relation \eqref{uniqferm} and then simplifying the spin structure with the help of \eqref{tracesigma} and \eqref{scalarproduct}, one can identify a divergent integral of the same kind as in \eqref{intI}, together with a finite remainder integral. Computing the integral $\mathcal{I}$ for small distances $y_{12}\rightarrow 0$, we can extract the following pole
\begin{equation}
G^{(0,0,4)}_{\phi_1\phi_{2}^\dagger}=
\frac{2\pi^8c_B^4c_F^4}{x_{12}^2x_{34}^2}\,\frac{1}{\epsilon}\,u\,\Phi^{(1)}(u,v)+\text{finite\,.}
\end{equation}
Combining all this results, we obtain that the expansion \eqref{expfixpoint2} takes the expected form \eqref{4pointgeneral} with the function 
$\mathcal{G}_{\phi_1\phi_2^\dagger}(u,v)$ given by
\begin{equation}\label{mathG12}
\mathcal{G}_{\phi_1\phi_2^\dagger}(u,v)=u+\tilde{\alpha}_2^2\,u\,\Phi^{(1)}(u,v)+\frac{\tilde{\alpha}_2^4+\lambda^4+2\lambda^2\mu^2}{\epsilon}\,u\,\Phi^{(1)}(u,v)+\text{finite}(\lambda^4,\mu^4)+\dots
 \end{equation}
where finite$(\lambda^4,\mu^4)$ stands for the finite part of $\mathcal{G}_{\phi_1\phi_{2}^\dagger}$ at two-loop.
Finally, imposing the UV finiteness of the correlation function and recalling the definition \eqref{alphatildedef}, we obtain the first order of the fixed point as follows
\begin{equation}
\alpha^2_{2\star}=-\mu^2\pm\, i\lambda \sqrt{\lambda^2+2\mu^2}+\dots\,.
\end{equation}
Replacing the effective coupling in \eqref{mathG12} with its value $\tilde{\alpha}^2_{2\star}=i\lambda\sqrt{\lambda^2+2\mu^2}$, \footnote{This choice is coherent with the sign convention used in Sec.\ref{sec:spectrumphi1phi2}.} we obtain the one-loop expansion of the correlation function in the following form
\begin{equation}
\mathcal{G}_{\phi_1\phi_2^\dagger}(u,v)=u+i\lambda\sqrt{\lambda^2+2\mu^2}\,u\,\Phi^{(1)}(u,v)+\dots\, ,
 \end{equation}
that perfectly matches the same quantity computed via OPE with conformal data fixed by the Bethe-Salpeter method \eqref{4pointfinal2}.

\section{Conclusion and discussion}

This paper represents an attempt of a deeper  understanding of  physical properties and   analytic structure  of the four-dimensional, three-coupling chiral CFT~-- the \(\chi\)CFT \footnote{ the name \(\chi\)CFT was suggested in~\cite{Caetano:2016ydc}} -- proposed by \"{O}.~G\"{u}rdo\u{g}an and one of the authors in~\cite{Gurdogan:2015csr} as a double scaling limit of \(\gamma\)-deformed \sym theory, combining the weak coupling with the strong imaginary \(\gamma\)-twist.  We study here two aspects of this \(\chi\)CFT with   three effective couplings, given by the Lagrangian~\eqref{fullL}: i)~the explicit description of the Feynman graph content of the perturbative expansion, partially uncovering their integrability properties; ii)~the exact computation, via conformal symmetry,  of two four-point correlation functions of shortest protected scalar operators of the theory.

As concerns the planar Feynman graphs content of the theory, we found here the complete description of possible Feynman graphs in the ``bulk" -- inside a generic big Feynman graph, far from its boundaries defined by a particular studied physical quantity. These graphs can be dubbed as ``dynamical fishnet", since, unlike the usual regular fishnet of the bi-scalar model~\eqref{bi-scalarL}  they have a certain  dynamics (summation over many of such graphs) preserving at the same time a kind of irregular fishnet structure shown on Figg.~\ref{fig:patched_fishnet},\ref{fig:bulk_DSlimit2}. Interestingly, this bulk structure is neatly realized as Feynman graphs describing arbitrary single-trace correlation functions of all elementary fields, as shown on  Figg.~\ref{fig:bulk_disc},\ref{fig:multA}. It would be very interesting to find the realisation of the Yangian symmetry of these correlators, and of the related planar amplitudes  (with disc topology), generalizing the results of~\cite{Chicherin:2017cns,Chicherin:2017frs} for the bi-scalar CFT. It would be the neatest demonstration of the integrability of the full model. In Sec.\ref{sec:chiCFT} we demonstrate such  integrability in the two-coupling reduction of the full \(\chi\)CFT, having a much simpler fishnet structure (combination of regular ``brick wall'' graphs with Yukawa vertices and regular square lattice fishnets). A considerably more involved  analysis of the integrability of the  full dynamical fishnet of \(\chi\)CFT, in particular, via the Yangian symmetry of single-trace correlators, is underway. We believe that it will be another important step to the understanding of integrability of the mother theory -- the \sym. It is worth noticing here that \(\gamma\)-deformation represents a rather mild, ``topological" modification of the planar graphs of original \sym, altering only the boundaries of these graphs, and not the bulk.

In the second  part of our paper, we managed to compute two non-trivial four point correlation functions of elementary fields of the full three coupling \(\chi\)CFT, generalizing the bi-scalar fishnet CFT results of~\cite{Grabner:2017pgm,Gromov:2018hut}. As in these papers, we employed the Bethe-Salpeter method and the conformal symmetry to do the computations, but the procedure is more complicated and the corresponding analytic structures, both in coordinate and in the coupling spaces,  are considerably richer,  due to a more ``dynamical" nature of the involved Feynman graphs. A new phenomenon presented in the correlators of the full theory is the non-perturbative behavior of certain individual OPE data  -- anomalous dimensions and structure constants of exchange operators, in the weak coupling limit. But the perturbative behavior of the four point correlator is restored in the sum over all OPE terms. The  equations on the anomalous dimensions, obtained from the pole structure of integrands in spectral decomposition of these correlators, appear to have a few interesting singularities in the coupling space, whose physical significance for the theory is left to understand. We also demonstrate the relevance of the double-trace terms for the correct Feynman graph interpretation of our results obtained via Bethe-Salpeter conformally symmetric procedure.

To get a further insight to these intereting chiral CFTs, we have to compute more complicated correlation functions, involving the exchanged operators of higher R-charges, such as \(\tr\phi_1^L\), or even more complicated multi-magnon operators. For the moment, only the exact anomalous dimensions of \(L=3\) case of such operators and of some related operators with the same R-charge have been computed for bi-scalar fishnet CFT in~\cite{Gromov:2017cja}\footnote{In the sense that the exact Baxter equation, together with its quantisation condition was obtained and studied perturbatively, to many loops, and numerically, to a veryally arbitrary precision } via the double-scaling limit of the QSC equations. Similar results on \(L=4,5\) and magnon operators will be reported in~\cite{GrabnerGromovKazakovKorchemsky}.  Not much is  yet done in this direction for the full \(\chi\)CFT, apart from the ABA approach of ~\cite{Caetano:2016ydc} to long operators \(L\gg 1\) and the  one-loop study of~\cite{Ipsen:2018fmu}, as well as the results of the current paper on the shortest exchange operators.  
As concerns the study of the structure constants, the first all-loop results for multi-magnon operators in bi-scalar fishnet CFT have been obtained in the very recent paper~\cite{Basso:2018cvy}. The generalization to four-point functions and to more complicated operators, and to the full \(\chi\)CFT, will necessitate a considerable new insight into integrability properties of these models. 

The generalization of bi-scalar fishnet CFT to any dimension  \(D\)~\cite{Kazakov:2018qbr} poses a natural question whether the \(D\)-dimensional generalization of the full 3-coupling \(\chi\)CFT exists. A related question: can we generalize the Basso-Dixon type fishnet integrals -- the four-point single-trace correlators of scalar fields in bi-scalar CFT -- explicitly computed in \(D=4\)~\cite{Basso:2017jwq} and \(D=2\)~\cite{Derkachov:2018rot}, to the case of dynamical fishnets of \(\chi\)CFT?   

It would be also interesting to understand the behavior of large Feynman graphs in the full \(\chi\)CFT, in-line with the early results of~\cite{Zamolodchikov:1980mb} and the recent observations of~\cite{Basso:2018agi} for the fishnet reduction of the \(\chi\)CFT. In particular, if the \(\sigma\)-model interpretation of the latter paper can be generalized to the full \(\chi\)CFT, it could be a big step in the explicit construction of the AdS dual of this chiral CFT, if such one exists at all after the double scaling limit of \(\gamma\)-deformed \sym\ theory. 

As a final comment:   it would be interesting to find a  realisation of these non-unitary theories in physical systems, if not in the fundamental quantum field theory (at least as an effective theory) than my be for certain statistical-mechanical and, presumably non-equilibrium, condensed matter models. The beautiful mathematical structures behind the \(\chi\)CFT promises  more of such applications in the future.

\section*{Acknowledgements}

We are thankful to J.~Caetano, S.~Derkachov, N.~Gromov, \"{O}.~G\"{u}rdo\u{g}an, G.~Korchemsky, O.~Veretin,  F.~Levkovich-Maslyuk for discussions. Our work  was supported by the European Research Council (Programme
``Ideas" ERC-2012-AdG 320769 AdS-CFT-solvable). The work of E.O. is supported by the German Science Foundation (DFG) under the Collaborative Research Center (SFB) 676 Particles, Strings and the Early Universe and the Research Training Group 1670. 

\section*{\Large Appendices}

\appendix

\section{Notation and conventions}\label{app:appendA}

In this paper, the metric tensor of the four dimensional Euclidean space is taken to be
\begin{equation}
g_{\mu\nu}=\delta_{\mu\nu}=\text{diag}(\;1,1,1,1\;)\,,
\end{equation}
where $\mu,\nu=0,1,2,3$ are spacetime vector indices. The massless scalar propagators are defined in configuration and momentum space as follows
\begin{equation}\label{FourierS}
\frac{1}{(x_{12}^2)^\alpha}=\frac{1}{4^\alpha\pi^{D/2}}\frac{\Gamma\left(2-\alpha\right)}{\Gamma\left(\alpha\right)}\int d^Dk\frac{e^{ik\cdot x_{12}}}{(k^2)^{D/2-\alpha}}\,,
\end{equation}
and the same for the fermionic propagators
\begin{equation}\label{FourierF}
\frac{\slashed{x}_{12}}{(x_{12}^2)^{\alpha+1/2}}=\frac{-i}{4^\alpha\pi^{D/2}}\frac{\Gamma\left(\tfrac{5}{2}-\alpha\right)}{\Gamma\left(\tfrac{1}{2}+\alpha\right)}\int d^Dk\frac{e^{ik\cdot x_{12}}\slashed{k}}{(k^2)^{D/2-\alpha+1/2}}\,,
\end{equation}
where $\slashed{x}$ stands for the position $x$ contracted with the spin structure matrix, and the same for $\slashed{k}$.
The positions satisfies the following identity
\begin{equation}\label{scalarproduct}
x_{ij}\cdot x_{kl}=\frac12(x_{il}^2+x_{jk}^2-x_{ik}^2-x_{jl}^2)\,.
\end{equation}

We can represent the four-dimensional gamma-matrices to have the off-block
diagonal form
\begin{equation}
\gamma^\mu=\begin{pmatrix} 
0 & (\sigma^\mu)_{\alpha\dot{\beta}} \\
(\bar{\sigma}^\mu)^{\dot{\alpha}\beta} & 0 
\end{pmatrix}\,,
\end{equation}
by introducing the $2\times 2$ Euclidean $\sigma$ matrices
\begin{equation}
\sigma^\mu=(-i\vec{\boldsymbol{\sigma}},\mathbb{I}_{2\times 2})\qquad\text{and}\qquad
\bar{\sigma}^\mu=(i\vec{\boldsymbol{\sigma}},\mathbb{I}_{2\times 2})\,,
\end{equation}
where $\vec{\boldsymbol{\sigma}}$ are the \textit{Pauli matrices}. We use the standard convention for raising/lowering of two-component spinor indices \(\alpha\), \(\dot \alpha\)
\begin{align}
&\psi_\alpha = \epsilon_{\alpha \beta} \psi^{\beta},\quad\psi^\alpha = \epsilon^{\alpha \beta} \psi_{\beta},\quad\bar \psi_{\dot \alpha} = \epsilon_{\dot\alpha \dot \beta} \bar \psi^{\dot \beta},\quad \bar \psi^{\dot \alpha} = \epsilon^{\dot\alpha \dot \beta} \bar \psi_{\dot \beta}\, ,
\end{align}
where we introduced the tensors \(\epsilon\) as
\begin{align}
 &\epsilon_{\alpha \beta}=\epsilon_{\dot \alpha \dot \beta} = i \sigma^2\, ,\qquad \epsilon^{\alpha \beta}=\epsilon^{\dot \alpha \dot \beta} = -i \sigma^2\, ,
 \end{align}
and the following relations hold 
\begin{align}
& (\bar \sigma^{\mu})^{\dot \alpha \alpha}= \epsilon^{\dot\alpha \dot \beta} \epsilon^{\alpha \beta} \sigma^{\mu}_{\dot \beta \beta},\qquad \epsilon^{\dot\alpha \dot \beta}\epsilon_{\dot \beta \dot \gamma}=\delta^{\dot \alpha}_{\dot \gamma}\qquad \epsilon^{\alpha \beta}\epsilon_{\beta \gamma}=\delta^{ \alpha}_{\gamma}\, .
\end{align}
The $\sigma$ matrices satisfy
\begin{equation}\label{symferm}
\bar{\sigma}_\mu\sigma_\nu +\bar{\sigma}_\nu \sigma_\mu=2\delta_{\mu\nu} \mathbb{I}_{2\times 2} \qquad\text{and}\qquad
\sigma_\mu\bar{\sigma}_\nu +\sigma_\nu \bar{\sigma}_\mu=2\delta_{\mu\nu} \mathbb{I}_{2\times 2} \,,
\end{equation}
and the trace identities are
\begin{equation}\begin{split}\label{tracesigma}
&\text{tr}(\text{\textit{odd number of $\sigma$'s}})= 0,\\
&\text{tr}(\sigma^\mu\bar{\sigma}^\nu)=\text{tr}(\bar{\sigma}^\mu\sigma^\nu)=2\delta^{\mu\nu},\\
&\tr\left(\sigma_{\mu}\overline{\sigma}_{\rho}\sigma_{\eta}\overline{\sigma}_{\nu}\right)=2\left(\delta_{\mu\rho}\delta_{\eta\nu}-\delta_{\mu\eta}\delta_{\rho\nu}+\delta_{\mu\nu}\delta_{\rho\eta}-\epsilon_{\mu\rho\eta\eta}\right),\\
&\tr\left(\overline{\sigma}_{\mu}{\sigma}_{\rho}\overline{\sigma}_{\eta}{\sigma}_{\nu}\right)=2\left(\delta_{\mu\rho}\delta_{\eta\nu}-\delta_{\mu\eta}\delta_{\rho\nu}+\delta_{\mu\nu}\delta_{\rho\eta}+\epsilon_{\mu\rho\eta\eta}\right)\,.
\end{split}\end{equation}

\section{The $\gamma$-deformed $\mathcal{N}=4$ SYM theory}
\label{app:gdef}
The Lagrangian of \(\gamma\)-deformed \({\cal\ N}=4\) 
SYM reads  (see e.g.\cite{Fokken:2013aea})
\begin{equation}
\label{N=4SYMlagrangian}
  {\cal L}=N_c\Tr\biggl[
  -\frac{1}{4} F_{\mu\nu}F^{\mu\nu}
  -\frac{1}{2}D^\mu\phi^\dagger_iD_\mu\phi^i
  +i\bar \psi_{\dot\alpha \;  A}D^{\dot\alpha\alpha}\psi^A_{\alpha }\biggr]
+{\cal L}_{\rm int}\,,
\end{equation}
where \(i=1,2,3\)\,\, \(A =1,2,3,4\),
\(D^{\dot \alpha \alpha}= D_\mu
(\bar \sigma^{\mu})^{\dot \alpha \alpha}\) and 
 \begin{equation}
   \begin{aligned}[y]
     &\mathcal{L}_{\rm int} =N_cg\,\,\Tr\bigl[\frac{g}{4} \{\phi^\dagger_i,\phi^i\}
     \{\phi^\dagger_j,\phi^j\}-g\,e^{-i\epsilon^{ijk}\gamma_k}
     \phi^\dagger_i\phi^\dagger_j\phi^i\phi^j\\
     &-e^{-\frac{i}{2}\gamma^-_{j}}\bar\psi^{}_{ j}\phi^j\bar\psi_{ 4}
     +e^{+\frac{i}{2}\gamma^-_{j}}\bar\psi^{}_{ 4}\phi^j\bar\psi_{ j}
     + i\epsilon_{ijk} e^{\frac{i}{2} \epsilon_{jkm} \gamma^+_m} \psi^k \phi^i \psi^{ j}\\
     &-e^{+\frac{i}{2}\gamma^-_{j}}\psi^{}_{ 4}\phi^\dagger_j\psi_{
       j}
     +e^{-\frac{i}{2}\gamma^-_{j}}\psi^{}_{j}\phi^\dagger_j\psi_{
       4}
     + i\epsilon^{ijk} e^{\frac{i}{2} \epsilon_{jkm} \gamma^+_m} \bar\psi_{ k} \phi^\dagger_i \bar\psi_j\bigr]\hfill\, .
   \end{aligned}
 \end{equation}
where the summation is assumed w.r.t. doubly and triply repeating indices. We suppress the  Lorentz indices of fermions, assuming the contractions
  \(\psi_i^\alpha \psi_{j,\alpha}\) 
 and  \(\bar\psi_{i,\dot\alpha} \bar\psi_j^{\dot\alpha}\). We also use the notations
 \begin{equation}\label{gammapm}
\gamma_1^{\pm}=-\frac{\gamma_3\pm\gamma_2}{2},\quad\!\!
\gamma_2^{\pm}=-\frac{\gamma_1\pm\gamma_3}{2},\quad\!\!
\gamma_3^{\pm}=-\frac{\gamma_2\pm\gamma_1}{2}\, .
\end{equation}
The parameters of the $\gamma$-deformation 
{$q_j=e^{-\frac{i}{2}\gamma_{j}}$ $j=1,2,3$} are related to
the Cartan subalgebra  \(\mathfrak{u}(1)^3\subset \mathfrak{su}(4) \cong \mathfrak{so}(6)\) and the related field charges are represented in Tab.\ref{charges_table}.

\begin{table}[!t]
\begin{center}
\begin{tabular}{ | m{2em} | m{1cm}| m{1cm} | m{1cm}| m{1cm}  | m{1cm}| m{1cm} | } 
\hline
 & \(\phi_1\) & \(\phi_2\) & \(\phi_3\) & \(\psi_1\) & \(\psi_2\) & \(\psi_3\) \\ 
\hline
\(Q_1\) & +1 & 0 & 0 & \(+\frac{1}{2}\) & \(-\frac{1}{2}\) & \(-\frac{1}{2}\) \\ 
\hline
\(Q_2\) & 0 & +1 & 0 & \(-\frac{1}{2}\) & \(+\frac{1}{2}\) & \(-\frac{1}{2}\) \\
 \hline
\(Q_3\) & 0 & 0 & +1 & \(-\frac{1}{2}\) & \(-\frac{1}{2}\) & \(+\frac{1}{2}\)\\ 
\hline
\end{tabular}
\caption{Charges of scalar and fermionic fields under the Cartan charges of R-symmetry \(SU(4)\). These charges generate the symmetry group \(U(1)\otimes U(1)\otimes U(1)\)~left over after the breaking  of \(R\)-symmetry by twisting.}
\label{charges_table}
\end{center}
\end{table}

\section{The uniqueness relations}\label{sec:appendixC}
We present here some useful formulas for integrals of the star-type, namely three propagators of various types linked in one integration point. Under the condition of conformal invariance, the  integrals considered below can be reduced to the computation of simple convolutions.\\
If the three propagator are scalar, the well-known formula (star-triangle relation)~\cite{DEramo:1971hnd, Vasiliev:1981yc} reads
\begin{equation}\label{unique}
\int\!\!\frac{d^4 y\;}{(x_1-y)^{2\alpha}(x_3-y)^{2\beta}(x_2-y)^{2\gamma}}=\pi^2 \frac{\Gamma(\alpha')\Gamma(\beta')\Gamma(\gamma ')}{\Gamma(\alpha)\Gamma(\beta)\Gamma(\gamma)}\frac{1}{(x_{13}^2)^{\gamma '}(x_{32}^2)^{\alpha '}(x_{21}^2)^{\beta'}}\, ,
\end{equation}
where \(a'=2-a\), and the scale-invariance condition \(\alpha+\beta+\gamma =4\) should be fulfilled. Another well-known identity is the star-triangle relation for Yukawa-like vertices, involving two propagators of spin \(\frac{1}{2}\):
\begin{equation}\label{uniqferm}
\int\!\!\frac{d^4 y\;(x_1-y)^\mu(\bar{\sigma}_\mu\sigma_\nu)^{\dot{\alpha}\alpha}(y\!-\!x_2)^\nu}{((x_1\!-\!y)^2)^\alpha((x_3\!-\!y)^2)^\beta((x_2\!-\!y)^2)^\gamma}=\pi^2\frac{\Gamma(1\!+\!\alpha ')\Gamma(\beta ')\Gamma(1\!+\!\gamma ')}{\Gamma(\alpha)\Gamma(\beta)\Gamma(\gamma)}\frac{x_{13}^\mu(\bar{\sigma}_\mu\sigma_\nu)^{\dot{\alpha}\alpha} x_{32}^\nu}{(x_{13}^2)^{1+\!\gamma '}(x_{32}^2)^{1+\!\alpha '}(x_{21}^2)^{\beta '}}\, ,
\end{equation}
under the condition \(\alpha+\beta+\gamma=5\), and being.\\
Such formulas are actually particular reductions of a more general one derived in \cite{Chicherin:2012yn}. This generalization involves two propagators in the representation of traceless symmetric tensors of integer rank \(S\), namely
\begin{align}
G_S(x-y)=\frac{[n_\mu(x-y)^{\mu}]^S}{(x-y)^{2\alpha}}\qquad \text{with}\quad n_{\mu} \,n^{\mu} =0\, .
\end{align}
Remarkably, the null vector \(n_{\mu}\) in \(4\) dimensions can be realized in terms of Pauli matrices and two 2-spinors as \(n^{\mu}=\lambda \,\sigma^{\mu} \tilde \lambda\) or \(n^{\mu}=\tilde\eta\, \overline{\sigma}^{\mu}  \eta\). 
In such a context two \(S\)-tensor propagators can be contracted to form a tensorial loop as follows:
\begin{align}
G_{\alpha,S}(x_1-x_2)*G_{\gamma,S}(x_2-x_3)=\frac{(x_{12}^{\mu} A_{\mu\nu}\,x_{23}^{\nu})^S}{(x_{12})^{2\alpha}(x_{23})^{2\gamma}}\qquad \text{where} \quad A_{\mu\nu}\equiv(\lambda \sigma_{\mu} \bar \sigma_{\nu}   \eta)\,.
\end{align}
Under the condition \(\alpha+\beta+\gamma=4+S\) we can write
\begin{equation}\label{uniqfermS}
\int\!\!\frac{d^4 y\;((x_1\!-\!y)^\mu A_{\mu\nu}(y\!-\!x_2)^\nu)^S}{[(x_1\!-\!y)^2]^\alpha[(x_3\!-\!y)^2]^\beta [(x_2\!-\!y)^2]^\gamma}=\pi^2 \frac{\Gamma(S\!+\!\alpha ')\Gamma(\beta ')\Gamma(S\!+\!\gamma ')}{\Gamma(\alpha)\Gamma(\beta)\Gamma(\gamma)}\frac{(x_{13}^\mu A_{\mu\nu} x_{32}^\nu)^S}{(x_{13}^2)^{S+\!\gamma '}(x_{32}^2)^{S+\!\alpha '}(x_{21}^2)^{\beta '}}\, ,
\end{equation}
which for \(S=0\) reduces to \eqref{unique} and for \(S=1\) is equivalent to \eqref{uniqferm}. In the same formalism we can encode the tensor structure of \eqref{eigenfunction}, that is \(\left[ n_{\mu} \left(\frac{x_{10}^{\mu}}{x_{10}^2}-\frac{x_{20}^{\mu}}{x_{20}^2}\right)\right]^S\) in the expression 
\begin{align}
\left[\frac{(\lambda \sigma_{\mu} \bar \sigma_{\nu}\sigma_{\rho} \tilde \lambda) x_{10}^{\mu}x_{12}^{\nu}x_{20}^{\rho}}{x_{10}^2 x_{20}^2}\right]^S\,,
\end{align} 
allowing the application of formula \eqref{uniqfermS} to the direct computation of \eqref{hbS}.

\section{Eigenvalue of the fermionic graph-building operator \(\hat{\mathcal{H}}_t\)}\label{appendix:ht}

In this section, we compute the relevant part of the fermionic eigenvalue $\hat{\mathcal{H}}_F^{(n)}$ defined in \eqref{hfop} which is involved in the computation of the four-point function $G_{\phi_1\phi_{2}^\dagger}$ \eqref{phi1phi2dagger}. The eigenfunction of the fermionic operator is determined, as usual, by the conformal representation of external fields, thus it corresponds to \eqref{Phi-def}. As shown in Sec.\ref{sec:eigen2}, the action of the sub-kernels $\hat{\mathcal{H}}_t$, $\hat{\mathcal{H}}_r$ and $\hat{\mathcal{H}}_b$ transforms the conformal triangle to a new one, adding a factor \(\slashed{x}/x^2\), with chirality depending on the number $n$ of sub-kernels we apply. Moreover, since the spectrum of $\hat{\mathcal{H}}_b$ is the same as of the bosonic operator \eqref{Hb} and the spectrum of  $\hat{\mathcal{H}}_r$ is expressed through the one for   $\hat{\mathcal{H}}_t$ \eqref{Hr}, the latter is the only operator we have to study.

The goal of this Appendix is to compute the eigenvalue ${h_t}{}_{\Delta,S}$ defined by the first  equation of \eqref{subeigen} at any $S$. 
First we send \(x_0 \to \infty \) in order to simplify the spin structures in both sides of the equation. In this limit the state \eqref{Phi-def} is $\Phi_{\Delta,S,x_0}(x_1,x_2)\rightarrow (n\cdot x_{12})^S/(x_{12}^2)^{1-p}$ where $p=(\Delta-S)/2$. Using the integral representation given in \eqref{intht} and contracting both sides of the equation with $(\sigma_\eta)_{\alpha\dot{\alpha}}x_{12}^\eta$ we obtain
\begin{equation}\label{eqhtS}
{h_t}{}_{\Delta,S}=-\frac{c_F^3}{2}\!\!\int\!\! d^4 y_1 d^4 y_2\frac{\text{tr}(\sigma_\eta\bar{\sigma}_\mu\sigma_{\nu}\bar{\sigma}_\rho)x_{12}^\eta(x_2-y_2)^\mu y_{12}^\nu (y_1-x_1)^\rho}{(x_2-y_2)^4(y_{12}^2)^{3-p}(y_1-x_1)^4}(n\cdot y_{12})^S\,,
\end{equation}
where we used \eqref{symferm} to compute the trace of two $\sigma$'s and we set $x_{12}^2=(n\cdot x_{12})=1$ for simplicity. The dependence on $y_{12}$ in \eqref{eqhtS} can be written as follows
\begin{equation}\label{propspin}
\frac{y_{12}^\nu (n \cdot y_{12})^S }{(y_{12}^2)^{3-p}} =\frac{1}{2(p-2)} \left[\partial_{y_1}^\nu \frac{(n \cdot y_{12})^S }{(y_{12}^2)^{2-p}}-S \frac{ n^\nu (n \cdot y_{12})^{S-1}}{(y_{12}^2)^{2-p}} \right]. 
\end{equation}
Using the Fourier transform \eqref{FourierS}, one can rewrite the combinations of propagators appearing in \eqref{propspin} in the following way
\begin{equation}\begin{split}\label{propspin2}
\frac{(n \cdot y_{12})^S }{(y_{12}^2)^{2-\tfrac{\Delta-S}{2}}}&=\frac{(-1)^S}{4^{2-\tfrac{\Delta}{2}}\pi^2}\;\frac{\Gamma\left(\tfrac{\Delta+S}{2}\right)}{\Gamma\left(2-\tfrac{\Delta-S}{2}\right)}
\;(n\cdot \partial_{y_1})^S\int d^4k\frac{e^{ik\cdot y_{12}}}{(k^2)^{\tfrac{\Delta+S}{2}}}\,,\\
\frac{(n \cdot y_{12})^{S-1} }{(y_{12}^2)^{2-\tfrac{\Delta-S}{2}}}&=\frac{(-1)^{S-1}}{4^{\tfrac{5-\Delta}{2}}\pi^2}\;\frac{\Gamma\left(\tfrac{\Delta+S}{2}-1\right)}{\Gamma\left(2-\tfrac{\Delta-S}{2}\right)}
\;(n\cdot \partial_{y_1})^{S-1}\int d^4k\frac{e^{ik\cdot y_{12}}}{(k^2)^{\tfrac{\Delta+S}{2}-1}}\,.
\end{split}\end{equation}
Substituting \eqref{propspin} together with \eqref{propspin2} into \eqref{eqhtS} and transforming the integrals in momentum space by means of \eqref{FourierF}, one can compute the  trivial integrations obtaining
\begin{equation}\begin{split}\label{eqhtS2}
{h_t}{}_{\Delta,S}=c_F^3\frac{(-i)^{S+1}\pi^2}{4^{2-\tfrac{\Delta}{2}}}\left[\frac{\Gamma\left(\tfrac{\Delta+S}{2}\right)}{\Gamma\left(3-\tfrac{\Delta-S}{2}\right)}
\int d^4k e^{ik\cdot x_{21}}\frac{\text{tr}(\sigma_\eta\bar{\sigma}_\mu\sigma_{\nu}\bar{\sigma}_\rho)x_{12}^\eta k^\mu k^\nu k^\rho}{(k^2)^{2+\tfrac{\Delta+S}{2}}}(n\cdot k)^S\right.\\
\left.-\frac{S}{2}\frac{\Gamma\left(\tfrac{\Delta+S}{2}-1\right)}{\Gamma\left(3-\tfrac{\Delta-S}{2}\right)}
\int d^4k e^{ik\cdot x_{21}}\frac{\text{tr}(\sigma_\eta\bar{\sigma}_\mu\sigma_{\nu}\bar{\sigma}_\rho)x_{12}^\eta k^\mu n^\nu k^\rho}{(k^2)^{1+\tfrac{\Delta+S}{2}}}(n\cdot k)^{S-1}\right]\,.
\end{split}\end{equation}
Let's focus on the integral in the first line of \eqref{eqhtS2}. Using the trace of four $\sigma$-matrices given in \eqref{tracesigma} and transforming the integral back to position space by means of \eqref{FourierS}, we have
\begin{equation}\begin{split}\label{intF1}
\int d^4k e^{ik\cdot x_{21}}&\frac{\text{tr}(\sigma_\eta\bar{\sigma}_\mu\sigma_{\nu}\bar{\sigma}_\rho)x_{12}^\eta k^\mu k^\nu k^\rho}{(k^2)^{2+\tfrac{\Delta+S}{2}}}(n\cdot k)^S\\
&=i^{-S-1}4^{\tfrac{3}{2}-\tfrac{\Delta+S}{2}}\pi^2\;\frac{\Gamma\left(1-\tfrac{\Delta+S}{2}\right)}{\Gamma\left(1+\tfrac{\Delta+S}{2}\right)}\;
(x_{12}\cdot \partial_{y_2})(n\cdot \partial_{y_2})^S\frac{1}{(x_{12}^2)^{1-\tfrac{\Delta+S}{2}}}\,.
\end{split}\end{equation}
Repeating the same procedure for the second line of \eqref{eqhtS2} we obtain
\begin{equation}\begin{split}\label{intF2}
&\int d^4k e^{ik\cdot x_{21}}\frac{\text{tr}(\sigma_\eta\bar{\sigma}_\mu\sigma_{\nu}\bar{\sigma}_\rho)x_{12}^\eta k^\mu n^\nu k^\rho}{(k^2)^{1+\tfrac{\Delta+S}{2}}}(n\cdot k)^{S-1}=\frac{4^{2-\tfrac{\Delta+S}{2}}\pi^2}{i^{S+1}}\left[\frac{\Gamma\left(1-\tfrac{\Delta+S}{2}\right)}{\Gamma\left(1+\tfrac{\Delta+S}{2}\right)}\right.\\
&\quad\times\left.(x_{12}\cdot \partial_{y_2})(n\cdot \partial_{y_2})^S\frac{1}{(x_{12}^2)^{1-\tfrac{\Delta+S}{2}}}+
2\frac{\Gamma\left(2-\tfrac{\Delta+S}{2}\right)}{\Gamma\left(\tfrac{\Delta+S}{2}\right)}\,
(n\cdot \partial_{y_2})^{S-1}\frac{n\cdot x_{12}}{(x_{12}^2)^{2-\tfrac{\Delta+S}{2}}}\right].
\end{split}\end{equation}
Finally, plugging the integrals \eqref{intF1} and \eqref{intF2} into \eqref{eqhtS2} and performing the derivatives we arrive at the following expression for the eigenvalue
\begin{equation}\label{htopS}
{h_t}{}_{\Delta,S}=4\pi^4c_F^3\frac{(\Delta-S)^2+S(S+2)}{(\Delta+S)(\Delta-S-4)(\Delta-S-2)(\Delta+S-2)}\,,
\end{equation}
where we  set $x_{12}^2=(n\cdot x_{12})=1$. Notice that setting $S=0$ we find \eqref{Ht}, as expected. 

\section{Cancellation of the spurious poles}\label{appendix:sp}

In order to confirm the validity of equation \eqref{OPE} we should show that the \emph{physical poles} given by the zeroes of the spectral equation \eqref{spectraleq} are the only contributions to the four-point correlators under study. This fact, well known for the bi-scalar reduction of our theory (see Appendix B in~\cite{Gromov:2018hut}), needs a proof for the full \(\chi\)CFT. It appears that additional possible contributions could come from the extra poles in $g_{\Delta,S}(u,v)$ and the measure factor $1/c_2(\Delta,S)$. In this appendix we will show that these contributions cancel each other thanks to a symmetry relation fulfilled by the eigenvalues $h_{i\,\Delta,S}$ of the Bethe-Salpeter kernels.

The conformal block $ g_{\Delta,S}(u,v)$   has simple poles at $\Delta_{S-n}=S+3-n$ (with $n=1,2,\dots,S$), namely
$2i\nu_n=S+1-n$.  Its residue at the  pole $\nu=\nu_n$ is given by
$r_n\, g_{S+3,S-n}(u,v)$ where (see for example Appendix B in \cite{Simmons-Duffin:2017nub}):
\begin{align}\label{rn}
r_n=(-1)^n \frac{i n \Gamma^2 \left(\frac{1}{2} \left(n-\Delta _{1}+\Delta_2+1\right)\right)
}{2 \Gamma
        (n+1)^2 \Gamma^2 \left(\frac{1}{2} \left(-n-\Delta _{1}+\Delta_2+1\right)\right)
}\,.
\end{align}
This results in the following extra contribution to \eqref{OPE}:
\begin{equation}
R^{g}_{S,m}=\left(\frac{r_m}{c_2(\Delta_{S-m},S)}\frac{h_{b\,\Delta_{S-m},S}}{1-\chi_b h_{b\,\Delta_{S-m},S}-\chi_f h_{f\,\Delta_{S-m},S}}\right)g_{S+3,S-m}(u,v)\;\;,\;\;1\leq n\leq S < \infty\;.
\end{equation}
In addition to that, the measure factor \(1/c_2(
\Delta,S)\) develops poles at $\Delta=S + 3 + k,\;k=0,1,2,\dots$. The corresponding contribution can be expressed as
\begin{equation}
R^{c_2}_{S,k}=-\left(\frac{r_k}{c_2(\Delta_{S},S+k)}
\frac{h_{b\,\Delta_{S+k},S}}{1-\chi_b h_{b\,\Delta_{S+k},S}-\chi_f h_{f\,\Delta_{S+k},S}}\right)g_{S+3+k,S}(u,v)
\;\;,\;\;0\leq S,k <\infty\;.
\end{equation}
The overall contribution of these terms  is the sum over all non-negative integers \(S,k\) of the generic term 
\begin{eqnarray}
R_{S,k}^{c_2}+R_{S+k,k}^g=&&
-\left(\frac{r_k}{c_2(\Delta_{S},S+k)}\,g_{S+3+k,S}(u,v)\right) \times\\&&\times
\left[\frac{h_{b\,\Delta_{S+k},S}}{1-\chi_b h_{b\,\Delta_{S+k},S}-\chi_f h_{f\,\Delta_{S+k},S}}-\frac{h_{b\,\Delta_{S},S+k}}{1-\chi_b h_{b\,\Delta_{S},S+k}-\chi_f h_{f\,\Delta_{S},S+k}}\right]\;.\notag
\end{eqnarray}
A possible vanishing condition for the full contribution is then
\begin{equation}
\label{vanpol}
r_k [h_{b\,\Delta_{S+k},S}(1-\chi_b h_{b\,\Delta_{S+k},S}-\chi_f h_{f\,\Delta_{S+k},S})-h_{b\,\Delta_{S+k},S}(1-\chi_b h_{b\,\Delta_{S+k},S}-\chi_f h_{f\,\Delta_{S+k},S})]=0
\end{equation}
for any \(k\,\in \,\mathbb{N}\).
We can actually verify  in both sectors under study that the following set of stronger conditions is fulfilled
\begin{eqnarray}
\label{vanpolb}
r_k (h_{b\,{3+S+k},S}-h_{b\,{3+S},S+k})&&=0 \;\;,\;\;k=0,1,2,\dots\;\\
\label{vanpolf}
r_k (h_{f\,{3+S+k},S}-h_{f\,{3+S},S+k})&&=0 \;\;,\;\;k=0,1,2,\dots\;.
\end{eqnarray}
It is easy to check that plugging \eqref{vanpolb} into \eqref{vanpol}, one is left with the condition \eqref{vanpolf}, which means that \eqref{vanpolb} together with \eqref{vanpolf} are a sufficient condition for \eqref{vanpol}.
To prove these equations hold, we notice first of all that at \(\Delta_1=\Delta_2=1\) \eqref{rn} vanishes at odd \(n\), so it would be sufficient to prove \eqref{vanpolb}, \eqref{vanpolf} at even \(k \,\in 2\mathbb{N}\). Moreover, equation \eqref{vanpolb} has been checked in \cite{Gromov:2018hut}, where it was enough to state the cancelation of spurious poles in \(\Tr[\phi_1^2]\) sector. Let us verify the second condition \eqref{vanpolf} at even integer \(k\). Starting from the sector \(\Tr[\phi_1^2]\) it is equivalent to
\begin{equation}
\label{vanf1}
\tilde h_{f\,{3+S+k},S}-\tilde h_{f\,{3+S},S+k}=0\,,
\end{equation}
where we recalled the definition of \(\tilde h_{f\,\Delta,S}\) \eqref{hffinale}.  Equation \eqref{vanf1} actually coincides with the vanishing condition for spurious contribution in the ``one-magnon" \(\Tr[\phi_1^2\phi_2]\) sector of bi-scalar theory, and is verified in \cite{Gromov:2018hut}. We can finally check \eqref{vanpolf} for the sector \(\Tr[\phi_1\phi_{2}^{\dagger}]\). Recalling that in this case
\begin{equation}
{h_F}_{\Delta,S}=-\frac{c_F^4[(\Delta-2)^2+S(S+2)]{{h_B}_{\Delta,S}}^2}{2c_B^6-8c_B^3c_F^2\pi^2[(\Delta-2)^2+S(S+2)]\xi_{j}\xi_{j_+}{h_B}_{\Delta,S}}\,,
\end{equation}
and that \({h_B}_{3+S+k,S}={h_B}_{3+S,S+k}\),  as we already know from \cite{Gromov:2018hut}, we are left to verify that\begin{equation}
[(\Delta-2)^2+S(S+2)]_{\Delta=3+s+k,\,S=s}=[(\Delta-2)^2+S(S+2)]_{\Delta=3+s,\,S=s+k}\, ,
\end{equation}
which is trivially true for any integer \(k\).

\section{Operator mixing and logarithmic multiplet}
In both sectors \(\Tr[\phi_1^2]\) and \(\Tr[\phi_1\phi_2^{\dagger}]\) of our theory the exchanged physical operators in the OPE s-channel of the 4-point correlators under analysis present mixing. Namely, due to the wide matter content of the theory, the renormalized operators are not just rescaled and normal-ordered monomials of elementary fields and derivatives, but linear coupling-dependant combination of several such terms which share the same symmetries. Concretely, in our theory we deal with single trace primary operators as
\begin{align}
\mathcal{O}_1(x)=\tr[\chi_{i_1}\chi_{i_2}\cdots \chi_{i_L}](x)\,,
\end{align}
made up of elementary fields of the theory \(\chi_{i_k}(x)\) eventually dressed by tensor structures and derivatives. Given the quantum numbers of such a term \(\mathcal{O}_1\), that is Cartan's \(U(1)^{\otimes 3}\) charge, twist and tensor rank \(S\), it is usually possible to write a few other conformal primaries with the same numbers, say \(\{\mathcal{O}_2,\, \mathcal{O}_3\,\dots\}\). This allows in general some of the two-point functions \(\langle \mathcal{O}_i(x) \mathcal{O}_j(0)^{\dagger}\rangle\) to not vanish at \(i\neq j\), that is to have transitions \(\mathcal{O}_i \rightarrow \mathcal{O}_j\). We define the anomalous dimension matrix \(\gamma_{ij}\) as 
\begin{align}
\label{mixmat}
 -\mu \frac{d}{d\mu}\, Z_{{O}_i} \,=\, \gamma_{ij}\, Z_{\mathcal{O}_j}\, ,
\end{align}
being \(Z_{O_i}\) the renormalization of operator \(\mathcal{O}_i\) and \(\mu\) the scale.
In absence of transitions, namely \(\gamma_{ij} = \delta_{ij}\gamma_i\), mixing does not happen and each operator \(\mathcal{O}_i\) has anomalous dimension \(\gamma_i\). Otherwise, one has to bring the \emph{mixing matrix} \(\gamma_{ij}\) into diagonal form via a rotation over the basis of local primaries \(\{\mathcal{O}_1,\, \mathcal{O}_2\,\dots\}\). The operators of the new basis are linear combinations of the kind
\begin{align}
\mathcal{O}'_i(x)\,=\,c_{1,i}(\xi)\,\mathcal{O}_1 +c_{2,i}(\xi)\,\mathcal{O}_2+\dots
\end{align}
and they do not mix among each other. The anomalous dimension of \(\mathcal{O}'_i(x)\) is the corresponding eigenvalue of the matrix, namely \(\gamma'_i\). The existence of a basis of eigenvectors for \(\gamma_{ij}\)-matrix is ensured by its hermiticity in unitary theories. The absence of invariance under hermitian conjugation of \eqref{fullL} prevent to come to similar conclusions for \(\chi\)CFT theory. In particular, performing the planar limit can lead to ``one-way" transitions
\begin{align}
\langle \mathcal{O}_i(x) \mathcal{O}_j(0)^{\dagger}\rangle \neq 0 \qquad
\langle \mathcal{O}_i(x) \mathcal{O}_j(0)^{\dagger}\rangle =0\,,
\end{align}
and the correspondent mixing matrix can be only brought into Jordan canonical form, e.g. for the mixing of four primaries:
\begin{align}
\label{exmix}
\gamma_{ij}\,\longrightarrow\, (S\gamma S^{-1})_{ij}\,=\,
\begin{bmatrix}
0 && 1 && 0 && 0\\
0 && 0 && 0 && 0\\
0 && 0 && \gamma'_3 && 0\\
0 && 0 && 0 && \gamma'_4
\end{bmatrix}\,.
\end{align} 
The matrix \eqref{exmix} contains a \(2\times 2\) Jordan block, together with two diagonal terms \(\gamma'_3\) and \(\gamma'_4\), corresponding to two renormalized operators with such anomalous dimensions. 
The physical interpretation of Jordan blocks leads to the formulation of logarithmic CFT (see ~\cite{Gurarie:1993xq},\cite{Hogervorst:2016itc}). In the example \eqref{exmix} the block corresponds to a  rank-\(2\) logarithmic multiplet.
This means that the corresponding operators  of the new basis, \(\mathcal{O}'_1,\mathcal{O}'_2\) show 2-point functions of the kind
\begin{align}
\langle \mathcal{O}'_1(x)\mathcal{O}_1^{'\dagger}(0)\rangle &=\frac{k\, \ln (\mu^2 x^2)}{(x)^{2\Delta_0}}\qquad &\langle \mathcal{O}'_1(x)\mathcal{O}_2{'^\dagger}(0)\rangle =\frac{k\,}{(x)^{2\Delta_0}}\\
\langle \mathcal{O}'_2(x)\mathcal{O}_1^{'\dagger}(0)\rangle &=\frac{k}{(x)^{2\Delta_0}}\qquad &\langle \mathcal{O}'_2(x)\mathcal{O}_2^{'\dagger}(0)\rangle=0\,,
\end{align}
where \(\Delta_0\) is the bare dimension of \(\mathcal{O}_i\) operators, and \(\mu\) the energy scale. This phenomenon, the presence of log-multiplets in \(\chi\)CFT has first be noticed by J.Caetano \cite{Caetano:TBP} for its bi-scalar reduction \eqref{bi-scalarL} and some examples of its occurance in the context of Fishnet CFT have been presented in \cite{Gromov:2017cja}. Despite such logarithmic operators appear in our theory, we are mostly interested in selecting the non-logarithmic ones: indeed these are the only exchanged in the OPE of the correlators under study, as the solutions of spectral equations \eqref{spec11} and \eqref{spec22} correspond to non-protected operators \(\Delta(\xi)\neq \Delta_0\).

 \subsection{\(\tr[\phi_1^2]\) sector}
\label{mix1}
This first sector is characterized  by the Cartan R-charge of two \(\phi_1\) fields \((2,0,0)\). The equation \eqref{spec11} shows physical solutions for every even twist. In particular there is only one solution at twist-2 \eqref{D2S1}, and two at twist-4 \eqref{D4S1} and higher \eqref{DtS1} both for spin \(S=0\) and \(S>0\). The twist-2 solution is easily interpreted as the scaling dimension of \begin{align}
\tr[\phi_1(n \cdot \partial)^S \phi_1] + \text{permutations}\qquad S=0,2,\dots\,,
\end{align}
indeed for any \(S\) there is no other twist-2 conformal primary with charge \((2,0,0)\).
On the other hand for \(\Delta_0-S=t \geq 4\) we can list several primaries with the right set of Cartan's charges. 
Let us concentrate on the scalar case \(S=0\) of twist four; we find \(9\) scalar conformal primaries which have the right set of charges\begin{align*}
&\mathcal{O}_1=\tr[\phi_1^3 \phi_1^\dagger]\qquad \mathcal{O}_j=\tr[\phi_1^2 \phi_j\phi_j^\dagger]\qquad \mathcal{O}_{2+j}=\tr[\phi_1^2 \phi_j^\dagger \phi_j]\qquad\mathcal{O}_{4+j}=\tr[\phi_1 \phi_j\phi_1 \phi_j^\dagger]\\ &\mathcal{O}_{8}=\tr[\bar \psi_{2}\bar \psi_3 \phi_1]\qquad \mathcal{O}_{9}=\tr[\bar \psi_{3}\bar \psi_2\phi_1],\,\quad j=2,3.
\end{align*}
As also the structure of \eqref{phi1phi1} shows, this sector is fully described in terms of the \(\chi_0\)CFT, thus the mixing transitions are realized by the vertices of \eqref{DSprime}.
At any coupling \(\mathcal{O}_1\) shows no planar transitions, and we deal with a set of \(8\) conformal primaries which at non-zero couplings \(\xi_2,\xi_3\) mix among themselves. This fact is apparently in contrast with the presence of only two twist-\(4\) exchanged operators in the OPE expansion of Sec.\eqref{fpoint1}, and can be explained with the arising of logarithmic multiplets of operators with \(\Delta=\Delta_0=4\), not being solutions of \eqref{spec11}. Indeed, for instance, the following planar transitions
\begin{align}
\label{npl1}
\mathcal{O}_{4+j} \longrightarrow \mathcal{O}_{2},\mathcal{O}_{5} \qquad \mathcal{O}_8  \longrightarrow \mathcal{O}_{4},\mathcal{O}_{3} ,
\end{align}
can happen respectively starting from order \(\xi^2\) and \(\xi^3\), while they lack the hermitian conjugate due to chirality of \eqref{DSprime}. One can actually check that there is no conjugate transition to \eqref{npl1} at any order. This suggest that matrix \(\gamma_{ij}\) won't be diagonalizable and presents Jordan blocks in its canonical form, i.e. logarithmic operators.

\subsection{\(\tr[\phi_1 \phi_{2}^{\dagger}]\) sector}
\label{appendix:sec2}
In this sector the Cartan's \(U(1)^{\otimes 3}\) charge is the difference between those of \(\phi_1\) and \(\phi_{2}\), that is \((1,-1,0)\).
The solutions of equation \eqref{spec22} are three in the scalar case \(S=0\), one of twist-\(2\) \eqref{D2S20} and two of twist-\(4\) \eqref{D4S20}. Interestingly, for tensors \(S>0\) there is an additional solution of twist-\(2\) \eqref{D2S2}. 
The explanation of such difference is that there is a unique way to realize a scalar \(\Delta_0=2\) operator with \((1,-1,0)\) charge, that is \(\tr[\phi_1 \phi_{2}^{\dagger}]\), while for higher rank tensors \(S>0\) two operators can mix up.  Starting from \(S=1\) we find that in addition to \(\Tr[\phi_1 (n \cdot \partial)  \phi_2^{\dagger}]\), the operator \(\Tr[\psi_1 (n \cdot \sigma) \bar \psi_2]=-\Tr[\bar \psi_2 (n \cdot \bar \sigma) \psi_1]\) satisfy the constraint of quantum numbers and behave as an \(SO(4)\) vector.  Therefore operator mixing happens among two conformal primaries
\begin{align}
\mathcal{O}_1=\tr[\phi_1 (n \cdot \partial)^S \phi_2^{\dagger}]\qquad\text{and}\qquad \mathcal{O}_2=\tr[\psi_1 (n \cdot \sigma)(n\cdot \partial)^{S-1} \bar\psi_2]\,,
\end{align} 
at any \(S\,\in\,2\mathbb{N}_+\). Indeed, being interested in symmetric traceless tensors (type I~\cite{Dobrev:1977qv}), no other bilinear covariant made of fermions can enter the game at higher \(S\). 
The allowed planar transitions are:
\begin{align}
\mathcal{O}_1 \longrightarrow \mathcal{O}_2 \qquad \mathcal{O}_2 \longrightarrow \mathcal{O}_1\,,
\end{align}
together with the ``diagonal" terms \(\langle \mathcal{O}_i(x) \mathcal{O}_i^{\dagger}(0)\rangle\), for \(i=1,2\). Dealing with single trace operators made of two elementary fields the effect of chiral interaction is cancelled by the trace cyclicity. The resulting transitions are symmetric and the anomalous dimension matrix is diagonalizable. Recalling the couplings  redefinition \eqref{coupdef2} and \eqref{zetakappa}, its first perturbative order \(\xi^2\) should coincide with:
\begin{align}
\gamma'_{ij}\,=\,(S \gamma S^{-1})_{ij}\,=\,\left(
\begin{array}{cc}
 -\frac{\lambda}{2}\left[\lambda-\frac{\zeta}{\sqrt{S(S+1)}}\right] & 0\\
 0 &\frac{\lambda}{2}\left[\lambda+\frac{\zeta}{\sqrt{S(S+1)}}\right] \end{array}
\right)\,.
\end{align}
As a result the two physical operators exchanged in the OPE of \eqref{sector2} have the form
\begin{align}\label{Oprimepm}
\mathcal{O'}_{\pm}\,=\, \pm c_{1,S}(\mu,\lambda) \mathcal{O}_1 + \left(\frac{1}{2}\pm c_{2,S}(\mu,\lambda)\right) \mathcal{O}_2\,,
\end{align}
and correspond to solutions \eqref{D2S2}.\\
At twist-\(4\), there are already for the scalar case \(S=0\) several local operators satisfying the constraints of bare dimension \(\Delta_0=4\) and Cartan's charge \((1,-1,0)\). We can compactly list them
\begin{align*}
&\mathcal{O}_j=\tr[\phi_1 \phi_2^\dagger \phi_j \phi_j^\dagger]\qquad  \mathcal{O}_{j+3}=\tr[\phi_1 \phi_2^\dagger \phi_j^\dagger \phi_j]\qquad \mathcal{O}_{j+6}=\tr[\phi_2^\dagger \phi_1 \phi_j \phi_j^\dagger] \quad j=1,2,3\\
 &\mathcal{O}_{10}=\tr[\phi_1 \phi_3 \phi_2^\dagger \phi_3^\dagger]\qquad  \mathcal{O}_{11}=\tr[\phi_1 \phi_3^\dagger \phi_2^\dagger \phi_3] \qquad  \mathcal{O}_{12}=\tr[\phi_2^\dagger\phi_1 \phi_3^\dagger \phi_3]\\ &\mathcal{O}_{13}=\tr[\bar \psi_2 \bar \psi_3\phi_1^\dagger] \quad \mathcal{O}_{14}=\tr[\bar \psi_3 \bar \psi_2 \phi_1^\dagger] \qquad \mathcal{O}_{15}=\tr[{\psi}_1 \psi_3 \phi_1]\qquad \mathcal{O}_{16}=\tr[{\psi}_3 \psi_1 \phi_1]\, .
\end{align*}
At non-zero couplings \(\{\xi_1,\xi_2,\xi_3\}\) these \(16\) operators mix into linear combinations obtained by the analysis of their transitons. As it happens in the first sector (\ref{mix1}), one can check that the effect of chiral interaction breaks the hermiticity of \(\gamma_{ij}\) matrix, preventing its diagonalization. Its associated Jordan canonical form \(\gamma_{ij}'\) will contain several log-multiplets of various rank, together with two diagonal elements, the anomalous dimensions \eqref{D4S20}.

\bibliographystyle{nb}

\bibliography{biblio_ChiCFT}
\end{document}